\batchmode
\documentclass[english,A4,12pt,superscriptaddress, nofootinbib, prd]{revtex4-2}
\usepackage[T1]{fontenc}
\usepackage[latin9]{inputenc}
\usepackage{color}
\usepackage{ulem}
\usepackage{xcolor}
\usepackage{babel}
\usepackage{amssymb}
\usepackage{amsmath,bm}
\usepackage{esint}
\usepackage[pdftex,unicode=true,pdfusetitle,
 bookmarks=true,bookmarksnumbered=false,bookmarksopen=false,
 breaklinks=false,pdfborder={0 0 0},pdfborderstyle={},backref=false,colorlinks=true]
 {hyperref}
\hypersetup{
 linkcolor=blue, citecolor=cyan}

\newcommand{\modSF}[1]{{{\color{violet}{#1}}}}
\makeatletter
\usepackage{slashed}
\allowdisplaybreaks[2]
\usepackage{braket}
\usepackage[a4paper,scale=0.8]{geometry}

\makeatother

\begin{document}
\title{Spin polarization and spin alignment from quantum kinetic theory with self-energy corrections}
\author{Shuo Fang}
\email{fangshuo@mail.ustc.edu.cn}

\affiliation{Department of Modern Physics, University of Science and Technology
of China, Anhui 230026}

\author{Shi Pu}
\email{shipu@ustc.edu.cn}
\affiliation{Department of Modern Physics, University of Science and Technology
of China, Anhui 230026}
\affiliation{Southern Center for Nuclear-Science Theory (SCNT), Institute of Modern Physics, Chinese Academy of Sciences, Huizhou 516000, Guangdong Province}

\author{Di-Lun Yang}
\email{dlyang@gate.sinica.edu.tw}

\affiliation{Institute of Physics, Academia Sinica, Taipei 11529}
\begin{abstract}
We derive the quantum kinetic theory
for massive fermions with collision terms and self-energy corrections based on quantum field theory.
We adopt an effective power counting scheme with $\hbar$ expansion to obtain the leading-order perturbative
solutions of the vector and axial Wigner functions and the corresponding kinetic equations. 
We observe that both the onshell relation and the structure of Wigner functions, along with the kinetic equations, are modified due to the presence of self-energies and their space-time gradients.
We further apply our formalism to investigate the spin polarization phenomena in relativistic heavy ion collisions and derive
the modification to the spin polarization spectrum of massive quarks. We find that the gradient of vector self-energy
plays a similar role to the background electromagnetic fields, which induces a more dominant contribution than the collisional effects by a naive power counting in the gradient expansion and weak coupling.
Our findings could further modify the spin polarization of strange quarks and spin alignment of $\phi$ mesons beyond local thermal equilibrium.
\end{abstract}
\maketitle 

\section{Introduction }

Nowadays, the quantum transport of relativistic matter with chirality or spin under
extreme conditions has attracted enormous interest because of its connection with fundamental properties of the underlying
quantum field theory such as the chiral anomaly
and spin-orbital coupling. A strongly coupled hot dense matter, known
as quark-gluon plasmas (QGP), produced in relativistic heavy ion collisions
\citep{Rischke:2003mt,Gyulassy:2004vg,Shuryak:2004cy} provides us
 a platform to investigate these intriguing phenomena \citep{Kharzeev:2015znc} such as the chiral magnetic effect (CME) \citep{Vilenkin:1980fu,Nielsen:1983rb,Kharzeev:2004ey,Kharzeev:2007jp,Fukushima:2008xe,Huang:2015oca,Kharzeev:2020jxw, Gao:2020vbh, Hidaka:2022dmn} for electric currents propagating along the strong magnetic fields from non-central collisions \citep{Deng:2012pc,Tuchin:2013ie, Roy:2015kma, Pu:2016ayh, Pu:2016bxy, Siddique:2019gqh, Peng:2022cya} in the presence of local chirality imbalance led by topological fluctuations of quantum-chromodynamics (QCD) vacuum at finite temperature \citep{Manton:1983nd,Klinkhamer:1984di,McLerran:1990de,Arnold:1996dy}.
On the other hand, large orbital angular momenta generated in non-central collisions could also produce strong vortical fields \citep{Liang:2004ph,Liang:2004xn,Gao:2007bc,Becattini:2007sr}, which may further induce spin polarization of QGP or the chiral vortical effect (CVE) \citep{Vilenkin:1979ui,Banerjee:2008th,Erdmenger:2008rm,Torabian:2009qk,Son:2009tf, Pu:2010as, Gao:2012ix} for electric currents propagating along the vortical fields. However, in realistic conditions, one has to study such transport phenomena beyond thermal equilibrium, for which the novel quantum transport theories were developed.

The chiral kinetic theory (CKT) in terms of one-particle distribution
functions can microscopically describe transport phenomena of massless chiral
fermions. It can be derived based on the Hamiltonian
approach \citep{Son:2012zy,Son:2012wh,Lin:2019ytz}, the path integrals \citep{Stephanov:2012ki,Chen:2013iga,Chen:2014cla,Chen:2015gta},
Wigner function formalism with the $\hbar$ expansion \citep{Gao:2012ix,Chen:2012ca,Hidaka:2016yjf, Hidaka:2017auj, Huang:2018wdl, Hidaka:2022dmn},
world-line formalism \citep{Mueller:2017arw,Mueller:2017lzw, Copinger:2020nyx}, and
effective field theories \citep{Son:2012zy,Manuel:2013zaa,Manuel:2014dza,Carignano:2018gqt,Carignano:2019zsh,Lin:2018aon,Lin:2019ytz}.
Using CKT, one can study the aforementioned phenomena in and out of equilibrium
\citep{Pu:2014fva,Pu:2014cwa, Chen:2016xtg,Gorbar:2017toh,Hidaka:2017auj,Dayi:2017xrr,Ebihara:2017suq,Huang:2017tsq,Hidaka:2018ekt,Yang:2018lew,Liu:2020ymh}.
Furthermore, the numerical simulations for chiral transport far from equilibrium were also conducted in the context of heavy ion collisions  \citep{Sun:2016nig,Sun:2016mvh,Sun:2017xhx,Sun:2018bjl,Liu:2019krs,Zhou:2018rkh,Zhou:2019jag}.
There have been the extension of the CKT to curved spacetime \citep{Liu:2018xip,Hayata:2020sqz,Yamamoto:2020zrs}
and to incorporate the second-order quantum corrections \citep{Yang:2020mtz,Hayata:2020sqz,Mameda:2023ueq}. 
Besides these applications in relativistic heavy ion collisions, the CKT is also applied to astrophysics
\citep{Kamada:2022nyt} for chiral transport of leptons in compact stars \citep{Yamamoto:2015gzz,Yamamoto:2020zrs,Yamamoto:2021hjs,Yamamoto:2022yva,Yamamoto:2023okm} and also to Weyl and Dirac semi-metals \citep{Armitage:2017cjs,Gorbar:2021ebc}.
A generalization of CKT to the massive fermions in terms of
quantum kinetic theory (QKT) for spin transport has also been developed these years.
QKT for covariant Wigner functions can be derived from the underlying
quantum field theory in terms of Wigner functions \citep{Gao:2019znl,Weickgenannt:2019dks,Weickgenannt:2020aaf,Hattori:2019ahi,Yang:2020hri,Liu:2020flb,Weickgenannt:2021cuo,Sheng:2021kfc,Wang:2019moi,Manuel:2021oah,Lin:2021mvw,Yang:2021fea,Ma:2022ins}
and its extension to the photons has also been discussed \citep{Huang:2020kik,Hattori:2020gqh,Mameda:2022ojk}. More details can be found in recent review \citep{Hidaka:2022dmn} and references therein.

Nevertheless, in the previous studies of QKT, only the imaginary part of the retarded self-energy contributing to the collision term for kinetic equations was considered,
but the real part and one-point potential associated with the dispersion relations were dropped. From now on, we will refer the former as the \textit{collisional effect} and the latter as the \textit{self-energy correction} for clarity. Although the
role of the self-energy correction has 
been discussed in the conventional kinetic theory in terms of Wigner functions \citep{Mrowczynski:1992hq,Blaizot:1999xk,Blaizot:2001nr}, it has been recently shown that its involvement with the quantum corrections up to $\mathcal{O}(\hbar)$ in CKT lead to further modifications upon the Wigner functions and kinetic equations and result in novel transport phenomena, such as the neutrino spin Hall effect for chiral fermions \citep{Yamamoto:2023okm}. On the theoretical side, it is curious how similar corrections could affect the QKT and spin transport phenomena for massive fermions. 
It serves as one of the motivations in this work on theoretical side. 


On the phenomenological side, the spin polarization effects in relativistic heavy ion collisions \citep{Becattini:2020ngo,Gao:2020vbh,Becattini:2022zvf}
has been recently observed by e.g. STAR and ALICE collaborations, and their
experimental results show the global and local polarization of hyperons
\citep{STAR:2017ckg,STAR:2019erd,ALICE:2019aid,STAR:2020xbm} and
spin alignment of vector mesons \citep{STAR:2022fan,ALICE:2019aid,ALICE:2023jad}.
As a unified framework, Wigner functions in the context of QKT are
widely applied to these spin polarization effects. The spin polarization
spectrum in terms of the modified Cooper-Frye formula can be derived using the quantum statistical model with the help of covariant Wigner functions
\citep{Becattini:2007nd,Becattini:2007sr,Becattini:2013fla,Becattini:2020sww}
or directly derived from the formal solution of Wigner function \citep{Fang:2016vpj,Yi:2021ryh,Liu:2021nyg}.
The experimental data of  global
polarization for $\Lambda$ and $\overline{\Lambda}$ hyperons can be successfully described by numerical simulations
\citep{Karpenko:2016jyx,Becattini:2017gcx,Xie:2017upb,Pang:2016igs,Li:2017slc,Wei:2018zfb,Ryu:2021lnx,Shi:2017wpk,Fu:2020oxj,Sun:2017xhx,Wu:2022mkr,Alzhrani:2022dpi,Xu:2022hql}
based on this formula. 
In addition, the local polarization of $\Lambda$ and $\overline{\Lambda}$ hyperons along
the longitudinal direction of beam line is not fully understood \citep{Becattini:2017gcx,Xia:2018tes} ------
although it is found that the effects beyond global equilibrium such as
the shear tensor and gradient of chemical potential over temperature play a role
\citep{Liu:2020dxg,Liu:2021uhn,Becattini:2021suc,Hidaka:2017auj,Becattini:2021iol,Fu:2021pok,Yi:2021ryh}, also see the discussion on the choice of parameters and equations of state \citep{Yi:2021ryh,Florkowski:2021xvy,Sun:2021nsg,Wu:2022mkr,Alzhrani:2022dpi}. 
Moreover, the spin alignment of vector mesons remains an open question albeit the alignment of $\phi$ mesons
at relativistic heavy ion collisions energies could be qualitatively 
explained by the vector-meson fields \citep{Sheng:2019kmk,Sheng:2020ghv,Sheng:2022ffb,Sheng:2022wsy,Sheng:2023urn} and
the spin alignment at LHC energies might be related to color fields from anisotropic QGP or glasma \citep{Muller:2021hpe,Yang:2021fea,Kumar:2022ylt,Kumar:2023ghs}, 
also see the studies based on NJL models \citep{Sheng:2022ssp}, spin hydrodynamics for vector mesons \citep{Wagner:2022gza}, light front quarks \citep{Fu:2023qht} and discussion on shear tensor induced spin alignment \citep{Wagner:2022gza,Li:2022vmb}.

There have also been persistent studies for off-equilibrium (more precisely near local-equilibrium) corrections upon local spin polarization from the collisional effects of QKT \citep{Wang:2020pej,Wang:2021qnt,Sheng:2021kfc,Weickgenannt:2022zxs,Weickgenannt:2022qvh,Wagner:2022amr,Fang:2022ttm,Wang:2022yli,Lin:2022tma}, but the self-energy corrections have not been considered and they may be more dominant according to the simple power counting from the gradient expansion in weak coupling as inferred by Ref.~\citep{Yamamoto:2023okm}. It is hence important to investigate the self-energy effects upon local spin polarization for the phenomenological purpose.

In this work, starting from quantum field theory with background $\mathrm{U}(1)$
field and using Schwinger-Keldysh (SK) formalism, we derive the 
leading-order QKT tracking the entangled dynamics of the vector-charge and spin transport with collisions and self-energy corrections under a prescribed power counting as the generalization of Ref.~\cite{Yang:2020hri}. We find
the additional self-energies not only modify the dispersion relations
of Wigner functions but also their structure and kinetic equations. Notably, the axial
Wigner function of our interest receives quantum corrections from self-energy gradients as new sources for spin polarization. Our main results for the vector and axial-vector Wigner functions and corresponding quantum kinetic equations are shown in Eqs.~(\ref{eq:Vector WF formal solution}), (\ref{eq:fV kinetic eq}), (\ref{eq:Axial WF formal solution}), and (\ref{eq:afA_kinetic eq}). See also Eq.~(\ref{eq:Rest_frame_AKE}) for a simplified version as the replacement of Eq.~(\ref{eq:afA_kinetic eq}) for practical applications. 
Focusing on just non-dynamical contributions, we further investigate the self-energy gradients in field theories for phenomenological applications of spin polarization and spin alignment in relativistic heavy ion collisions.
We find the one-point potential $\Sigma^{\delta}(x)$
can dynamically generate the spin polarization of quarks from background
meson fields shown in Eq.~(\ref{eq:MF_Modif_spin_vector})
as a theoretical support for the model proposed in Refs. ~\citep{Sheng:2019kmk,Sheng:2020ghv,Sheng:2022wsy,Sheng:2022ffb, Sheng:2023urn}.
On the other hand, the retarded self-energy $\Sigma^{\mathrm{r}}(q,x)$ from a thermal QCD background
modifies the spin Cooper-Frye formula as Eq.~(\ref{eq:Modified_C-F_formula}),
where the novel interaction-dependent corrections are of the same order in gradient as those in local equilibrium like the thermal-shear correction. Such non-equilibrium corrections are hence important for local spin polarization.

The paper is organized as follows: In Sec.\ref{sec:Quantum-kinetic-theory},
we derive the Kadanoff-Baym equation for the Wigner functions and
further derive the master equations including the self-energies for
the components in Clifford basis. Then we obtain the perturbative solutions
with self-energy corrections up to the first order under the same
$\hbar$ power counting adopted in Refs.~\citep{Yang:2020hri,Yamamoto:2023okm} and we further derive the scalar kinetic equation (SKE) and axial kinetic equation (AKE) based on these solutions.
In Sec.\ref{sec:Self-energy-corrections-to}, we discuss the physical
consequences of these self-energy corrections including the tadpole
part and retarded self-energy part and obtain the corresponding modification
to the spin polarization pseudovector, and we find the one-loop self-energy
correction is of first order in gradient and with coupling dependence. Simple numerical estimation has shown that these self-energy corrections are important at low transverse momenta in the $s$ equilibrium scenario. We also comment on the potential application to the spin alignment based on the 
coalescence models. Finally, we present our conclusions and outlook in Sec.~\ref{sec:Conclusions-and-outlook}.
Some technical details are presented in Append.~\ref{sec:Master-equations-of},
\ref{sec:The- mean-field- contributions-in } and \ref{sec:Feynman-vector-Wigner}.

Throughout this paper, several conventions and notations below are adopted. 
We use the nature unit $k_{B}=c=1$ but keep $\hbar$ in \textit{all}
physical quantities if not emphasized. We apply the most minus Minkowski
metric as $\eta^{\mu\nu}=\eta_{\mu\nu}=\mathrm{diag}(+,-,-,-)$. We
work in the Weyl representation of $\gamma$ matrices with $\gamma^{\mu}=\left(\begin{array}{cc}
 & \sigma^{\mu}\\
\overline{\sigma}^{\mu}
\end{array}\right)$ and 
$\gamma^5 = \textrm{diag}\{-1,-1,+1,+1\}$.
Here the Pauli matrices are denoted as $\sigma^{\mu}=(1,\mathbf{\sigma})$
and $\overline{\sigma}^{\mu}=(1,-\mathbf{\sigma})$ with $\sigma^{i}$
($i=1,2,3$) being the Pauli matrices. 
Next, we introduce the conventions for the tensor decomposition. For a rank-$2$ tensor $A^{\mu\nu}$,
we introduce the symmetric
and anti-symmetric symbols as,
\begin{equation}
    A_{(\mu\nu)}=(A_{\mu\nu}+A_{\nu\mu})/2,\; A_{[\mu\nu]}=(A_{\mu\nu}-A_{\nu\mu})/2,
\end{equation}
respectively. 
We define
the projection operator $\Delta_{\mu\nu}=\eta_{\mu\nu}-u_{\mu}u_{\nu}$,
which is orthogonal to a time-like vector $u^{\mu}$, and 
\begin{equation}
    A^{\langle\mu\rangle}=A_{\perp}^{\mu}=\Delta^{\mu\nu}A_{\nu}.
\end{equation}
A symmetric and traceless operator orthogonal to $u^{\mu}$ is defined
as $\Delta^{\mu\nu\alpha\beta}=(\Delta^{\mu\alpha}\Delta^{\nu\beta}+\Delta^{\mu\beta}\Delta^{\nu\alpha})/2-(\Delta^{\mu\nu}\Delta_{\alpha\beta})/3$
and
\begin{equation}
    A^{\langle\mu\nu\rangle}=\Delta_{\alpha\beta}^{\mu\nu}A^{\alpha\beta}.
\end{equation}
The decomposition of $\partial_{\mu}u_{\nu}$
is
\begin{eqnarray}
\partial_{\mu}u_{\nu} & = & \sigma_{\mu\nu}+\omega_{\mu\nu}+u_{\mu}Du_{\nu}+\frac{1}{3}\Delta_{\mu\nu}\theta,\label{eq:fluid derivative}
\end{eqnarray}
where we define the following hydrodynamical symbols, $\theta=\partial_{\mu}u^{\mu}$
the expansion scalar, $D=u_{\mu}\partial^{\mu}$ the comoving derivative,
$\sigma^{\mu\nu}=\Delta_{\alpha\beta}^{\mu\nu}\partial^{\alpha}u^{\beta}$
the shear tensor, $\omega^{\mu\nu}=\Delta^{\mu\alpha}\Delta^{\nu\beta}\partial_{[\alpha}u_{\beta]}\equiv\epsilon^{\mu\nu\rho\sigma}u_{\rho}\omega_{\sigma}$
the fluid vorticity tensor with $\omega_{\sigma}=\frac{1}{2}\epsilon_{\sigma\alpha\beta\gamma}u^{\alpha}\partial^{\beta}u^{\gamma}$
being the vorticity vector.

\section{Quantum kinetic theory with self-energy corrections\label{sec:Quantum-kinetic-theory}}
In this section, we first review the derivation of the Kadanoff-Baym (KB) equations of Dirac fermions from the real-time formalism in Sec.~\ref{subsec:real-time}. Then we further construct the gauge-invariant expressions in phase space in Sec.~\ref{subsec:KB-equations-of}. In Sec.~\ref{subsec:master_eq}, by making the decomposition of Wigner functions in terms of Clifford basis with the prescribed power counting, we derive the master equations governing the vector and axial-vector components of Wigner functions, which are responsible for the energy or charge transport and dynamical spin polarization, respectively. In Sec.~\ref{subsec:Solutions-to-Wigner}, the perturbative solutions for the vector and axial-vector Wigner functions with self-energy corrections are obtained and the corresponding kinetic equations are derived in Sec.~\ref{subsec:Axial-kinetic-equation}.

\subsection{Kadanoff-Baym equations from the real time formalism\label{subsec:real-time}}
Let us start from the Lagrangian,
\begin{eqnarray}
\mathcal{L}_{j} & = & \overline{\psi}(i\hbar\gamma^{\mu}D_{\mu}-m)\psi+ Qe\overline{\psi}\gamma^{\mu}a_{\mu}\psi+g\overline{\psi}\gamma^{\mu}\mathfrak{a}_{\mu}^{a}t^{a}\psi-\frac{1}{4}(\mathcal{F}_{\mu\nu}^{a}[\mathfrak{a}])^{2}-\frac{1}{4}(F_{\mu\nu}[A+a])^{2}\nonumber \\
 &  & \qquad+\mathcal{L}_{\mathrm{FP}}-\overline{\eta}\psi-\overline{\psi}\eta-j^{\mu}a_{\mu}-\mathfrak{j}^{a,\mu}\mathfrak{a}_{\mu}^{a},\label{eq:Lagrangian}
\end{eqnarray}
where $\psi$ is the fermionic field operator, $m$ is the fermionic mass, $e$ and $g$ are the electric and strong coupling constants, respectively, and $Q$ is the number of electric charge carried by the fermions. Here we have chosen the background field gauge \citep{Peskin:1995ev,Blaizot:2001nr}
by decomposing the gauge fields into the classical part $A_{\mu},\mathfrak{A}_{\mu}^{a}$
and the quantum part $a_{\mu},\mathfrak{a}^{a}_{\mu}$ with $\langle a_{\mu}\rangle=\langle\mathfrak{a}_{\mu}^{a}\rangle=0$ for the U(1) and gluonic gauge fields, respectively. Here the superindices $a$ represent colors and $\langle\hat{O}\rangle$ represents the ensemble average of operator $\hat{O}$. 
We have set the background gluon field $\mathfrak{A}_{\mu}^{a}$ to
be zero in this work for simplicity and the covariant derivative reads
$D_{\mu}=\partial_{\mu}-i\hbar^{-1}QeA_{\mu}$. $\mathcal{F}^a_{\mu\nu}$ and $F_{\mu\nu}$ are the gluonic and electromagnetic field strengths, respectively. The $\mathcal{L}_{\mathrm{FP}}$
is the Fadeev-Popov Lagrangian containing the gauge fixing term and
the ghost part. The last four terms are the external sources describing
the non-equilibrium effects of the system, with $\overline{\eta}$ ($\eta$), $j^{\mu}$, $\mathfrak{j}^{a,\mu}$ being the external currents coupled to $\psi$ ($\overline{\psi}$),  $a_\mu$ and $\mathfrak{a}^{a,\mu}$, respectively. We can further define the
generating functional along the Schwinger-Keldysh (SK) contour $\mathrm{C}$ \citep{Blaizot:1999xk,Blaizot:2001nr},
\begin{eqnarray}
Z[j] & = & \int[\mathcal{D}\phi]\exp\left[\frac{i}{\hbar}\int_{\mathrm{C}}\mathrm{d}^{4}z\mathcal{L}_{j}\right],
\end{eqnarray}
where $j$ denotes the external sources and $\mathcal{D}\phi$ represents
the integral measure of the fields. In this work, we shall only focus on the transport of fermions. Now, the fermion two-point function
can be defined as, 
\begin{eqnarray}
\langle\widetilde{T}_{\mathrm{C}}\psi(x)\overline{\psi}(y)\rangle & = & \frac{1}{Z[j]}\left.\frac{(i\hbar)^{2}\delta^{(2)}Z[j]}{\delta\overline{\eta}(x)\delta\eta(y)}\right|_{j=0}=\theta_{\mathrm{C}}(x_{0},y_{0})S^{>}(x,y)+\theta_{\mathrm{C}}(y_{0},x_{0})S^{<}(x,y),
\end{eqnarray}
with $\widetilde{T}_{\mathrm{C}}$ and $\theta_{\mathrm{C}}$ being
the time-ordering operator and unit step function along $\mathrm{C}$.
Here the operators are taken as the ensemble average. The connected
Green's function is defined as
\begin{eqnarray}
G(x,y) & = & \left.\frac{(i\hbar)^{2}\delta^{(2)}\ln Z[j]}{\delta\overline{\eta}(x)\delta\eta(y)}\right|_{j=0}=\theta_{\mathrm{C}}(x_{0},y_{0})G^{>}(x,y)+\theta_{\mathrm{C}}(y_{0},x_{0})G^{<}(x,y).\label{eq:GF_S-K_contour}
\end{eqnarray}

We further define the U(1) gauge invariant fermion Wigner functions
as the Wigner transform of the connected two-point Green's functions
w.r.t. the relevant spacetime difference,
\begin{eqnarray}
S_{ab}^{\lessgtr}(q,X) & = & \int\mathrm{d}^{4}Ye^{i\frac{q\cdot Y}{\hbar}}S_{ab}^{\lessgtr}(x,y);
\end{eqnarray}
with 
\begin{equation}
    X=\frac{x+y}{2},\; Y=x-y,
\end{equation}
the connected lessor and greater
Green's functions are defined as,
\begin{eqnarray}
S_{\alpha\beta}^{<}(x,y) & = & -\langle\overline{\psi}_{\beta}(y)U(y,x)\psi_{\alpha}(x)\rangle_{\mathrm{c}},\label{eq:Lessor_2-point_function}\\
S_{\alpha\beta}^{>}(x,y) & = & \langle\psi_{\alpha}(x)U^{\dagger}(x,y)\overline{\psi}_{\beta}(y)\rangle_{\mathrm{c}},\label{eq:Greater_2-point_function}
\end{eqnarray}
where  $\langle...\rangle_\mathrm{c}$ denotes the ensemble average of operators which subtracts the disconnected diagrams. The gauge link for the U(1) field is defined as 
\begin{equation}
    U(A;y,x)=\mathcal{P}\exp\left[i\frac{Qe}{\hbar}\int_{x}^{y}\mathrm{d}z^{\mu}A_{\mu}(z)\right],
\end{equation}
with $Qe$ being the conserved charge. We follow Refs.~\citep{Elze:1986qd,Vasak:1987um} to choose
the straight line for the gauge link. 
Note that,  there is an
extra minus sign in the definition
for lessor function in Eq.~(\ref{eq:Lessor_2-point_function}),compared to greater one in Eq.~(\ref{eq:Greater_2-point_function})
due to the anti-commutation of fermion field operators as a convention.

Making variation w.r.t. $\eta,\overline{\eta}$ to the equations of motion (EoMs) for mean fermionic fields, we get the Dyson-Schwinger equations for the Green's functions
\begin{eqnarray}
(i\hbar\gamma^{\mu}D_{x,\mu}-m)G(x,y)+i\hbar\int_{\mathrm{C}}\mathrm{d}^{4}z\Sigma(x,z)G(z,y) & = & i\hbar\delta_{\mathrm{C}}^{(4)}(x-y),\label{eq:GF_D-S eq 1}\\
G(x,y)(-i\hbar\gamma^{\mu}D_{y,\mu}^{\dagger}-m)+i\hbar\int_{\mathrm{C}}\mathrm{d}^{4}zG(x,z)\Sigma(z,y) & = & i\hbar\delta_{\mathrm{C}}^{(4)}(y-x),\label{eq:GF_D-S eq 2}
\end{eqnarray}
where the fermion self-energy is defined as, 
\begin{eqnarray}
\Sigma(x,z) & = &-\frac{i}{\hbar}\frac{\delta}{\delta\langle\psi(z)\rangle}\left\langle\frac{\partial\mathcal{L}_{\mathrm{int}}}{\partial\overline{\psi}}(x)\right\rangle,\label{eq:Def1_SE}\\
\Sigma(z,y) & = & -\frac{i}{\hbar}\frac{\delta}{\delta\langle\overline{\psi}(z)\rangle}\left\langle\frac{\partial\mathcal{L}_{\mathrm{int}}}{\partial\psi}(y)\right\rangle,\label{eq:Def2_SE}
\end{eqnarray}
and $\mathcal{L}_{\mathrm{int}}$ represents the interaction terms between the fermionic field and quantum gauge fields in the Lagrangian (\ref{eq:Lagrangian}). The self-energies along the SK contour reads
\begin{eqnarray}
\Sigma(x,y) & = & -i\hbar^{-1}\Sigma^{\delta}(x)\delta_{\mathrm{C}}^{(4)}(x,y)+\theta_{C}(x_{0},y_{0})\Sigma^{>}(x,y)+\theta_{C}(y_{0},x_{0})\Sigma^{<}(x,y),\label{eq:SE_S-K_contour}
\end{eqnarray}
where we have extracted an $\hbar^{-1}$ in $\Sigma^{\delta}$ to make
the one point potential starts from $\hbar^{0}$. The $\Sigma^{\lessgtr}(x,y)$ are defined as, 
\begin{eqnarray}
\Sigma^{<}(x,y) & = & \Sigma^{+-}(x,y),\qquad\Sigma^{>}(x,y)=\Sigma^{-+}(x,y),
\end{eqnarray}
following the convention
in Ref.~\citep{Blaizot:2001nr}.
Furthermore, for a two point function $\mathcal{O}$, the retarded and advanced functions are defined as
\begin{eqnarray}
\mathcal{O}_{\mathrm{r}}(x,y) & = & i\theta_{\mathrm{C}}(x_{0},y_{0})\left[\mathcal{O}^{>}(x,y)-\mathcal{O}^{<}(x,y)\right]=i(\mathcal{O}^{++}-\mathcal{O}^{+-}),\label{eq:Retarded_quantities}\\
\mathcal{O}_{\mathrm{a}}(x,y) & = & -i\theta_{\mathrm{C}}(y_{0},x_{0})\left[\mathcal{O}^{>}(x,y)-\mathcal{O}^{<}(x,y)\right]=i(\mathcal{O}^{++}-\mathcal{O}^{-+}),\label{eq:Advanced_quantities}
\end{eqnarray}
so that
\begin{eqnarray}
\mathcal{O}_{\mathrm{r}}(x,y)-\mathcal{O}_{\mathrm{a}}(x,y) & = & i\left[\mathcal{O}^{>}(x,y)-\mathcal{O}^{<}(x,y)\right].
\end{eqnarray}

Inserting Eq.~(\ref{eq:SE_S-K_contour}) and Eq.~(\ref{eq:GF_S-K_contour})
into the EoMs of Green's functions, we obtain the following KB equations for $G^{<}$,
\begin{eqnarray}
 &  & \left[i\hbar\gamma^{\mu}D_{x,\mu}-m+\Sigma^{\delta}(x)\right]G^{<}(x,y)\nonumber \\
 & = & -\hbar\int_{-\infty}^{+\infty}\mathrm{d}^{4}z\left[\Sigma^{\mathrm{r}}(x,z)G^{<}(z,y)+\Sigma^{<}(x,z)G^{\mathrm{a}}(z,y)\right],\label{eq:KB eq 1}\\
\nonumber \\
 &  & G^{<}(x,y)\left[-i\hbar\gamma^{\mu}D_{y,\mu}^{\dagger}-m+\Sigma^{\delta}(y)\right]\nonumber \\
 & = & -\hbar\int_{-\infty}^{+\infty}\mathrm{d}^{4}z\left[G^{\mathrm{r}}(x,z)\Sigma^{<}(z,y)+G^{<}(x,z)\Sigma^{\mathrm{a}}(z,y)\right],\label{eq:KB eq 2}
\end{eqnarray}
and, for $G^{>}$,
\begin{eqnarray}
 &  & \left[i\hbar\gamma^{\mu}D_{x,\mu}-m+\Sigma^{\delta}(x)\right]G^{>}(x,y)\nonumber \\
 & = & -\hbar\int_{-\infty}^{+\infty}\mathrm{d}^{4}z\left[\Sigma^{>}(x,z)G^{\mathrm{a}}(z,y)+\Sigma^{\mathrm{r}}(x,z)G^{>}(z,y)\right],\label{eq:KB eq 3}\\
\nonumber \\
 &  & G^{>}(x,y)\left[-i\hbar\gamma^{\mu}D_{y,\mu}^{\dagger}-m+\Sigma^{\delta}(y)\right]\nonumber \\
 & = & -\hbar\int_{-\infty}^{+\infty}\mathrm{d}^{4}z\left[G^{>}(x,z)\Sigma^{\mathrm{a}}(z,y)+G^{\mathrm{r}}(x,z)\Sigma^{>}(z,y)\right].\label{eq:KB eq 4}
\end{eqnarray}



Before going one step further, let us introduce the $\hbar$ power counting
of the KB equations. 
When making a scaling
transformation $a_{\mu}\to a_{\mu}\hbar^{-1}$, $e\to\hbar e$,
and $j_{\mu}\to\hbar j_{\mu}$ in the Lagrangian (\ref{eq:Lagrangian}),
we find each interaction vertex in the Feynman rules is accompanied
with an $\hbar^{-1}$ factor, while each propagator is with $\hbar^{1}$ \citep{Itzykson:1980rh}.
Accordingly, the tadpole diagrams are of $\mathcal{O}(\hbar^{-1})$ and the one-loop (bubble) diagrams
are of $\mathcal{O}(\hbar^{0})$, while two-loop diagrams are of $\mathcal{O}(\hbar^{1})$. 
Here the tadpole and tree-level diagrams related to
the mean field approximation are dominant in the power counting. For
example, the tadpole self-energy reads 
\begin{eqnarray}
\hbar^{-1}\Sigma^{\delta} =   -i\hbar^{-1}Qe\gamma^{\mu}\langle a_{\mu}\rangle\sim\mathcal{O}(\hbar^{-1}),
\end{eqnarray}
whereas the one-loop self-energy is given by 
\begin{eqnarray}
\hbar^{0}\Sigma^{(1)} & = & -i\hbar^{-1}Qe\frac{\delta}{\delta\langle\psi(z)\rangle}\left\langle\widetilde{T}_\mathrm{C}\gamma^{\mu}a_{\mu}(x)\psi(x)\left(\frac{-i}{\hbar}\int\mathrm{d}^{4}yQe\overline{\psi}(y)\gamma^{\nu}a_{\nu}(y)\psi(y)\right)\right\rangle\nonumber \\
 & \sim & (-iQe)^2\hbar^{-2}\gamma^{\mu}\langle\widetilde{T}_\mathrm{C}\psi(x)\overline{\psi}(z)\rangle\gamma^{\nu}\langle \widetilde{T}_\mathrm{C}a_{\mu}(x)a_{\nu}(z)\rangle\nonumber \\
 & \sim & \mathcal{O}(\hbar^{0}),
\end{eqnarray}
where the $\hbar^{-2}$ prefactor is implicitly canceled by the $\hbar^2$ contribution led by two propagators above.
Generally speaking, the $\hbar$
power counting is somewhat different from the gradient expansion.

In this work, we will focus on the effects led by gradient corrections, so all $\hbar$ factors
originating from Dyson series and Feynman rules will be neglected for simplicity. That is, the $\hbar$ terms hereafter only come from the gradient corrections, which are also referred as the quantum corrections while the higher-loop corrections will not be discussed.  
In such a case, our power counting scheme is the same as Refs.~
\citep{Hidaka:2016yjf,Yang:2020hri,Yamamoto:2023okm}.

\subsection{Gauge invariant expressions in phase space\label{subsec:KB-equations-of}
}
For the U(1) gauge invariant Green's functions, we have
\begin{eqnarray}
S_{\alpha\beta}^{<}(x,y)=U(y,x)G_{\alpha\beta}^{<}(x,y), & \quad & S_{\alpha\beta}^{>}(x,y)=U^{\dagger}(x,y)G_{\alpha\beta}^{>}(x,y),
\end{eqnarray}
and insert them into the KB Eqs.~(\ref{eq:KB eq 1},\ref{eq:KB eq 2}),
the EoMs for the gauge invariant Wigner functions become 
\begin{eqnarray}
 &  & \left\{i\hbar\gamma^{\mu}\left[\partial_{x,\mu}+\frac{iQe}{\hbar}(x^{\nu}-y^{\nu})\int_{0}^{1}\mathrm{d}ssF_{\mu\nu}(z(s))\right]-m+\Sigma^{\delta}(x)\right\}S^{<}(x,y)\nonumber \\
 & = & -\hbar\int_{-\infty}^{+\infty}\mathrm{d}^{4}zP(y,x,z)\left[\Sigma_{\mathrm{g}}^{\mathrm{r}}(x,z)S^{<}(z,y)+\Sigma_{\mathrm{g}}^{<}(x,z)S^{\mathrm{a}}(z,y)\right],\label{eq:EoM_Gauge_inv_GF_1}
\end{eqnarray}
and 
\begin{eqnarray}
 &  & S^{<}(x,y)\left\{-i\hbar\gamma^{\mu}\left[\overleftarrow{\partial}_{y,\mu}+\frac{iQe}{\hbar}(x^{\nu}-y^{\nu})\int_{0}^{1}\mathrm{d}s(1-s)F_{\mu\nu}(z(s))\right]-m+\Sigma^{\delta}(y)\right\}\nonumber \\
 & = & -\hbar\int_{-\infty}^{+\infty}\mathrm{d}^{4}zP(y,x,z)\left[S^{\mathrm{r}}(x,z)\Sigma_{\mathrm{g}}^{<}(z,y)+S^{<}(x,z)\Sigma^{\mathrm{a}}(z,y)\right],\label{eq:EoM_Gauge_inv_GF_2}
\end{eqnarray}
where $z(s)=(1-s)y+sx$ and $P(y,x,z)$ is the U(1) Wilson loop 
\begin{eqnarray}
P(y,x,z) & \equiv & U(y,x)U(x,z)U(z,y)=\exp \left[i\frac{Qe}{\hbar} \oint_\mathcal{C}\mathrm{d}\xi^{\mu}A_{\mu}(\xi)\right],
\end{eqnarray}
with $\mathcal{C}$ being the triangle area circled by the straight lines connecting the points $x,y,z$. 
Similarly, we define the gauge-invariant self-energy 
as,
\begin{eqnarray}
\Sigma_{\mathrm{g}}(x,z) & \equiv & U(z,x)\Sigma(x,z).
\end{eqnarray}

Making the Wigner transform, we obtain \citep{Yang:2020hri,Hidaka:2022dmn}
\begin{eqnarray}
\left[\frac{i\hbar}{2}\gamma^{\mu}\nabla_{\mu}+\gamma^{\mu}\Pi_{\mu}-m+\Sigma^{\delta}(X)\star\right]S^{<}(q,X) & = & -\hbar\left[\Sigma_{\mathrm{g}}^{\mathrm{r}}\star S^{<}+\Sigma_{\mathrm{g}}^{<}\star S^{\mathrm{a}}\right],\label{eq:KB_Gauge_inv_fermion_WF_1}\\
S^{<}\left(-\frac{i\hbar}{2}\gamma^{\mu}\overleftarrow{\nabla}_{\mu}+\gamma^{\mu}\overleftarrow{\Pi}_{\mu}-m\right)+S^{<}\star\Sigma^{\delta}(X) & = & -\hbar\left[S^{\mathrm{r}}\star\Sigma_{\mathrm{g}}^{<}+S^{<}\star\Sigma_{\mathrm{g}}^{\mathrm{a}}\right],
\label{eq:KB_Gauge_inv_fermion_WF_2}
\end{eqnarray}
where $\star$ denotes the Moyal product for the gauge invariant quantities 
\begin{eqnarray}
 &  & f(q,X)\star g(q,X)\nonumber \\
 & = & f(q,X)\exp\left[i\hbar\frac{\overrightarrow{\partial}_{X}\cdot\overleftarrow{\partial}_{q}-\overleftarrow{\partial}_{X}\cdot\overrightarrow{\partial}_{q}}{2}+Qe\overleftarrow{\partial}_{q}^{\mu}\overrightarrow{\partial}_{q}^{\nu}G(\overleftarrow{\partial}_{q}\cdot\partial_{X},\overrightarrow{\partial}_{q}\cdot\partial_{X})F_{\mu\nu}(X)\right]g(q,X)\nonumber \\
 & = & f(q,X)g(q,X)+\frac{i\hbar}{2}f(q,X)\left(\overrightarrow{\partial}_{X}\cdot\overleftarrow{\partial}_{q}-\overleftarrow{\partial}_{X}\cdot\overrightarrow{\partial}_{q}+QeF_{\mu\nu}(X)\overleftarrow{\partial}_{q}^{\mu}\overrightarrow{\partial}_{q}^{\nu}\right)g(q,X)+\mathcal{O}(\hbar^{2}), \nonumber \\
\end{eqnarray}
with
\begin{eqnarray}
G(a,b) & = & \frac{e^{-\frac{i\hbar}{2}(a+b)}\left[(b-e^{i\hbar a}(a+b-ae^{i\hbar b})\right]}{i\hbar ab(a+b)}.
\end{eqnarray}
Here the covariant derivative and kinetic momentum are given by
\begin{eqnarray}
\nabla_{\mu} & = & \partial_{X,\mu}+j_{0}(\Delta)QeF_{\mu\nu}(X)\partial_{q}^{\nu},\qquad\overleftarrow{\nabla}_{\mu}=\overleftarrow{\partial}_{X,\mu}+j_{0}(\overleftarrow{\Delta})QeF_{\mu\nu}(X)\overleftarrow{\partial}_{q}^{\nu},\label{eq:Covariant_derivative}\\
\Pi_{\mu} & = & q_{\mu}+\frac{\hbar}{2}j_{1}(\Delta)QeF_{\mu\nu}(X)\partial_{q}^{\nu},\qquad\overleftarrow{\Pi}_{\mu}=q_{\mu}+\frac{\hbar}{2}j_{1}(\overleftarrow{\Delta})QeF_{\mu\nu}(X)\overleftarrow{\partial}_{q}^{\nu},\label{eq:Covariant_kinetic_momentum}
\end{eqnarray}
where 
\begin{equation}
    \Delta=\frac{\hbar}{2}\partial_{q}\cdot\partial_{X}, \nonumber
\end{equation}
and, $j_{0}(x)=\frac{\sin x}{x}$,
$j_{1}(x)=\frac{\sin x-x\cos x}{x^{2}}$ are the spherical Bessel functions. 
Note that the spacetime derivatives
in $\nabla_{\mu},\Pi_{\mu},\Delta$ only act on the external gauge fields and the one-point potential $\Sigma^{\delta}(X)$
here has no momentum dependence. 

Then utilizing the definition of retarded and advanced quantities defined as shown in Eqs.~(\ref{eq:Retarded_quantities},\ref{eq:Advanced_quantities}), one
can derive the following relations \citep{Mrowczynski:1992hq,Blaizot:2001nr},
\begin{eqnarray}
\mathcal{O}^{\mathrm{r}}(q,X) & = & \mathrm{Re}\mathcal{O}^{\mathrm{r}}(q,X)+i\frac{\mathcal{O}^{>}(q,X)-\mathcal{O}^{<}(q,X)}{2},\\
\mathcal{O}^{\mathrm{a}}(q,X) & = & \mathrm{Re}\mathcal{O}^{\mathrm{r}}(q,X)-i\frac{\mathcal{O}^{>}(q,X)-\mathcal{O}^{<}(q,X)}{2}.
\end{eqnarray}
Here the real part of the Wigner function is related to the
applicability of the quasi-particle approximation, which is valid when setting $\mathrm{Re}S^{\mathrm{r}}=0$ \citep{Blaizot:2001nr}.
On the other hand, the dispersion relation
of the quasi-particle is modified by the interaction for $\mathrm{Re}\Sigma^{\mathrm{r/a}}\neq0$ \citep{Blaizot:2001nr,Yamamoto:2023okm}.
In this work, we limit our discussions under the quasi-particle approximation and incorporate the modified dispersion relation to the spin polarization.
That is, we assume $\mathrm{Re}S^{\mathrm{r}}=0$ and $\mathrm{Re}\Sigma^{\mathrm{r/a}}\neq0$.
Then Eqs.~(\ref{eq:KB_Gauge_inv_fermion_WF_1},\ref{eq:KB_Gauge_inv_fermion_WF_2})
become
\begin{eqnarray}
\left[\frac{i\hbar}{2}\gamma^{\mu}\nabla_{\mu}+\gamma^{\mu}\Pi_{\mu}-m+\overline{\Sigma}_{\mathrm{g}}\star\right]S^{<}(q,X) & = & -\frac{i\hbar}{2}(\Sigma_{\mathrm{g}}^{>}\star S^{<}-\Sigma_{\mathrm{g}}^{<}\star S^{>}),\label{eq:KB_WF_QPApprox_1}\\
S^{<}\left(-\frac{i\hbar}{2}\gamma^{\mu}\overleftarrow{\nabla}_{\mu}+\gamma^{\mu}\overleftarrow{\Pi}_{\mu}-m\right)+S^{<}\star\overline{\Sigma}_{\mathrm{g}} & = & -\frac{i\hbar}{2}(S^{>}\star\Sigma_{\mathrm{g}}^{<}-S^{<}\star\Sigma_{\mathrm{g}}^{>}),\label{eq:KB_WF_QPApprox_2}
\end{eqnarray}
where 
\begin{eqnarray}
\overline{\Sigma}_{\mathrm{g}}(q,X) & = & \Sigma^{\delta}(X)+\mathrm{Re}\Sigma_{\mathrm{g}}^{\mathrm{r}}
\end{eqnarray}
with $\Sigma_{\mathrm{g}}^{\mathrm{r}}$
being the retarded one-loop self-energy.


\subsection{Master equations of Wigner functions\label{subsec:master_eq}}

We make sum and difference of Eqs.~(\ref{eq:KB_WF_QPApprox_1},\ref{eq:KB_WF_QPApprox_2})
and obtain
\begin{eqnarray}
\{(\overrightarrow{\slashed{\Pi}}-m),S^{<}\}+\{\overline{\Sigma}_{\mathrm{g}},S^{<}\}_{\star}+\frac{i\hbar}{2}\left(\frac{}{}[\gamma^{\mu},\nabla_{\mu}S^{<}]+[\Sigma_{\mathrm{g}}^{>},S^{<}]_{\star}-[\Sigma_{\mathrm{g}}^{<},S^{>}]_{\star}\right) & = & 0,\label{eq:Addition KB equation}\\{}
[(\overrightarrow{\slashed{\Pi}}-m),S^{<}]+[\overline{\Sigma}_{\mathrm{g}},S^{<}]_{\star}
+\frac{i\hbar}{2}\left(\frac{}{}\{\gamma^{\mu},\nabla_{\mu}S^{<}\}+\{\Sigma_{\mathrm{g}}^{>},S^{<}\}_{\star}-\{\Sigma_{\mathrm{g}}^{<},S^{>}\}_{\star}\right) & = & 0,\label{eq:Difference KB equation}
\end{eqnarray}
where $\{\overrightarrow{A},B\}=\overrightarrow{A}B+B\overleftarrow{A}$
and we have defined
\begin{eqnarray}
\{A,B\}_{\star}=A\star B+B\star A, & \quad & [A,B]_{\star}=A\star B-B\star A.
\end{eqnarray}
One may also introduce a similar decomposition for $\bar{\Sigma}$. 
From now on we will omit the subscripts '$\mathrm{g}$' in self-energies
for notational simplicity.

In our following calculations, we only keep terms with one $\hbar$
higher than the leading order and Eqs.~(\ref{eq:Covariant_derivative},\ref{eq:Covariant_kinetic_momentum})
hence reduce to
\begin{eqnarray}
\nabla_{\mu} & = & \partial_{X,\mu}+QeF_{\mu\nu}(X)\partial_{q}^{\nu}+\mathcal{O}(\hbar^{2}), \nonumber \\
\Pi_{\mu} & = & q_{\mu}+\mathcal{O}(\hbar^{2}).
\end{eqnarray}
We will focus on the Wigner functions and self-energies in the Clifford basis $\{1,i\gamma^{5},\gamma^{\mu},
\gamma^{5}\gamma^{\mu}$, $\gamma^{\mu\nu}=i[\gamma^{\mu},\gamma^{\nu}]/2\}$,
\begin{eqnarray}
S^{\lessgtr} & = & \mp\Big[\mathcal{F}^{\lessgtr}+i\mathcal{P}^{\lessgtr}\gamma^{5}+\gamma^{\mu}\mathcal{V}_{\mu}^{\lessgtr}+\gamma^{5}\gamma^{\mu}\mathcal{A}_{\mu}^{\lessgtr}+\frac{1}{2}\mathcal{S}_{\mu\nu}^{\lessgtr}\gamma^{\mu\nu}\Big],\label{eq:Clifford_decomposition_WF}\\
\Sigma^{\lessgtr} & = &\mp\Big[\Sigma_{\mathrm{F}}^{\lessgtr}+i\Sigma_{\mathrm{P}}^{\lessgtr}\gamma^{5}+\gamma^{\mu}\Sigma_{\mathrm{V},\mu}^{\lessgtr}+\gamma^{5}\gamma^{\mu}\Sigma_{\mathrm{A},\mu}^{\lessgtr}+\frac{1}{2}\Sigma_{\mathrm{T},\mu\nu}^{\lessgtr}\gamma^{\mu\nu}\Big].\label{eq:Clifford_decomposition_SE}
\end{eqnarray}
Here we have defined the lessor quantities with an extra minus sign due to our definition in Eqs.~(\ref{eq:Lessor_2-point_function}, \ref{eq:SE_S-K_contour}), and under such definition, the components of the decomposed Wigner functions and self-energies above are the same as those in Refs.~\citep{Hidaka:2016yjf,Hattori:2019ahi,Yang:2020hri}. We
can also introduce the Poisson bracket, 
\begin{eqnarray}
[AB]_{\mathrm{P.B.}}\equiv    \partial_{q}A\cdot\partial_{X}B-\partial_{X}A\cdot\partial_{q}B=-[BA]_{\mathrm{P.B.}},
\end{eqnarray}
if $[A,B]=0$ and a modified Poisson bracket with external fields will also be useful, 
\begin{eqnarray}
[AB]_{\mathrm{P.B.}}^{F} & = & [AB]_{\mathrm{P.B.}}+QeF_{\mu\nu}(X)\partial_{q}^{\mu}A\partial_{q}^{\nu}B.
\end{eqnarray}

Inserting the decomposed Wigner functions and self-energies like in Eqs.~(\ref{eq:Clifford_decomposition_WF},\ref{eq:Clifford_decomposition_SE})
into the KB Eqs.~(\ref{eq:Addition KB equation},\ref{eq:Difference KB equation}),
we can reduce the KB equations to $10$ independent equations by matching
the coefficients for the Clifford basis.
One can refer to Appendix.\ref{subsec:General-equations} for computational
details. For convenience, we introduce the short-hand notations, 
\begin{eqnarray}
\widehat{XY}=X^{>}Y^{<}-X^{<}Y^{>},\quad 
\{\widehat{\Sigma_{\mathrm{V},\mu},\mathcal{V}_{\nu}}\}_{\star} & = & \{\Sigma_{\mathrm{V},\mu}^{>},\mathcal{V}_{\nu}^{<}\}_{\star}-\{\Sigma_{\mathrm{V},\mu}^{<},\mathcal{V}_{\nu}^{>}\}_{\star}. 
\label{eq:def_widehat}
\end{eqnarray} 
We also introduce the following modified mass and kinetic
momentum with self-energy corrections,  
\begin{eqnarray}
\widetilde{m} =  m-\overline{\Sigma}_{\mathrm{F}},\qquad
\widetilde{q}_{\mu}  =  q_{\mu}+\overline{\Sigma}_{\mathrm{V},\mu},
\label{eq:SE modified Mass-momentum-1}
\end{eqnarray}
for $\widetilde{\Pi}_{\mu}=\Pi_{\mu}+\overline{\Sigma}_{\mathrm{V},\mu}$
and the generalized derivatives,
\begin{eqnarray}
\mathcal{D}_{\rho}A^{<} & = & \nabla_{\rho}A^{<}+\Sigma_{\mathrm{V},\rho}^{>}A^{<}-\Sigma_{\mathrm{V},\rho}^{<}A^{>}=\nabla_{\rho}A^{<}+\widehat{\Sigma_{\mathrm{V},\rho}A},\\
\widetilde{\mathcal{D}}_{\rho}A^{<} & = & \mathcal{D}_{\rho}A^{<}+[\overline{\Sigma}_{\mathrm{V},\rho}A^{<}]_{\mathrm{P.B.}}^{F}\nonumber \\
 & = & \mathcal{D}_{\rho}A^{<}-(\nabla_{\alpha}\overline{\Sigma}_{\mathrm{V},\rho})\partial_{q}^{\alpha}A^{<}+(\partial_{q,\alpha}\overline{\Sigma}_{\mathrm{V},\rho})\partial_{X}^{\alpha}A^{<}.\label{eq:curly_D-1}
\end{eqnarray}

As shown in Appendices.\ref{subsec:Keep-up-to} and \ref{subsec:Linear-order-of},
the full expressions of the master equations are rather complicated. For practical purposes, we now adopt
the power counting by assuming
\begin{eqnarray}
\mathcal{V}^{\mu}\sim\mathcal{O}(\hbar^{0}) & , & \mathcal{A}^{\mu}\sim\mathcal{O}(\hbar^{1}),\label{eq:V-A power counting}
\end{eqnarray}
based on the quantum nature of the spin polarization delineated by  $\mathcal{A}^{\mu}$ \cite{Hattori:2019ahi,Yang:2020hri}.
One may relate other components to $\mathcal{V}^{\mu}$ and $\mathcal{A}^{\mu}$ through the master equations. In the free case, one easily obtains
\begin{eqnarray}
\mathcal{F}^{<} & \sim & \frac{\widetilde{\Pi}_{\mu}}{\widetilde{m}}\mathcal{V}^{<,\mu}, \nonumber\\
\mathcal{P}^{<} & \sim & -\frac{\hbar}{2\widetilde{m}}\widetilde{\mathcal{D}}^{\mu}\mathcal{A}_{\mu}^{<}, \nonumber \\
\mathcal{S}_{\mu\nu}^{<} & \sim & -\frac{1}{\widetilde{m}}\epsilon_{\mu\nu\alpha\beta}\widetilde{\Pi}^{\alpha}\mathcal{A}^{<,\beta}+\frac{\hbar}{\widetilde{m}}\widetilde{\mathcal{D}}_{[\mu}\mathcal{V}_{\nu]}^{<},
\end{eqnarray}
which yields the hierarchy, 
\begin{eqnarray}
\mathcal{F}\sim\mathcal{O}(\hbar^{0}) & , & \mathcal{P}\sim\mathcal{O}(\hbar^{2}),\mathcal{S}_{\mu\nu}\sim\mathcal{O}(\hbar^{1}).
\end{eqnarray}
More generic relations are shown in Appendix.~\ref{subsec:Linear-order-of}, whereas the hierarchy in the $\hbar$ power counting remains unchanged.
For brevity, we will sometimes denote $\overline{\Sigma}$ and $\Sigma^{\lessgtr}$ as $\Sigma$ in the following context if not specified. Then up to the lowest order in $\hbar$, from the expression
of $\mathcal{P}^{<}$ in Eq.~(\ref{eq:Master eq 1-2}),
\begin{eqnarray}
\mathcal{P}^{<} & \sim & \frac{1}{\widetilde{m}}\left[\overline{\Sigma}_{\mathrm{P}}\frac{\widetilde{\Pi}_{\mu}}{\widetilde{m}}\mathcal{V}^{<,\mu}{\color{blue}-}\frac{\hbar}{4\widetilde{m}}\epsilon^{\mu\nu\alpha\beta}\overline{\Sigma}_{\mathrm{T},\alpha\beta}\widetilde{\mathcal{D}}_{[\mu}\mathcal{V}_{\nu]}^{<}\right],
\end{eqnarray}
and thus, we have,
\begin{eqnarray}
\Sigma_{\mathrm{P}}\sim\mathcal{O}(\hbar^{2}) & , & \Sigma_{\mathrm{T},\mu\nu}\sim\mathcal{O}(\hbar^{1}).
\end{eqnarray}
From the kinetic equation of $\mathcal{V}_{\mu}$ in Eq.~(\ref{eq:Master eq 2-1})
and the constraint equation of $\mathcal{A}_{\mu}$ in Eq.~(\ref{eq:Master eq 2-2}), we can read out
\citep{Yang:2020hri} 
\begin{eqnarray}
\Sigma_{\mathrm{F}} & \sim & \Sigma_{\mathrm{V}}^{\mu}\sim\mathcal{O}(\hbar^{0}),\quad\Sigma_{\mathrm{A}}^{\mu}\sim\mathcal{O}(\hbar^{1}).
\end{eqnarray}


Our goal is to obtain EoMs for Wigner functions up to the leading non-vanishing order in the $\hbar$ expansion.
As for some technical details,
we need to work out both sides up to $\mathcal{O}(\hbar^{1})$ for Eqs.~(\ref{eq:Master eq 1-1}, \ref{eq:Master eq 1-3}, \ref{eq:Master eq 2-1}, \ref{eq:Master eq 2-5}), 
up to $\mathcal{O}(\hbar^{2})$ for Eqs.~(\ref{eq:Master eq 1-5}, \ref{eq:Master eq 2-2}, \ref{eq:Master eq 2-3}), and 
up to $\mathcal{O}(\hbar^{3})$ for Eqs.~(\ref{eq:Master eq 1-2}, \ref{eq:Master eq 1-4}, \ref{eq:Master eq 2-4}). 
We can rewrite $\{\mathcal{F}^<,\mathcal{P}^<,\mathcal{S}^<_{\mu\nu}\}$
in terms of $\mathcal{V}^{<}_{\mu}$ and $\mathcal{A}^{<}_{\mu}$ as
\begin{eqnarray}
\mathcal{F}^{<} & = & \frac{1}{\widetilde{m}}\widetilde{q}_{\mu}\mathcal{V}^{<,\mu}+\mathcal{O}(\hbar^{2}),\label{eq:F(V,A)}\\
\mathcal{P}^{<} & = & -\frac{\hbar}{2\widetilde{m}}(\widetilde{\mathcal{D}}^{\mu}\mathcal{A}_{\mu}^{<}-\widehat{\Sigma_{\mathrm{A}}^{\mu}\mathcal{V}_{\mu}})+\frac{1}{\widetilde{m}^{2}}(\overline{\Sigma}_{\mathrm{P}}\widetilde{q}\cdot\mathcal{V}^{<}-\overline{\Sigma}_{\mathrm{T},\rho\sigma}\widetilde{q}^{\rho}\mathcal{A}^{<,\sigma}{\color{blue}-}\frac{\hbar}{4}\epsilon^{\mu\nu\alpha\beta}\overline{\Sigma}_{\mathrm{T},\alpha\beta}\widetilde{\mathcal{D}}_{\mu}\mathcal{V}_{\nu}^{<})\nonumber \\
 &  & +\frac{\hbar}{2\widetilde{m}}[\overline{\Sigma}_{\mathrm{A}}^{\mu}\mathcal{V}_{\mu}^{<}]_{\mathrm{P.B.}}^{F}+\mathcal{O}(\hbar^{4}),\label{eq:P(V,A)}\\
\mathcal{S}_{\mu\nu}^{<} & = & -\frac{1}{\widetilde{m}}\epsilon_{\mu\nu\alpha\beta}\widetilde{q}^{\alpha}\mathcal{A}^{<,\beta}+\frac{\hbar}{\widetilde{m}}\widetilde{\mathcal{D}}_{[\mu}\mathcal{V}_{\nu]}^{<}+\frac{1}{\widetilde{m}}\left(\frac{1}{\widetilde{m}}\overline{\Sigma}_{\mathrm{T},\mu\nu}\widetilde{q}_{\alpha}\mathcal{V}^{<,\alpha}+\epsilon_{\mu\nu\alpha\beta}\overline{\Sigma}_{\mathrm{A}}^{\alpha}\mathcal{V}^{<,\beta}\right)+\mathcal{O}(\hbar^{3}),\label{eq:S(V,A)}
\end{eqnarray}
where we have neglected the nonlinear terms 
of $\mathcal{O}(\Sigma^2)$ in the weak coupling limit. Under our power counting scheme, the master
equations are greatly simplified to
\begin{eqnarray}
0 & = & \widetilde{q}_{\mu}\mathcal{F}^{<}-\widetilde{m}\mathcal{V}_{\mu}^{<}+\mathcal{O}(\hbar^{2}),\\
0 & = & 
-\epsilon_{\mu\nu\alpha\beta}\widetilde{q}^{\nu}\mathcal{S}^{<,\alpha\beta}-2\widetilde{m}\mathcal{A}_{\mu}^{<}+
2\overline{\Sigma}_{\mathrm{A},\mu}\mathcal{F}^{<}-\epsilon_{\mu\nu\alpha\beta}\overline{\Sigma}_{\mathrm{T}}^{\alpha\beta}\mathcal{V}^{<,\nu}+\mathcal{O}(\hbar^{3}), \\
0 & = & 2\widetilde{q}^{\nu}\mathcal{S}_{\nu\mu}^{<}+\hbar\widetilde{\mathcal{D}}_{\mu}\mathcal{F}^{<}+\hbar\widehat{\Sigma_{\mathrm{F}}\mathcal{V}_{\mu}}+2\overline{\Sigma}_{\mathrm{T},\mu\nu}\mathcal{V}^{<,\nu}+\hbar\left[\overline{\Sigma}_{\mathrm{F}}\mathcal{V}_{\mu}^{<}\right]_{\mathrm{P.B.}}^{F}+\mathcal{O}(\hbar^{3}),\\
0 & = & 2\widetilde{q}_{\mu}\mathcal{P}^{<}+\frac{\hbar}{2}\epsilon_{\mu\nu\rho\sigma}(\widetilde{\mathcal{D}}^{\nu}\mathcal{S}^{<,\rho\sigma}+\widehat{\Sigma_{\mathrm{T}}^{\rho\sigma}\mathcal{V}^{\nu}})-\hbar\left(\widehat{\Sigma_{\mathrm{F}}\mathcal{A}_{\mu}}+\widehat{\Sigma_{\mathrm{A},\mu}\mathcal{F}}\right)\nonumber \\
 & &-\hbar\left[\overline{\Sigma}_{\mathrm{F}}\mathcal{A}_{\mu}^{<}+\overline{\Sigma}_{\mathrm{A},\mu}\mathcal{F}^{<}-\frac{1}{2}\epsilon_{\mu\nu\alpha\beta}\overline{\Sigma}_{\mathrm{T}}^{\alpha\beta}\mathcal{V}^{<,\nu}\right]_{\mathrm{P.B.}}^{F} \nonumber \\
  &  & -2\left(\overline{\Sigma}_{\mathrm{P}}\mathcal{V}_{\mu}^{<}+\overline{\Sigma}_{\mathrm{A}}^{\nu}\mathcal{S}_{\nu\mu}^{<}+\overline{\Sigma}_{\mathrm{T},\mu\nu}\mathcal{A}^{<,\nu}\right)
 +\mathcal{O}(\hbar^{4}),  \\
0 & = & 
\widetilde{\mathcal{D}}^{\mu}\mathcal{V}_{\mu}^{<}
+\widehat{\Sigma_{\mathrm{F}}\mathcal{F}}+\left[\overline{\Sigma}_{\mathrm{F}}\mathcal{F}^{<}\right]_{\mathrm{P.B.}}^{F}+\mathcal{O}(\hbar^{2}),\\
0 & = & \widetilde{q}^{\mu}\mathcal{A}_{\mu}^{<}-\overline{\Sigma}_{\mathrm{A}}^{\mu}\mathcal{V}_{\mu}^{<}+\mathcal{O}(\hbar^{3}),\\
0 & = & 2\widetilde{q}_{[\mu}\mathcal{V}_{\nu]}^{<}+\mathcal{O}(\hbar^{2}).
\end{eqnarray}

We then substitute the expression of $\mathcal{F}^{<},\mathcal{P}^{<},\mathcal{S}_{\mu\nu}^{<}$
in Eqs.~(\ref{eq:F(V,A)}, \ref{eq:P(V,A)}, \ref{eq:S(V,A)}) to the master
equations, and get the following master equations in terms of $\{\mathcal{V}_{\mu}^{<},\mathcal{A}_{\mu}^{<}\}$
basis,
\begin{eqnarray}
\widetilde{q}_{[\mu}\mathcal{V}_{\nu]}^{<} & = & \mathcal{O}(\hbar^{2}),\label{eq:Vector constrain 1}\\
(\widetilde{q}^{2}-\widetilde{m}^{2})\mathcal{V}_{\mu}^{<} & = & \mathcal{O}(\hbar^{2}),\label{eq:Vector constrain 2}\\
\widetilde{q}^{\mu}\mathcal{A}_{\mu}^{<} & = & \overline{\Sigma}_{\mathrm{A}}^{\mu}\mathcal{V}_{\mu}^{<}+\mathcal{O}(\hbar^{3}),\label{eq:Axial constrain 1}\\
(\widetilde{q}^{2}-\widetilde{m}^{2})\mathcal{A}_{\mu}^{<} & = & \frac{\hbar}{2}\epsilon_{\mu\nu\alpha\beta}\widetilde{q}^{\nu}\widetilde{\mathcal{D}}^{\alpha}\mathcal{V}^{<,\beta}+\widetilde{m}\epsilon_{\mu\nu\alpha\beta}\mathcal{V}^{<,\nu}\overline{\Sigma}_{\mathrm{T}}^{\alpha\beta}\nonumber \\
 &  & +2(\overline{\Sigma}_{\mathrm{A}}\cdot\widetilde{q})\mathcal{V}_{\mu}^{<}-2\overline{\Sigma}_{\mathrm{A},\mu}(\widetilde{q}\cdot\mathcal{V}^{<})+\mathcal{O}(\hbar^{3})\label{eq:Axial constrain 2} \\
\widetilde{\mathcal{D}}^{\mu}\mathcal{V}_{\mu}^{<} & = & -\frac{1}{\widetilde{m}}\widetilde{q}_{\mu}\widehat{\Sigma_{\mathrm{F}}\mathcal{V}^{\mu}}-\frac{1}{\widetilde{m}}\left[\overline{\Sigma}_{\mathrm{F}}\left(\widetilde{q}_{\mu}\mathcal{V}^{<,\mu}\right)\right]_{\mathrm{P.B.}}^{F}+\mathcal{O}(\hbar^{2}),\label{eq:Vector kinetic eq}, 
\end{eqnarray}
and,
\begin{eqnarray}
 &  & \hbar\widetilde{q}_{\nu}\widetilde{\mathcal{D}}^{\nu}\mathcal{A}_{\mu}^{<}+\hbar(QeF_{\mu\nu}+2\nabla_{[\mu}\overline{\Sigma}_{\mathrm{V},\nu]})\mathcal{A}^{<,\nu}+\frac{\hbar}{\widetilde{m}}(\nabla^{\nu}\overline{\Sigma}_{\mathrm{F}})\left(\widetilde{q}_{\nu}\mathcal{A}_{\mu}^{<}-\widetilde{q}_{\mu}\mathcal{A}_{\nu}^{<}\right)\nonumber \\
 & = &-\hbar\widetilde{m}\left(\widehat{\Sigma_{\mathrm{F}}\mathcal{A}_{\mu}}+\left[\overline{\Sigma}_{\mathrm{F}}\mathcal{A}_{\mu}^{<}\right]_{\mathrm{P.B.}}^{F}-\frac{1}{2}\epsilon_{\mu\nu\alpha\beta}\widehat{\Sigma_{\mathrm{T}}^{\alpha\beta}\mathcal{V}^{\nu}}-\frac{1}{2}\epsilon_{\mu\nu\alpha\beta}\left[\overline{\Sigma}_{\mathrm{T}}^{\alpha\beta}\mathcal{V}^{<,\nu}\right]_{\mathrm{P.B.}}^{F}\right)\nonumber \\
 &  & +\hbar\left(-\widehat{\Sigma_{\mathrm{A},\mu}(\widetilde{q}\cdot\mathcal{V})}-\left[\overline{\Sigma}_{\mathrm{A},\mu}(\widetilde{q}\cdot\mathcal{V}^{<})\right]_{\mathrm{P.B.}}^{F}+\widetilde{q}_{\mu}\widehat{\Sigma_{\mathrm{A}}^{\lambda}\mathcal{V}_{\lambda}}+\widetilde{q}_{\mu}[\overline{\Sigma}_{\mathrm{A}}^{\lambda}\mathcal{V}_{\lambda}^{<}]_{\mathrm{P.B.}}^{F}\right)\nonumber \\
 &  & -2\frac{\widetilde{q}_{\mu}}{\widetilde{m}}\overline{\Sigma}_{\mathrm{T},\alpha\beta}\left[\widetilde{q}^{\alpha}\mathcal{A}^{<,\beta}+\frac{\hbar}{4}\epsilon^{\rho\sigma\alpha\beta}\nabla_{\rho}\mathcal{V}_{\sigma}^{<}\right]+\frac{\hbar}{2}\epsilon_{\mu\nu\alpha\beta}\nabla^{\nu}\left[\frac{1}{\widetilde{m}}\overline{\Sigma}_{\mathrm{T}}^{\alpha\beta}(\widetilde{q}\cdot\mathcal{V}^{<})+\epsilon^{\alpha\beta\rho\sigma}\overline{\Sigma}_{\mathrm{A},\rho}\mathcal{V}_{\sigma}^{<}\right]\nonumber \\
 &  & -2\overline{\Sigma}_{\mathrm{A}}^{\nu}\left[\epsilon_{\mu\nu\alpha\beta}\widetilde{q}^{\alpha}\mathcal{A}^{<,\beta}-\hbar\nabla_{[\mu}\mathcal{V}_{\nu]}^{<}\right]-2\widetilde{m}\overline{\Sigma}_{\mathrm{T},\mu\nu}\mathcal{A}^{<,\nu}
 + \frac{\hbar^{2}}{2}\widetilde{m}\epsilon_{\mu\nu\alpha\beta}\widetilde{\mathcal{D}}^{\nu}\left[\frac{1}{\widetilde{m}}\widetilde{\mathcal{D}}^{\alpha}\mathcal{V}^{<,\beta}\right]
 +\mathcal{O}(\hbar^{4}),\label{eq:Axial kinetic eq} \nonumber \\ 
 \\
& & \widetilde{q}\cdot\widetilde{\mathcal{D}}\mathcal{V}_{\mu}^{<}+\widetilde{m}\widetilde{\mathcal{D}}_{\mu}(\frac{1}{\widetilde{m}}\widetilde{q}\cdot\mathcal{V}^{<})-\widetilde{q}^{\nu}\widetilde{\mathcal{D}}_{\mu}\mathcal{V}_{\nu}^{<}  =  -\widetilde{m}\widehat{\Sigma_{\mathrm{F}}\mathcal{V}_{\mu}}-\widetilde{m}\left[\overline{\Sigma}_{\mathrm{F}}\mathcal{V}_{\mu}^{<}\right]_{\mathrm{P.B.}}^{F}+\mathcal{O}(\hbar^{2}),\label{eq:Redundant master eq}
\end{eqnarray}
Here Eq.(\ref{eq:Redundant master eq}) is redundant as shown in Appendix.~\ref{app:redundancy}.
Now, Eqs.~(\ref{eq:Vector kinetic eq},\ref{eq:Axial kinetic eq}) contribute to
the SKE and AKE, respectively. Eqs.~(\ref{eq:Vector constrain 1},\ref{eq:Vector constrain 2})
are the constraint equations for the vector Wigner function, while Eqs.~(\ref{eq:Axial constrain 1},\ref{eq:Axial constrain 2})
are those for the axial Wigner function, which lead to the perturbative solutions as will be obtained in the next subsection. It is straightforward to reproduce the results in Ref.~\citep{Yang:2020hri} by simply setting $\overline{\Sigma}=0$.

From Eqs.~(\ref{eq:Vector constrain 1},\ref{eq:Vector constrain 2}),
we find that the vector Wigner function is on-shell with the modified
momentum and modified mass up to $\mathcal{O}(\hbar^{1})$ and we have
\begin{eqnarray}
\mathcal{V}_{\mu}^{<} & = & \frac{\widetilde{q}_{\mu}}{\widetilde{m}^{2}}(\widetilde{q}\cdot\mathcal{V}^{<}).
\end{eqnarray}
Also, from the kinetic equations and constraint equations, we notice that $\mathcal{V}_{\mu}$
does not couple with $\Sigma_{\mathrm{A}},\Sigma_{\mathrm{T}},\Sigma_{\mathrm{P}}$
and $\mathcal{A}_{\mu}$ does not couple with $\Sigma_{\mathrm{P}}$
up to one $\hbar$ higher w.r.t. their leading order solutions.
Notably, the self-energy corrections to the AKE in Eq.~(\ref{eq:Axial kinetic eq}) modify both
the collision terms and the free-streaming part. For example, the
term associated with external electromagnetic fields, $QeF_{\mu\nu}$, on the left-hand side of Eq.~(\ref{eq:Axial kinetic eq}), is further accompanied by the
gradient of $\overline{\Sigma}_{\mathrm{V},\nu}$ and, the gradient of $\tilde{m}$ is also involved for the collisionless AKE. 


\subsection{Solutions to Wigner functions}\label{subsec:Solutions-to-Wigner}

To find perturbative solutions of the Wigner functions, we start
from solving the constraint equations. We firstly discuss the vector
Wigner functions, for which the constraint equations (\ref{eq:Vector constrain 1},\ref{eq:Vector constrain 2}) form a close set of equations themselves.
We can directly read out the solution,
\begin{eqnarray}
\mathcal{V}_{\mu}^{<}(q,X) & = & 2\pi\epsilon(q_{0})\delta(\widetilde{q}^{2}-\widetilde{m}^{2})\widetilde{q}_{\mu}f_{\mathrm{V}}^{<}(q,X)+\mathcal{O}(\hbar^{2}),\label{eq:Vector WF formal solution}
\end{eqnarray}
with $f^<_{\mathrm{V}}(q,X)$ being the vector distribution function and $\epsilon(q_0)$ being the sign of energy for incorporating the contributions from both particles and anti-particles.  
As a generalization of classical Boltzmann equation, the SKE reads 
\begin{eqnarray}
0 & = & 2\pi\epsilon(q_{0})\delta(\widetilde{q}^{2}-\widetilde{m}^{2})\left\{ \widetilde{q}_{\mu}\widetilde{\nabla}^{\mu}f_{\mathrm{V}}^{<}+\widetilde{q}_{\mu}\widehat{\Sigma_{\mathrm{V}}^{\mu}f_{\mathrm{V}}}+\widetilde{m}\widehat{\Sigma_{\mathrm{F}}f_{\mathrm{V}}}\right.\nonumber \\
 &  & \qquad\left.-\widetilde{m}\left[(\nabla_{\alpha}\overline{\Sigma}_{\mathrm{F}})\partial_{q}^{\alpha}f_{\mathrm{V}}^{<}-(\partial_{q,\alpha}\overline{\Sigma}_{\mathrm{F}})\partial_{X}^{\alpha}f_{\mathrm{V}}^{<}\right]\right\} +\mathcal{O}(\hbar^{2}),\label{eq:fV kinetic eq}
\end{eqnarray}
where we have used 
\begin{eqnarray}
(\nabla_{\alpha}\overline{\Sigma}_{\mathrm{F}})\partial_{q}^{\alpha}\widetilde{m}-(\partial_{q,\alpha}\overline{\Sigma}_{\mathrm{F}})\partial_{X}^{\alpha}\widetilde{m} & = & -(\nabla_{\alpha}\widetilde{m})\partial_{q}^{\alpha}\widetilde{m}+(\partial_{q,\alpha}\widetilde{m})\partial_{X}^{\alpha}\widetilde{m}=0.
\end{eqnarray}
Here the Wigner functions as the Wigner transform of two-point functions
should be of $\mathcal{O}(\hbar^{1})$, but as shown in Eq.~(\ref{eq:V-A power counting})
and discussions in Sec.~\ref{subsec:KB-equations-of}, an overall
$\hbar$ prefactor of Wigner functions and of self-energies have been dropped and
extracted explicitly, so all $\hbar$ here  are related to the gradient expansion only. 

As for the axial Wigner function, we consider the constraint equations in Eqs.~(\ref{eq:Axial constrain 1},\ref{eq:Axial constrain 2}). As in the case of vanishing $\overline{\Sigma}$  \cite{Hattori:2019ahi,Yang:2020hri}, $\mathcal{A}^<_{\mu}$ contains both the dynamical and non-dynamical components, where the former is governed by an effective spin four vector $\tilde{a}^{\mu}(q,X)$ as an extra dynamical variables in addition to $f_{\mathrm{V}}^<(q,X)$ and the latter is from explicit quantum contributions proportional to $\hbar f_{\mathrm{V}}^<(q,X)$ stemming from intertwined dynamics of the vector and spin polarization. One may further make the decomposition, 
\begin{equation}
    \tilde{a}^{\mu}(q,X)=a^{\mu}(q,X)f^<_{\rm A}(q,X),
\end{equation}
where $f^<_{\rm A}(q,X)$ denotes an axial-charge distribution function. Unlike $f^<_{\rm V}=1-f^>_{\rm V}$, it follows $f^>_{\rm A}=-f^<_{\rm A}$. 
In the massless limit, we have $a^{\mu}(q,X)=q^{\mu}$ due to the spin alignment along the momentum direction based on the equality of chirality and helicity. Also, one can rewrite $f^<_{\rm V/\rm A}$ in terms of $f_{\rm R/\rm L}$ as the distribution functions for right/left-handed fermions via $f^<_{\rm V}=(f_{\rm R}+f_{\rm L})/2$ and $f^<_{\rm A}=(f_{\rm R}-f_{\rm L})/2$. Now, the dynamical part of $\mathcal{A}^<_{\mu}$ can be simply obtained when setting $f_{\mathrm{V}}^<(q,X)=0$, which yields
\begin{eqnarray}
\widetilde{q}^{\mu}\mathcal{A}_{\mu}^{<} & = & 0,\\
(\widetilde{q}^{2}-\widetilde{m}^{2})\mathcal{A}_{\mu}^{<} & = & 0,
\end{eqnarray}
and the corresponding solution for the dynamical part reads
\begin{eqnarray}
\mathcal{A}_{(\rm dy)\mu}^{<}(q,X) & = & 2\pi\epsilon(q_{0})\delta(\widetilde{q}^{2}-\widetilde{m}^{2})\widetilde{a}_{\mu}(q,X),
\end{eqnarray}
with $\delta(\widetilde{q}^{2}-\widetilde{m}^{2})\widetilde{q}\cdot \widetilde{a}=0$. 

To derive the non-dynamical part, we have to retrieve $f^<_{\mathrm{V}}(q,X)$ in Eqs.~(\ref{eq:Axial constrain 1},\ref{eq:Axial constrain 2}),
\begin{eqnarray}\label{eq:Axial_constrain_1_fV}
\widetilde{q}^{\mu}\mathcal{A}_{\mu}^{<} & = & 2\pi\epsilon(q_{0})\delta(\widetilde{q}^{2}-\widetilde{m}^{2})(\widetilde{q}\cdot\overline{\Sigma}_{\mathrm{A}})f_{\mathrm{V}}^{<}+\mathcal{O}(\hbar^{3}), \\\label{eq:Axial_constrain_2_fV}
(\widetilde{q}^{2}-\widetilde{m}^{2})\mathcal{A}_{\mu}^{<} & = & 2\pi\epsilon(q_{0})\delta(\widetilde{q}^{2}-\widetilde{m}^{2})f_{\mathrm{V}}^{<}\left[\frac{\hbar}{2}\epsilon_{\mu\nu\alpha\beta}\widetilde{q}^{\nu}(QeF^{\alpha\beta}+2\nabla^{[\alpha}\overline{\Sigma}_{\mathrm{V}}^{\beta]})\right.\nonumber \\
 &  & \qquad\left.+\widetilde{m}\epsilon_{\mu\nu\alpha\beta}\widetilde{q}^{\nu}\overline{\Sigma}_{\mathrm{T}}^{\alpha\beta}+2(\overline{\Sigma}_{\mathrm{A}}\cdot\widetilde{q})\widetilde{q}_{\mu}-2\widetilde{m}^{2}\overline{\Sigma}_{\mathrm{A},\mu}\right]+\mathcal{O}(\hbar^{3}).
\end{eqnarray}
Similar to the case with $\overline{\Sigma}=0$ \cite{Yang:2020hri}, the non-dynamical part can be further decomposed into 
\begin{eqnarray}
\mathcal{A}_{(\rm non)\mu}^{<}=2\pi\epsilon(q_{0})\Big[\delta(\widetilde{q}^2-\widetilde{m}^2)\mathcal{A}_{(1)\mu}^{<}+\delta'(\widetilde{q}^2-\widetilde{m}^2)\mathcal{A}_{(2)\mu}^{<}\Big],
\end{eqnarray}
where $\delta'(x)= \mathrm{d}\delta(x)/\mathrm{d}x$. From Eq.~(\ref{eq:Axial_constrain_2_fV}), we obtain
\begin{eqnarray}
\mathcal{A}_{(2)\mu}^{<}=-f_{\mathrm{V}}^{<}(q,X)\left[\frac{\hbar}{2}\epsilon_{\mu\nu\alpha\beta}\widetilde{q}^{\nu}(QeF^{\alpha\beta}+2\nabla^{[\alpha}\overline{\Sigma}_{\mathrm{V}}^{\beta]})
+\widetilde{m}\epsilon_{\mu\nu\alpha\beta}\widetilde{q}^{\nu}\overline{\Sigma}_{\mathrm{T}}^{\alpha\beta}+2(\overline{\Sigma}_{\mathrm{A}}\cdot\widetilde{q})\widetilde{q}_{\mu}-2\widetilde{m}^{2}\overline{\Sigma}_{\mathrm{A},\mu}\right]. \nonumber \\
\end{eqnarray}
Given $\mathcal{A}_{(2)\mu}^{<}$ above, from Eq.~(\ref{eq:Axial_constrain_1_fV}), we further derive
\begin{eqnarray}
\mathcal{A}_{(1)\mu}^{<}=-\overline{\Sigma}_{\mathrm{A},\mu}f_{\mathrm{V}}^{<}(q,X),
\end{eqnarray}
where we have used $x\delta^{\prime}(x)=-\delta(x)$.


However, unlike the case for massless fermions, the constraints equations here cannot uniquely determine a magnetization current term as part of the non-dynamical solution that trivially satisfies Eqs.~(\ref{eq:Axial_constrain_1_fV}, \ref{eq:Axial_constrain_2_fV}) (also see e.g. Ref.~\citep{Hattori:2019ahi} for more detailed discussions). We may follow the approaches in Refs.~\citep{Hattori:2019ahi,Yang:2020hri} for $\overline{\Sigma}=0$ by generalizing the massless solution \citep{Yamamoto:2023okm} to introduce such a magnetization-current term,
\begin{eqnarray}
\mathcal{A}_{(\rm mag)\mu}^{<} =  2\pi\epsilon(q_{0})\delta(\widetilde{q}^{2}-\widetilde{m}^{2})\hbar S_{n,\mu\nu}(\widetilde{q},\widetilde{m})\widetilde{\mathcal{D}}^{\nu}f_{\mathrm{V}}^{<},
\end{eqnarray}
where we introduce the frame-dependent spin tensor for the massive
fermions \footnote{For the modes with negative energies, we should replace $\tilde{q}\cdot n$
to $|\tilde{q}\cdot n|$ in the spin tensor as noted in \citep{Hattori:2019ahi}. }, 
\begin{eqnarray}
S_{n}^{\mu\nu}(\widetilde{q},\widetilde{m}) & = & \frac{\epsilon^{\mu\nu\rho\sigma}\widetilde{q}_{\rho}n_{\sigma}}{2(\widetilde{q}\cdot n+\widetilde{m})},
\end{eqnarray}
with $n_{\mu}(q,X)$ being a timelike frame vector satisfying $n^2=1$. The choice of $n_{\mu}(q,X)$ is associated with the choice of a spin basis, which does not affect the results for physical observables \cite{Hidaka:2022dmn}. Combining all the contributions, 
we conclude our axial Wigner function as follows, 
\begin{eqnarray}
\mathcal{A}_{\mu}^{<}(q,X) & = &  \mathcal{A}_{(\rm dy)\mu}^{<}(q,X)+\mathcal{A}_{(\rm non)\mu}^{<}(q,X)+\mathcal{A}_{(\rm mag)\mu}^{<}(q,X) \nonumber \\
& = & 2\pi\epsilon(q_{0})\bigg\{\delta(\widetilde{q}^{2}-\widetilde{m}^{2})\left[a_{\mu}f_{\mathrm{A}}^{<}+\hbar S_{n,\mu\nu}(\widetilde{q},\widetilde{m})\widetilde{\mathcal{D}}^{\nu}f_{\mathrm{V}}^{<}-\overline{\Sigma}_{\mathrm{A},\mu}f_{\mathrm{V}}^{<}\right]
 -\delta^{\prime}(\widetilde{q}^{2}-\widetilde{m}^{2})f_{\mathrm{V}}^{<}
 \nonumber \\
 &  &\times\left[\frac{\hbar}{2}\epsilon_{\mu\nu\alpha\beta}\widetilde{q}^{\nu}(QeF^{\alpha\beta}+2\nabla^{[\alpha}\overline{\Sigma}_{\mathrm{V}}^{\beta]})
+\widetilde{m}\epsilon_{\mu\nu\alpha\beta}\widetilde{q}^{\nu}\overline{\Sigma}_{\mathrm{T}}^{\alpha\beta}+2(\overline{\Sigma}_{\mathrm{A}}\cdot\widetilde{q})\widetilde{q}_{\mu}-2\widetilde{m}^{2}\overline{\Sigma}_{\mathrm{A},\mu}\right]\bigg\}. \nonumber \\
\label{eq:Axial WF formal solution}
\end{eqnarray}
A smooth connection to the massless case \citep{Yamamoto:2023okm} 
can be easily obtained by setting $m=0$ and changing to the chiral basis. 

Apparently, the perturbative solutions of Wigner functions shown in Eqs.~(\ref{eq:Vector WF formal solution}, \ref{eq:Axial WF formal solution})
manifest the modifications upon the on-shell dispersion relations as discussed in Sec.\ref{subsec:KB-equations-of}.
Moreover, the modification on the axial Wigner function further incorporates the correction on the direction of spin polarization. For example, the term proportional to $\epsilon_{\mu\nu\alpha\beta}\tilde{q}^{\nu}\nabla^{[\alpha}\overline{\Sigma}_{\mathrm{V}}^{\beta]}$
plays a similar role to the external electromagnetic fields, which may induce the interaction-dependent spin polarization of fermions
up to the first-order in gradient expansion. Besides, given $\overline{\Sigma}_{\mathrm{V}}^{\mu}$
calculated from the weakly coupled quantum field theories, such spin polarization
is expected to be more dominant than the
polarization induced by collisions. The terms associated with $\overline{\Sigma}_{\mathrm{A},\mu}$ and $\overline{\Sigma}^{\alpha\beta}_{\rm T}$ may also modify the spin polarization, but their explicit forms require further studies on the quantum corrections for Wigner functions of polarized photons or gluons,  which is beyond the scope of this paper.


\subsection{AKE with self-energy\label{subsec:Axial-kinetic-equation}}

Equipped with the perturbative solutions to the vector and axial Wigner
functions in Eqs.~(\ref{eq:Vector WF formal solution}, \ref{eq:Axial WF formal solution}),
we are ready to derive the AKE for  $a_{\mu}f_{\mathrm{A}}^{<}$ from the EoMs
of $\mathcal{A}_{\mu}^{<}$ in Eq.~(\ref{eq:Axial kinetic eq}). We work
in a general frame vector $n^{\mu}(X)$ depending on spacetime
and mainly focus on the spin evolution of particles with positive
energies. 

We neglect the nonlinear order terms of self-energies as before,
namely, we drop the terms of $\mathcal{O}(\Sigma^{\lessgtr}\Sigma^{\lessgtr})$,
$\mathcal{O}(\overline{\Sigma}\Sigma^{\lessgtr})$ and $\mathcal{O}(\overline{\Sigma}^{2})$
for simplicity. It is helpful to introduce the modified field tensor
and its dual, 
\begin{eqnarray}
Qe\overline{F}^{\mu\nu} & = & QeF^{\mu\nu}+2\nabla^{[\mu}\overline{\Sigma}_{\mathrm{V}}^{\nu]},\qquad\widetilde{\overline{F}}^{\mu\nu}=\frac{1}{2}\epsilon^{\mu\nu\rho\sigma}\overline{F}_{\rho\sigma}.
\end{eqnarray}
Notably, though the field strength tensor is modified by the self-energy, we find under the differential operator $\widetilde{\nabla}^{\mu}$, the modified Bianchi identity still holds,
\begin{eqnarray}
\epsilon_{\mu\nu\alpha\beta}\widetilde{\nabla}^{\beta}(Qe\overline{F}^{\nu\alpha})	=	0.
\end{eqnarray}

Then we can simplify the first term on the r.h.s. of Eq.~(\ref{eq:Axial kinetic eq}),
\begin{eqnarray}
 &  & -\frac{\hbar^{2}}{2}\widetilde{m}\epsilon_{\mu\nu\alpha\beta}\widetilde{\mathcal{D}}^{\nu}\left[\frac{1}{\widetilde{m}}\widetilde{\mathcal{D}}^{\alpha}\mathcal{V}^{<,\beta}\right]\nonumber \\
 & = &  -\frac{\hbar^{2}}{2}\epsilon_{\mu\nu\alpha\beta}\frac{(\nabla^{\nu}\overline{\Sigma}_{\mathrm{F}})}{\widetilde{m}}\nabla^{\alpha}\mathcal{V}^{<,\beta}-\frac{\hbar^{2}}{2}\epsilon_{\mu\nu\alpha\beta}(\nabla^{\nu}\widehat{\Sigma_{\mathrm{V}}^{\alpha})\mathcal{V}^{\beta}}-\frac{\hbar^{2}}{4}\epsilon_{\mu\nu\alpha\beta}[(Qe\overline{F}^{\nu\alpha})\mathcal{V}^{<,\beta}]_{\mathrm{P.B.}}^{F},
\end{eqnarray}
with
\begin{eqnarray}
\epsilon_{\mu\nu\alpha\beta}\widetilde{\nabla}^{\nu}\widetilde{\nabla}^{\alpha}f & = & \epsilon_{\mu\nu\alpha\beta}\frac{1}{2}[\widetilde{\nabla}^{\nu},\widetilde{\nabla}^{\alpha}]f=\frac{1}{2}\epsilon_{\mu\nu\alpha\beta}[(Qe\overline{F}^{\nu\alpha})f]_{\mathrm{P.B.}}^{F}+\mathcal{O}({\Sigma}^{2}).\label{eq:Nable_commutator}
\end{eqnarray}
Also, the contribution from the magnetization-current term in the
first term of Eq.~(\ref{eq:Axial kinetic eq}) can be written as,
\begin{eqnarray}
 &  & \hbar^{2}2\pi\delta(\widetilde{q}^{2}-\widetilde{m}^{2})\widetilde{q}_{\rho}\widetilde{\nabla}^{\rho}(S_{n,\mu\nu}\widetilde{\nabla}^{\nu}f_{\mathrm{V}}^{<})+\hbar^{2}2\pi\delta(\widetilde{q}^{2}-\widetilde{m}^{2})Qe\overline{F}_{\mu\nu}S_{n}^{\nu\rho}\widetilde{\nabla}_{\rho}f_{\mathrm{V}}^{<}\nonumber \\
 & = &  \hbar^{2}2\pi\delta(\widetilde{q}^{2}-\widetilde{m}^{2})\Bigg[\widetilde{q}_{\mu}\left(\frac{\epsilon_{\rho\nu\alpha\beta}\widetilde{q}^{\alpha}\widetilde{\nabla}^{\rho}n^{\beta}}{2(\widetilde{q}\cdot n+\widetilde{m})}-\frac{\widetilde{\nabla}^{\rho}(\widetilde{q}\cdot n+\widetilde{m})}{\widetilde{q}\cdot n+\widetilde{m}}S_{n,\rho\nu}\right)\nonumber \\
 &  & \quad-\frac{\epsilon_{\alpha\rho\mu\nu}\widetilde{q}^{\alpha}\widetilde{\nabla}^{\rho}\widetilde{m}}{2(\widetilde{q}\cdot n+\widetilde{m})}-\widetilde{m}^{2}\frac{\epsilon_{\alpha\rho\mu\nu}\widetilde{\nabla}^{\rho}n^{\alpha}}{2(\widetilde{q}\cdot n+\widetilde{m})}+\widetilde{m}\frac{\epsilon_{\alpha\rho\mu\nu}(\widetilde{m}n^{\alpha}+\widetilde{q}^{\alpha})}{2(\widetilde{q}\cdot n+\widetilde{m})^{2}}\widetilde{\nabla}^{\rho}(\widetilde{q}\cdot n+\widetilde{m})\Bigg]\widetilde{\nabla}^{\nu}f_{\mathrm{V}}^{<}\nonumber \\
 &  & \quad+\hbar^{2}2\pi\delta(\widetilde{q}^{2}-\widetilde{m}^{2})\left(-\frac{\epsilon_{\rho\mu\alpha\beta}\widetilde{q}^{\alpha}\widetilde{\nabla}^{\rho}n^{\beta}}{2(\widetilde{q}\cdot n+\widetilde{m})}+\frac{\widetilde{\nabla}^{\rho}(\widetilde{q}\cdot n+\widetilde{m})}{\widetilde{q}\cdot n+\widetilde{m}}S_{n,\rho\mu}\right)(\widetilde{q}\cdot\widetilde{\nabla}f_{\mathrm{V}}^{<})\nonumber \\
 &  &\quad +\hbar^{2}2\pi\delta(\widetilde{q}^{2}-\widetilde{m}^{2})S_{n,\mu\nu}\widetilde{q}_{\rho}[(Qe\overline{F}^{\rho\nu})f_{\mathrm{V}}^{<}]_{\mathrm{P.B.}}^{F}+\hbar^{2}S_{n,\mu\nu}\widetilde{\nabla}^{\nu}\left(2\pi\delta(\widetilde{q}^{2}-\widetilde{m}^{2})\widetilde{q}\cdot\widetilde{\nabla}f_{\mathrm{V}}^{<}\right)\nonumber \\
 &  &\quad -\hbar^{2}2\pi\delta^{\prime}(\widetilde{q}^{2}-\widetilde{m}^{2})S_{n,\mu\nu}(2\widetilde{q}_{\alpha}Qe\overline{F}^{\nu\alpha}+2\widetilde{m}\nabla^{\nu}\overline{\Sigma}_{\mathrm{F}})\widetilde{q}\cdot\widetilde{\nabla}f_{\mathrm{V}}^{<},
\end{eqnarray}
where we have used the Schouten identity and Eq.~(\ref{eq:Nable_commutator}).

After complicated yet straightforward calculations, we finally derive
the AKE with self-energy corrections,
\begin{eqnarray}
\square_{a,\mu}^{(n)}\mathcal{A}^{<}+\widetilde{q}_{\mu}\square_{q}^{(n)}\mathcal{A}^{<}+\widetilde{m}\square_{m,\mu}^{(n)}\mathcal{A}^{<} & = & \mathcal{C}_{1,\mu}^{(n)}+\hbar(\widetilde{q}_{\mu}\mathcal{C}_{2,\mu}^{(n),q}+\widetilde{m}\mathcal{C}_{2,\mu}^{(n),m}).\label{eq:afA_kinetic eq}
\end{eqnarray}
The free-streaming part can be divided in three parts \citep{Hattori:2019ahi,Yang:2020hri}, which are proportional to $a_{\rho}f_{\mathrm{A}}$,
$\widetilde{q}_{\mu}$ and $\widetilde{m}$ separately,
\begin{eqnarray}
\square_{a,\mu}^{(n)}\mathcal{A}^{<} & = & \delta(\widetilde{q}^{2}-\widetilde{m}^{2})\Bigg\{\widetilde{q}\cdot\widetilde{\nabla}(a_{\mu}f_{\mathrm{A}}^{<})+Qe\overline{F}_{\mu\nu}(a^{\nu}f_{\mathrm{A}}^{<})+2\epsilon_{\mu\nu\alpha\beta}\frac{\overline{\Sigma}_{\mathrm{A}}^{\nu}}{\hbar}\widetilde{q}^{\alpha}a^{\beta}f_{\mathrm{A}}^{<}\nonumber \\
 &  & +\frac{\widetilde{q}\cdot\nabla\overline{\Sigma}_{\mathrm{F}}}{\widetilde{m}}(a_{\mu}f_{\mathrm{A}}^{<})-\widetilde{q}_{\mu}\frac{\nabla^{\nu}\overline{\Sigma}_{\mathrm{F}}}{\widetilde{m}}(a_{\nu}f_{\mathrm{A}}^{<})\Bigg\},\label{eq:AKE_afA}
\end{eqnarray}
The second and third terms in the r.h.s of Eq.~(\ref{eq:afA_kinetic eq}) read, 
\begin{eqnarray}
\square_{q}^{(n)}\mathcal{A}^{<} & = & \delta(\widetilde{q}^{2}-\widetilde{m}^{2})\Bigg\{2\widetilde{q}^{\alpha}\frac{\overline{\Sigma}_{\mathrm{T},\alpha\beta}}{\hbar\widetilde{m}}a^{\beta}f_{\mathrm{A}}^{<}+\frac{2\overline{\Sigma}_{\mathrm{T},\alpha\beta}}{\widetilde{m}}\widetilde{q}^{\alpha}S_{n}^{\beta\rho}\nabla_{\rho}f_{\mathrm{V}}^{<}+\epsilon^{\rho\sigma\alpha\beta}\frac{\overline{\Sigma}_{\mathrm{T},\alpha\beta}}{2\widetilde{m}}\widetilde{q}_{\sigma}\nabla_{\rho}f_{\mathrm{V}}^{<}\nonumber \\
 &  & -\widetilde{q}_{\lambda}[\overline{\Sigma}_{\mathrm{A}}^{\lambda}f_{\mathrm{V}}^{<}]_{\mathrm{P.B.}}^{F}+\frac{\hbar}{2}S_{n,\rho\sigma}[(Qe\overline{F}^{\rho\sigma})f_{\mathrm{V}}^{<}]_{\mathrm{P.B.}}^{F}-\frac{(\widetilde{q}\cdot n)\overline{\Sigma}_{\mathrm{A}}\cdot\nabla f_{\mathrm{V}}^{<}}{(\widetilde{q}\cdot n+\widetilde{m})}\nonumber \\
 &  & +\hbar\left(\frac{\epsilon_{\rho\nu\alpha\beta}\widetilde{q}^{\alpha}\widetilde{\nabla}^{\rho}n^{\beta}}{2(\widetilde{q}\cdot n+\widetilde{m})}-\frac{\widetilde{\nabla}^{\rho}(\widetilde{q}\cdot n+\widetilde{m})}{\widetilde{q}\cdot n+\widetilde{m}}S_{n,\rho\nu}\right)\widetilde{\nabla}^{\nu}f_{\mathrm{V}}^{<}\Bigg\}\nonumber \\
 &  & +\delta^{\prime}(\widetilde{q}^{2}-\widetilde{m}^{2})(\widetilde{q}\cdot\widetilde{\nabla}f_{\mathrm{V}}^{<})\left[\hbar\frac{\widetilde{q}^{\rho}n^{\sigma}Qe\widetilde{\overline{F}}_{\rho\sigma}}{(\widetilde{q}\cdot n+\widetilde{m})}-2(\overline{\Sigma}_{\mathrm{A}}\cdot\widetilde{q})\right],\label{eq:AKE_q}
\end{eqnarray}
and
\begin{eqnarray}
\square_{m,\mu}^{(n)}\mathcal{A}^{<} & = & \delta(\widetilde{q}^{2}-\widetilde{m}^{2})\Bigg\{2\overline{\Sigma}_{\mathrm{T},\mu\nu}\frac{a^{\nu}f_{\mathrm{A}}^{<}}{\hbar}+[\overline{\Sigma}_{\mathrm{F}}(a_{\mu}f_{\mathrm{A}}^{<})]_{\mathrm{P.B.}}^{F}+\widetilde{m}[\overline{\Sigma}_{\mathrm{A},\mu}f_{\mathrm{V}}^{<}]_{\mathrm{P.B.}}^{F}\nonumber \\
 &  & -\frac{1}{2}\epsilon_{\mu\nu\alpha\beta}\left((\nabla^{\nu}f_{\mathrm{V}}^{<})\overline{\Sigma}_{\mathrm{T}}^{\alpha\beta}+\widetilde{q}^{\nu}[\overline{\Sigma}_{\mathrm{T}}^{\alpha\beta}f_{\mathrm{V}}^{<}]_{\mathrm{P.B.}}^{F}\right)+2\overline{\Sigma}_{\mathrm{T},\mu\nu}S_{n}^{\nu\alpha}\nabla_{\alpha}f_{\mathrm{V}}^{<}\nonumber \\
 &  & -\frac{(\widetilde{q}\cdot\overline{\Sigma}_{\mathrm{A}})\nabla_{\mu}f_{\mathrm{V}}^{<}+\widetilde{m}(\overline{\Sigma}_{\mathrm{A}}\cdot n)\nabla_{\mu}f_{\mathrm{V}}^{<}-\widetilde{m}n_{\mu}(\overline{\Sigma}_{\mathrm{A}}\cdot\nabla f_{\mathrm{V}}^{<})}{(\widetilde{q}\cdot n+\widetilde{m})}\nonumber \\
 &  & +\hbar[\overline{\Sigma}_{\mathrm{F}}S_{n,\mu\rho}]_{\mathrm{P.B.}}^{F}\nabla^{\rho}f_{\mathrm{V}}^{<}-\hbar S_{n,\mu\nu}[(\nabla^{\nu}\overline{\Sigma}_{\mathrm{F}})f_{\mathrm{V}}^{<}]_{\mathrm{P.B.}}^{F}+\hbar\widetilde{\nabla}^{\rho}S_{n,\rho\mu}[\overline{\Sigma}_{\mathrm{F}}f_{\mathrm{V}}^{<}]_{\mathrm{P.B.}}^{F}\nonumber \\
 &  & +\hbar\frac{(\widetilde{m}n^{\nu}+\widetilde{q}^{\nu})}{2(\widetilde{q}\cdot n+\widetilde{m})}\left(\epsilon_{\mu\nu\beta\gamma}\frac{\widetilde{\nabla}^{\beta}(\widetilde{q}\cdot n+\widetilde{m})}{(\widetilde{q}\cdot n+\widetilde{m})}\widetilde{\nabla}^{\gamma}f_{\mathrm{V}}^{<}-[(Qe\widetilde{\overline{F}}_{\mu\nu})f_{\mathrm{V}}^{<}]_{\mathrm{P.B.}}^{F}\right)\nonumber \\
 &  & -\hbar\epsilon_{\mu\nu\alpha\beta}\frac{n^{\alpha}\nabla^{\beta}\overline{\Sigma}_{\mathrm{F}}+\widetilde{m}\widetilde{\nabla}^{\beta}n^{\alpha}}{2(\widetilde{q}\cdot n+\widetilde{m})}\widetilde{\nabla}^{\nu}f_{\mathrm{V}}^{<}\Bigg\}\nonumber \\
 &  & +\delta^{\prime}(\widetilde{q}^{2}-\widetilde{m}^{2})\Bigg\{(\widetilde{q}\cdot\widetilde{\nabla}f_{\mathrm{V}}^{<}) I_\mu
+2q_{\mu}f_{\mathrm{V}}^{<}Qe\widetilde{F}_{\alpha\beta}\overline{\Sigma}_{\mathrm{T}}^{\alpha\beta}
\nonumber \\
 &  & 
-\hbar[\overline{\Sigma}_{\mathrm{F}}f_{\mathrm{V}}^{<}]_{\mathrm{P.B.}}^{F}\left(\widetilde{q}^{\alpha}Qe\widetilde{F}_{\mu\alpha}+S_{n,\mu\nu}2\widetilde{q}_{\alpha}QeF^{\nu\alpha}\right)\Bigg\},\label{eq:AKE_m}
\end{eqnarray}
where 
\begin{equation}
    I_\mu \equiv -\hbar\frac{Qe\widetilde{\overline{F}}_{\mu\alpha}(\widetilde{m}n^{\alpha}+\widetilde{q}^{\alpha})}{(\widetilde{q}\cdot n+\widetilde{m})}-2\hbar S_{n,\mu\nu}\nabla^{\nu}\overline{\Sigma}_{\mathrm{F}}-\epsilon_{\mu\rho\alpha\beta}\widetilde{q}^{\rho}\overline{\Sigma}_{\mathrm{T}}^{\alpha\beta}+2\widetilde{m}\overline{\Sigma}_{\mathrm{A},\mu}. 
\end{equation}
The collision term on the other hand is not modified by the self-energies
and is the same as in Ref.~\citep{Yang:2020hri},
\begin{eqnarray}
\mathcal{C}_{1,\mu}^{(n)} & = & \delta(\widetilde{q}^{2}-\widetilde{m}^{2})\Bigg[-(\widetilde{q}\cdot\widehat{\Sigma_{\mathrm{V}})(a_{\mu}f_{\mathrm{A}})}-\widetilde{m}^{2}\widehat{\Sigma_{\mathrm{A},\mu}f_{\mathrm{V}}}+\widetilde{q}_{\mu}(\widetilde{q}\cdot\widehat{\Sigma_{\mathrm{A}})f_{\mathrm{V}}}\nonumber \\
 &  & -\widetilde{m}\left(\widehat{\Sigma_{\mathrm{F}}(a_{\mu}f_{\mathrm{A}})}-\frac{1}{2}\epsilon_{\mu\nu\alpha\beta}\widetilde{q}^{\nu}\widehat{\Sigma_{\mathrm{T}}^{\alpha\beta}f_{\mathrm{V}}}\right)\Bigg],\label{eq:CK_1} \\
\mathcal{C}_{2,\mu}^{(n),q} & = & \delta(\widetilde{q}^{2}-\widetilde{m}^{2})\Bigg\{-(\nabla^{\rho}S_{n,\rho\nu})\widehat{\Sigma_{\mathrm{V}}^{\nu}f_{\mathrm{V}}}-S_{n,\nu\rho}(\widetilde{\nabla}^{\nu}\widehat{\Sigma_{\mathrm{V}}^{\rho})f_{\mathrm{V}}}+\frac{n^{\alpha}Qe\widetilde{F}_{\alpha\nu}}{(\widetilde{q}\cdot n+\widetilde{m})}\widehat{\Sigma_{\mathrm{V}}^{\nu}f_{\mathrm{V}}}\Bigg\}\nonumber \\
 &  & -\delta^{\prime}(\widetilde{q}^{2}-\widetilde{m}^{2})\frac{Qe\widetilde{F}_{\rho\sigma}\widetilde{q}^{\rho}n^{\sigma}}{(\widetilde{q}\cdot n+\widetilde{m})}(\widetilde{q}\cdot\widehat{\Sigma_{\mathrm{V}})f_{\mathrm{V}}},\label{eq:CK_2_q} \\
\mathcal{C}_{2,\mu}^{(n),m} & = & \delta(\widetilde{q}^{2}-\widetilde{m}^{2})\Bigg\{-(\widetilde{\nabla}^{\rho}S_{n,\rho\mu})\widehat{\Sigma_{\mathrm{F}}f_{\mathrm{V}}}-\frac{Qe\widetilde{F}_{\mu\nu}n^{\nu}}{(\widetilde{q}\cdot n+\widetilde{m})}\widehat{\Sigma_{\mathrm{F}}f_{\mathrm{V}}}+S_{n,\mu\nu}(\widetilde{\nabla}^{\nu}\widehat{\Sigma_{\mathrm{F}})f_{\mathrm{V}}}\nonumber \\
 &  & -\frac{\epsilon_{\mu\nu\alpha\beta}(\widetilde{m}n^{\alpha}+\widetilde{q}^{\alpha})}{2(\widetilde{q}\cdot n+\widetilde{m})}(\widetilde{\nabla}^{\nu}\widehat{\Sigma_{\mathrm{V}}^{\beta})f_{\mathrm{V}}}+\frac{\epsilon_{\mu\nu\alpha\rho}(\widehat{\Sigma_{\mathrm{V}}^{\nu}f_{\mathrm{V}}})}{2(\widetilde{q}\cdot n+\widetilde{m})}\left(\widetilde{m}\nabla^{\rho}n^{\alpha}-\frac{(\widetilde{q}^{\alpha}+\widetilde{m}n^{\alpha})}{(\widetilde{q}\cdot n+\widetilde{m})}\nabla^{\rho}(\widetilde{q}\cdot n)\right)\Bigg\}\nonumber \\
 &  & +\delta^{\prime}(\widetilde{q}^{2}-\widetilde{m}^{2})\left(-\widetilde{q}_{\mu}\frac{Qe\widetilde{F}_{\rho\sigma}\widetilde{q}^{\rho}n^{\sigma}}{(\widetilde{q}\cdot n+\widetilde{m})}\widehat{\Sigma_{\mathrm{F}}f_{\mathrm{V}}}+\frac{(\widetilde{m}n^{\nu}+q^{\nu})Qe\widetilde{F}_{\mu\nu}}{(\widetilde{q}\cdot n+\widetilde{m})}\left[\widetilde{m}\widehat{\Sigma_{\mathrm{F}}f_{\mathrm{V}}}+(\widetilde{q}\cdot\widehat{\Sigma_{\mathrm{V}})f_{\mathrm{V}}}\right]\right),\label{eq:CK_2_m} 
 \nonumber \\
\end{eqnarray}
where we have used the SKE in Eq.~(\ref{eq:fV kinetic eq}) and the collision
kernel with explicit $\hbar$ is divided into parts proportional to
$\widetilde{q}_{\mu}$ and $\widetilde{m}$ separately, similar to
the free-streaming part.

The AKE (\ref{eq:afA_kinetic eq}) is a generalization of Bargman-Michel-Telegdi
(BMT) equation \citep{BLT_spin} of relativistic spin-$1/2$ fermions with collisions and self-energy corrections
for spin transport. It is found that different from the SKE in Eq.~(\ref{eq:fV kinetic eq}), the self-energies greatly complicate
the structure of the free-streaming part of the AKE. There
emerge terms like the spin relaxation rate proportional to the gradient
of scalar self-energy in Eq.~(\ref{eq:AKE_afA}) even in the absence of
collision terms. Here we have separated the free-streaming part and
collision part of the AKE into the pieces proportional to $a_{\rho}f_{\mathrm{A}}^{<}$,
$\widetilde{q}^{\mu}$, and $\widetilde{m}$, so that taking the massless
limit is straightforward. Nevertheless, the collision terms in the AKE
are not modified by the self-energies because we have only kept up
to the linear order of self-energies by dropping the $\mathcal{O}(\Sigma^{2})$
terms. A smooth connection to QKT without self-energy in Ref.~\citep{Yang:2020hri}
can be also easily checked.


 We can consider a special case by taking the frame vector as the particle's rest
frame, namely $n^{\mu}=n_{\mathrm{r}}^{\mu}=\widetilde{q}^{\mu}/\widetilde{m}$,
to simplify the AKE and axial Wigner function. In such case, the magnetization
current part of the axial Wigner function (\ref{eq:Axial WF formal solution}) vanishes and the AKE can be reduced to 
\begin{eqnarray}
\square_{a,\mu}^{(n_{\mathrm{r}})}\mathcal{A}^{<}+\widetilde{q}_{\mu}\square_{q}^{(n_{\mathrm{r}})}\mathcal{A}^{<}+\widetilde{m}\square_{m,\mu}^{(n_{\mathrm{r}})}\mathcal{A}^{<} & = & \mathcal{C}_{1,\mu}^{(n_{\mathrm{r}})}+\hbar\mathcal{C}_{2,\mu}^{(n_{\mathrm{r}})},\label{eq:Rest_frame_AKE} 
\end{eqnarray}
where
\begin{eqnarray}
\square_{q}^{(n_{\mathrm{r}})}\mathcal{A}^{<} & = & \delta(\widetilde{q}^{2}-\widetilde{m}^{2})\Bigg\{-\widetilde{q}_{\lambda}[\overline{\Sigma}_{\mathrm{A}}^{\lambda}f_{\mathrm{V}}^{<}]_{\mathrm{P.B.}}^{F}+2\widetilde{q}^{\rho}\frac{\overline{\Sigma}_{\mathrm{T},\rho\sigma}}{\hbar\widetilde{m}}a^{\sigma}f_{\mathrm{A}}^{<}+\epsilon^{\lambda\nu\alpha\beta}\widetilde{q}_{\nu}\frac{\overline{\Sigma}_{\mathrm{T},\alpha\beta}}{2\widetilde{m}}\nabla_{\lambda}f_{\mathrm{V}}^{<}\Bigg\}\nonumber \\
 &  & -\delta^{\prime}(\widetilde{q}^{2}-\widetilde{m}^{2})(\widetilde{q}\cdot\widetilde{\nabla}f_{\mathrm{V}}^{<})2(\overline{\Sigma}_{\mathrm{A}}\cdot\widetilde{q}),\label{eq:Rest_frame_AKE_q} \\
\square_{m,\mu}^{(n_{\mathrm{r}})}\mathcal{A}^{<} & = & \delta(\widetilde{q}^{2}-\widetilde{m}^{2})\Bigg\{-\frac{\hbar}{2}\frac{\widetilde{q}^{\beta}}{\widetilde{m}}[(Qe\widetilde{\overline{F}}_{\mu\beta})f_{\mathrm{V}}^{<}]_{\mathrm{P.B.}}^{F}-\frac{\hbar}{2}\epsilon_{\mu\nu\alpha\beta}\frac{\widetilde{q}^{\beta}\nabla^{\nu}\overline{\Sigma}_{\mathrm{F}}}{\widetilde{m}^{2}}\nabla^{\alpha}f_{\mathrm{V}}^{<}\nonumber \\
 &  & \qquad+[\overline{\Sigma}_{\mathrm{F}}(a_{\mu}f_{\mathrm{A}}^{<})]_{\mathrm{P.B.}}^{F}-\frac{(\widetilde{q}\cdot\overline{\Sigma}_{\mathrm{A}})}{\widetilde{m}}\nabla_{\mu}f_{\mathrm{V}}^{<}+\widetilde{m}[\overline{\Sigma}_{\mathrm{A},\mu}f_{\mathrm{V}}^{<}]_{\mathrm{P.B.}}^{F}\nonumber \\
 &  & \qquad+2\frac{\overline{\Sigma}_{\mathrm{T},\mu\nu}}{\hbar}a^{\nu}f_{\mathrm{A}}^{<}-\frac{1}{2}\epsilon_{\mu\nu\alpha\beta}\widetilde{q}^{\nu}[\overline{\Sigma}_{\mathrm{T}}^{\alpha\beta}f_{\mathrm{V}}^{<}]_{\mathrm{P.B.}}^{F}-\frac{1}{2}\epsilon_{\mu\nu\alpha\beta}\overline{\Sigma}_{\mathrm{T}}^{\alpha\beta}\nabla^{\nu}f_{\mathrm{V}}^{<}\Bigg\}\nonumber \\
 &  & +\delta^{\prime}(\widetilde{q}^{2}-\widetilde{m}^{2})\Bigg\{2q_{\mu}Qe\widetilde{F}_{\alpha\beta}\overline{\Sigma}_{\mathrm{T}}^{\alpha\beta}f_{\mathrm{V}}^{<}-\hbar\widetilde{q}^{\nu}Qe\widetilde{F}_{\mu\nu}[\overline{\Sigma}_{\mathrm{F}}f_{\mathrm{V}}^{<}]_{\mathrm{P.B.}}^{F}\nonumber \\
 &  & \qquad-(\widetilde{q}\cdot\widetilde{\nabla}f_{\mathrm{V}}^{<})\left[\hbar\frac{\widetilde{q}^{\nu}}{\widetilde{m}}Qe\widetilde{\overline{F}}_{\mu\nu}+\epsilon_{\mu\nu\alpha\beta}\widetilde{q}^{\nu}\overline{\Sigma}_{\mathrm{T}}^{\alpha\beta}-2\widetilde{m}\overline{\Sigma}_{\mathrm{A},\mu}\right]\Bigg\}.\label{eq:Rest_frame_AKE_m}
\end{eqnarray}
Here $\square_{a,\mu}^{(n_{\mathrm{r}})}\mathcal{A}^{<}$ and $\mathcal{C}_{1,\mu}^{(n_{\mathrm{r}})}$ are the same as Eqs.~(\ref{eq:AKE_afA}) and (\ref{eq:CK_1}) separately 
by setting $n^{\mu}=n_{\mathrm{r}}^{\mu}$, while $\mathcal{C}_{2,\mu}^{(n_{\mathrm{r}})}$
reads,
\begin{eqnarray}
\mathcal{C}_{2,\mu}^{(n_{\mathrm{r}})} & = & \delta(\widetilde{q}^{2}-\widetilde{m}^{2})\frac{1}{2}\epsilon_{\mu\nu\alpha\beta}\widetilde{q}^{\beta}(\nabla^{\nu}\widehat{\Sigma_{\mathrm{V}}^{\alpha})f_{\mathrm{V}}}+\delta^{\prime}(\widetilde{q}^{2}-\widetilde{m}^{2})Qe\widetilde{F}_{\mu\nu}\widetilde{q}^{\nu}\left[\widetilde{m}\widehat{\Sigma_{\mathrm{F}}f_{\mathrm{V}}}+(\widetilde{q}\cdot\widehat{\Sigma_{\mathrm{V}})f_{\mathrm{V}}}\right]. \label{eq:Rf_CK_2}
\end{eqnarray}
In such a case, the AKE is greatly simplified. Although the price is that we lose the direct connection to the massless limit since the choice of the particle's rest frame is invalid when $m\rightarrow 0$, such a simplified expression could be more useful to study the spin transport of strange(s) or heavy quarks in relativistic heavy ion collisions.

\section{Self-energy corrections to the spin polarization of massive quarks\label{sec:Self-energy-corrections-to}}

In this section, we apply the derived formalism in the previous section to analyze the spin polarization phenomenon of massive quarks. Nonetheless, due to complication for solving the AKE to obtain the dynamical component of the axial Wigner function, we concentrate on the non-dynamical component in the particle's rest  frame for simplicity. Albeit the contribution is incomplete, we are still able to estimate the order of magnitude for the self-energy gradient corrections. 
We also demonstrate how the spin polarization induced by $\phi$-meson fields proposed in Refs.~\citep{Sheng:2019kmk,Sheng:2022wsy,Sheng:2022ffb} is connected to the contribution from the tadpole one-point self-energy from a
systematic derivation. Then we investigate the spin polarization from the self-energy corrections with a thermal QCD background at weak coupling. In both cases, we will consider the probe limit by neglecting the backreaction of the massive-quark probe to the backgrounds.
At last, we discuss the possible application to spin polarization and spin alignment in the relativistic heavy ion collision.

\subsection{Mean-field contributions} \label{subsec:mean_field_contribution}

In vacuum, the one-point potential from the tadpole diagram vanishes
automatically, but in a thermal medium, it can make significant
contribution, e.g. to the nuclear equation of state, charge density
distribution, and binding energy in low-energy nuclear physics \citep{Serot:1992ti,Serot:1997xg}.
As found in Eq.~(\ref{eq:Axial WF formal solution}), the mean fields
and their gradients also contribute to the spin polarization pseudovector,
e.g. the external electromagnetic field can induce the difference of spin
polarization of $\Lambda$ and $\overline{\Lambda}$ hyperons \citep{Muller:2018ibh,Guo:2019joy,Xu:2022hql,Buzzegoli:2022qrr,Peng:2022cya}.
Such mean-field contribution to the nucleon transport has also been
realized and derived long ago in the framework of Walecka model \citep{Mrowczynski:1992hq}.
Recently, the spin alignment of $\phi$ mesons in heavy ion collisions
can be explained by the strong-force fields. The mean $\phi$ field in the medium induces a spin polarization of $s/\overline{s}$ quark, and their correlations can contribute to the spin alignment of final-state mesons \citep{Sheng:2022ffb,Sheng:2022wsy,Sheng:2023urn}. Such a mean-field contribution could be further manifested in our formalism.

In our present interaction in the Lagrangian (\ref{eq:Lagrangian}), since we have already separated the classical part and quantum part of the gauge fields and taken $\langle a_\mu\rangle=\mathfrak{a}^{a}_{\mu}=0$, the mean field contributions from QED/QCD interaction are zero. See Appendix.\ref{sec:The- mean-field- contributions-in } for more details.
More importantly, our theory can incorporate another kind of interaction such as the interaction between $s,\overline{s}$
quarks and the $\phi$ meson field in the quark-meson model,
\begin{eqnarray}
\mathcal{L}_{\mathrm{int}} & = & g_{\phi}\phi_{\mu}\overline{\psi}\gamma^{\mu}\psi,
\end{eqnarray}
and then we find 
\begin{eqnarray}
i\hbar\int_{\mathrm{C}}\mathrm{d}^{4}z\Sigma(x,z)G(z,y) & = & g_{\phi}\langle\phi_{\mu}\rangle\gamma^{\mu}G(x,y)=\int_{\mathrm{C}}\mathrm{d}^{4}z\delta_{\mathrm{C}}^{(4)}(x-z)\Sigma_{\delta}(x)G(z,y),
\end{eqnarray}
so that 
\begin{eqnarray}
\Sigma_{\delta}(x) & = & g_{\phi}\gamma^{\mu}\langle\phi_{\mu}\rangle.
\end{eqnarray}
If we neglect the collisional energies $\Sigma^{\lessgtr}$, we find
such a mean-field self-energy contributes to 
\begin{eqnarray}
\overline{\Sigma}_{\mathrm{V}}^{\mu} & = & g_{\phi}\langle\phi_{\mu}\rangle,
\end{eqnarray}
and the spin polarization of $s$ quark has a source from, 
\begin{eqnarray}
g_{\phi}F_{\mu\nu}^{\phi} & = & 2\nabla_{[\mu}\overline{\Sigma}_{\mathrm{V},\nu]}^{\phi}=g_{\phi}(\partial_{\mu}\langle\phi_{\nu}\rangle-\partial_{\nu}\langle\phi_{\mu}\rangle),
\end{eqnarray}
namely,
\begin{eqnarray}
\delta\mathcal{A}_{s}^{<,\mu} & = & -2\pi\delta^{\prime}(\widetilde{q}^{2}-\widetilde{m}^{2})\frac{\hbar}{2}\epsilon_{\mu\nu\alpha\beta}\widetilde{q}^{\nu}g_{\phi}F_{\phi}^{\alpha\beta}f_{\mathrm{V}}^{<},
\end{eqnarray}
by working with the particle's rest frame $n^\mu=n^\mu_{\mathrm{r}}$ and the modification to spin polarization density in phase space becomes
\begin{eqnarray}
\delta\mathcal{P}_{s}^{\mu}(\mathbf{q},x) & = & \frac{\int\frac{\mathrm{d}q_{0}}{2\pi}\delta\mathcal{A}_{s}^{<,\mu}(q,x)}{2\int\frac{\mathrm{d}q_{0}}{2\pi}\mathcal{F}_{s}^{<}(q,x)}=-\frac{\hbar g_{\phi} }{4mT}B_{\phi}^{\mu}(1-f_{\mathrm{V}}^{<}),
\label{eq:MF_Modif_spin_vector}
\end{eqnarray}
where $B_{\phi}^{\mu}=\epsilon^{\mu\nu\alpha\beta}\hat{t}_{\nu}F_{\phi\alpha\beta}/2$ with $\hat{t}^{\nu}=(1,\bm 0)$ being a temporal vector and the factor of $2$ in the denominator comes from the $(2s+1)$
spin degeneracy of fermions and we have dropped the non-linear terms
of self-energies. We have inserted the global equilibrium distribution function of quarks in a Fermi-Dirac form that only depends on $q_0$ with global equlibrium temperature $T$. We also set $E_{\phi}^{\mu}=F_{\phi}^{\mu\nu}\hat{t}_{\nu}=0$, for which a nonzero $E_{\phi}^{\mu}$ as an effective electric field should in principle drive the $f^<_{\rm V}$ out of global equilibrium. Eq.~(\ref{eq:MF_Modif_spin_vector}) is consistent with
the results in Refs. \citep{Sheng:2019kmk,Sheng:2022ffb,Sheng:2023urn} and obtained from the quantum
field theory systematically.

Notably, in the above derivation we have assumed the ensemble average of $\phi$ meson fields are nonzero in local equilibrium, that is, $\langle\phi(x)\rangle\neq0$ at different spacetime point $x$. When we try to apply Eq.~(\ref{eq:MF_Modif_spin_vector}) to the calculation of spin density matrix element  of $\phi$ meson, the phase space density of spin density matrix, e.g. $\rho_{00}(q,x)$, can acquire the linear terms related to the vector meson field strength tensor.
When we take the spacetime average along the freeze-out hypersurface, the linear term proportional to $B^\mu_\phi$ in Eq.~(\ref{eq:MF_Modif_spin_vector}) vanishes. However, the correlation terms of these meson fields, such as $(B_\phi)^2$, can survive similar to the discussion based on the coalescence models \citep{Sheng:2019kmk,Sheng:2022ffb,Sheng:2023urn}.
On the other hand, the authors in Refs.~\citep{Muller:2021hpe,Yang:2021fea,Kumar:2022ylt,Kumar:2023ghs} expect the vanish of ensemble average of the individual color fields but the existence of correlation functions of two color fields. Therefore, one could alternatively derive the spin polarization density in phase space in the operator level and keep only the correlation functions of the field operators in the practical calculation of spin density matrix.
More discussions on the possible corrections to the spin alignment based on the coalescence models are present in Sec.~\ref{subsec:application}. 


\subsection{Effects from the thermal QCD background}

Now we discuss another source of dynamical polarization from the Fock
diagrams, i.e. the one-loop quark self-energy. Since in our power counting,
the one-loop level self-energy promote the corresponding correction
$\delta\mathcal{A}^{\mu}$ to $\mathcal{O}(\hbar^{1})$, we only need
to concentrate the classical one-loop quark self-energy at finite
temperature. In such one-loop level, $\Sigma_{\mathrm{A}}=\Sigma_{\mathrm{T}}=0$ if the quantum corrections of gluonic Wigner functions are neglected,
and the result should be same as the self-energy calculated in the
standard textbook, e.g. see Ref.~\citep{Bellac:2011kqa}; and the Wigner function
that will be used are needed only up to $\mathcal{O}(\hbar^{0})$,
namely, the vector and scalar components are necessary. That is, we
need to calculate, 
\begin{eqnarray}
\Sigma^{\mathrm{r}}(q,X) & = & i\Sigma^{++}(q,X)-i\Sigma^{>}(q,X).
\end{eqnarray}
Since $\Sigma^{\lessgtr}$ are real, we have, 
\begin{eqnarray}
\mathrm{Re}\Sigma^{\mathrm{r}}(q,X) & = & -\mathrm{Im}\Sigma^{++}(q,X).
\end{eqnarray}
We can write down  
\begin{eqnarray}
-\Sigma^{++}(x,y) & = & (iQe\gamma_{\mu})S^{++}(x,y)(iQe\gamma_{\nu})G^{\mu\nu,++}(x,y),
\end{eqnarray}
from the Feynman diagram and obtain 
\begin{eqnarray}
-\Sigma^{++}(q,X) & = & -Q^{2}e^{2}\int\frac{\mathrm{d}^{4}q_{1}}{(2\pi\hbar)^{4}}\gamma_{\mu}S^{++}(q_{1},X)\gamma_{\nu}G^{\mu\nu,++}(q-q_{1},X),
\end{eqnarray}
by taking the Wigner transform. 
Thus, 
\begin{eqnarray}
\Sigma_{\mathrm{V},\alpha}^{++} & = & \frac{1}{4}\mathrm{Tr}(\gamma_{\alpha}\Sigma^{++})\nonumber \\
 & = & Q^{2}e^{2}\int\frac{\mathrm{d}^{4}q_{1}}{(2\pi\hbar)^{4}}(\eta_{\alpha\mu}\eta_{\beta\nu}-\eta_{\alpha\beta}\eta_{\mu\nu}+\eta_{\alpha\nu}\eta_{\beta\mu})\mathcal{V}^{++,\beta}(q_{1},X)G^{\mu\nu,++}(q-q_{1},X),
\end{eqnarray}
where we have simply dropped the second term related to axial Wigner
function and the overall $\hbar$ in the propagators and vertices
due to our power counting.

We remark that next we consider an $s$-quark probe propagating in a locally thermal equilibrium QCD background. In Sec.~\ref{subsubsection:self-energy}, the distribution functions of background quarks and gluons are in local equilibrium, denoted as $n_{\rm{V,leq}}$ and $\tilde{n}_{\rm{V,leq}}$, respectively. Later, in Sec.~\ref{subsubsection:probe-s}, the distribution of the $s$-quark probe, denoted as $f_\mathrm{V}^<$, can be either off-equilibrium or in-equilibrium.

\subsubsection{Calculation of self-energy}\label{subsubsection:self-energy}
For clarity, following the approach in Ref.~\cite{Bellac:2011kqa}, we first consider the case for electromagnetic interaction and then extend the result to the strong interaction.
Up to the leading order in coupling, from Appendix.\ref{sec:Feynman-vector-Wigner},
we can get,
\begin{eqnarray}
\Sigma_{\mathrm{V},\alpha}^{++} & = & -Q^{2}e^{2}\int\frac{\mathrm{d}^{4}q_{1}}{(2\pi)^{4}}2q_{1,\alpha}\left[\mathrm{P.V.}\frac{i}{q_{1}^{2}-m^{2}}-2\pi\delta(q_{1}^{2}-m^{2})\left(\widetilde{n}_{\mathrm{V}}^{<}(q_{1},X)-\frac{1}{2}\right)\right]\nonumber \\
 &  & \times\left[\mathrm{P.V.}\frac{-i}{(q-q_{1})^{2}}-2\pi\left(n(q-q_{1},X)+\frac{1}{2}\right)\delta\left((q-q_{1})^{2}\right)\right],
\end{eqnarray}
where $\mathrm{P.V.}$ stands for the principal value, and $\widetilde{n}_{\mathrm{V}},n_{\mathrm{V}}$ are the background fermion distribution
and boson distribution, respectively. They are taken to be in local equilibrium, namely, \begin{eqnarray}
n_{\mathrm{V,leq}}(q,X)=\frac{1}{e^{\beta|q_{0}|}-1},\quad
\widetilde{n}_{\mathrm{V}}^{<}(q,X) & = & \theta(q_{0})\widetilde{n}_{\mathrm{V},+,\mathrm{leq}}^{<}(q,X)+\theta(-q_{0})\widetilde{n}_{\mathrm{V},-,\mathrm{leq}}^{<}(q,X),\label{eq:thermal equilibrium distribution}
\end{eqnarray} where $\beta=1/T$ is the inverse temperature and
\begin{eqnarray}
\widetilde{n}_{\mathrm{V},+,\mathrm{leq}}^{<}(q,X)=\frac{1}{e^{\beta(|q_{0}|-\mu)}+1},\quad \widetilde{n}_{\mathrm{V},-,\mathrm{leq}}^{<}(q,X)=\frac{1}{e^{\beta(|q_{0}|+\mu)}+1}.
\end{eqnarray} 
We do not consider the possible quantum corrections for the background and thus these background distributions here are only up to $\mathcal{O}(\hbar^0)$ in our power counting. Here, we mainly focus on the imaginary part of the Feynman vector self-energy and get
\begin{eqnarray}\label{eq:Im_Sigma_V}
\mathrm{Im}\Sigma_{\mathrm{V},\alpha}^{++}(q,X) & = & -2Q^{2}e^{2}\int\frac{\mathrm{d}^{4}q_{1}}{(2\pi)^{3}}q_{1,\alpha}\left[\mathrm{P.V.}\frac{1}{(q-q_{1})^{2}}\delta(q_{1}^{2}-m^{2})\left(\widetilde{n}_{\mathrm{V}}^{<}(q_{1},X)-\frac{1}{2}\right)\right.\nonumber \\
 &  & \left.-\mathrm{P.V.}\frac{1}{q_{1}^{2}-m^{2}}\delta\left((q-q_{1})^{2}\right)\left(n_{\mathrm{V}}(q-q_{1},X)+\frac{1}{2}\right)\right].
\end{eqnarray}
We assume the distribution functions are in local equilibrium and
denote 
\begin{eqnarray}
    q_{1,0} &=&u\cdot q_{1}, \qquad q_{1}^{\alpha}=q_{1,0}u^{\alpha}+q_{1,\perp}^{\alpha}, \qquad |q_{\perp}|=\sqrt{-q_{\perp}\cdot q_{\perp}}
    \nonumber \\ 
    E_{1}&=&E_{\mathbf{q}_{1}}=\sqrt{m^{2}+|q_{1,\perp}|^{2}},  \qquad
    E_{2}=E_{\mathbf{q}-\mathbf{q}_{1}}=\sqrt{|q_{\perp}-q_{1,\perp}|^{2}} . 
    \label{eq:momentum_decomposition}
\end{eqnarray}
Here $u_{\mu}$ is
the fluid four-velocity and  we will work in the fluid comoving
frame (do not confuse it with the choice of a frame for Wigner functions $n^\mu$) of the QCD background when calculating the vector self-energy. Then, we find,
\begin{eqnarray}
 &  & \mathrm{Im}\Sigma_{\mathrm{V},\alpha}^{++}(q,X)\nonumber \\
 & = & -2Q^{2}e^{2}\int\frac{\mathrm{d}^{3}\mathbf{q}_{1}}{(2\pi)^{3}}u_{\alpha}\frac{1}{4E_{1}E_{2}}\left[E_{1}\left(\frac{1}{q_{0}-E_{1}-E_{2}}-\frac{1}{q_{0}-E_{1}+E_{2}}\right)\left(\widetilde{n}_{\mathrm{V,+,leq}}^{<}(E_{1},X)-\frac{1}{2}\right)\right.\nonumber \\
 &  & -E_{1}\left(\frac{1}{q_{0}+E_{1}-E_{2}}-\frac{1}{q_{0}+E_{1}+E_{2}}\right)\left(\widetilde{n}_{\mathrm{V,-,leq}}^{<}(E_{1},X)-\frac{1}{2}\right)\nonumber \\
 &  & -(q_{0}-E_{2})\left(\frac{1}{q_{0}-E_{1}-E_{2}}-\frac{1}{q_{0}+E_{1}-E_{2}}\right)\left(n_{\mathrm{V,leq}}(E_{2},X)+\frac{1}{2}\right)\nonumber \\
 &  & \left.-(q_{0}+E_{2})\left(\frac{1}{q_{0}-E_{1}+E_{2}}-\frac{1}{q_{0}+E_{1}+E_{2}}\right)\left(n_{\mathrm{V,leq}}(E_{2},X)+\frac{1}{2}\right)\right]\nonumber \\
 &  & -2Q^{2}e^{2}\int\frac{\mathrm{d}^{3}\mathbf{q}_{1}}{(2\pi)^{3}}\frac{q_{1,\perp,\alpha}}{4E_{1}E_{2}}\Bigg[\left(\frac{1}{q_{0}-E_{1}-E_{2}}-\frac{1}{q_{0}-E_{1}+E_{2}}\right)\left(\widetilde{n}_{\mathrm{V,+,leq}}^{<}(E_{1},X)-\frac{1}{2}\right)\nonumber \\
 &  & +\left(\frac{1}{q_{0}+E_{1}-E_{2}}-\frac{1}{q_{0}+E_{1}+E_{2}}\right)\left(\widetilde{n}_{\mathrm{V,-,leq}}^{<}(E_{1},X)-\frac{1}{2}\right)\nonumber \\
 &  & -\left(\frac{1}{q_{0}-E_{1}-E_{2}}-\frac{1}{q_{0}+E_{1}-E_{2}}+\frac{1}{q_{0}-E_{1}+E_{2}}-\frac{1}{q_{0}+E_{1}+E_{2}}\right)\left(n_{\mathrm{V,leq}}(E_{2},X)+\frac{1}{2}\right)\Bigg].
 \nonumber \\
\end{eqnarray}
We then adopt the hard thermal loop (HTL)
approximation, under which the external momentum $q\sim eT$ is soft and the internal
momentum $q_{1}\sim T$ is hard so that $q_{1}\gg q,m$, and we can
neglect the external momentum compared to the internal momenta and
extract the leading contributions in temperature $T$. It is valid
for the current quarks at high collision energy.

The quark mass can be neglected compared to the loop momenta $q_{1}$,
i.e.
\begin{equation}
    E_{2}\simeq E_{1}\simeq|q_{1,\perp}|\equiv E,\qquad E_{2}\simeq E-|q_{\perp}|\cos\theta \qquad 
    \cos\theta=\cos\langle\mathbf{q}_{\perp},\mathbf{q}_{1,\perp}\rangle,
\end{equation}
and thus, we extract the $T^{2}$ behavior terms,
\begin{eqnarray}
& & \mathrm{Im}\Sigma_{\mathrm{V},\alpha}^{++}(q,X) \nonumber \\
& \simeq & 2Q^{2}e^{2}\int\frac{\mathrm{d}^{3}\mathbf{q}_{1}}{(2\pi)^{3}}\frac{u_{\alpha}}{4E}\frac{1}{q_{0}+|q_{\perp}|\cos\theta}\left[\widetilde{n}_{\mathrm{V,+,leq}}^{<}(E,X)+\widetilde{n}_{\mathrm{V,-,leq}}^{<}(E,X)+2n_{\mathrm{V,leq}}(E_{2},X)\right]\nonumber \\
 &  & -2Q^{2}e^{2}\int\frac{\mathrm{d}^{3}\mathbf{q}_{1}}{(2\pi)^{3}}\frac{\hat{q}_{1,\perp,\alpha}}{4E}\frac{1}{q_{0}+|q_{\perp}|\cos\theta}\left[\widetilde{n}_{\mathrm{V,+,leq}}^{<}(E_{1},X)+\widetilde{n}_{\mathrm{V,-,leq}}^{<}(E_{1},X)+2n_{\mathrm{V,leq}}(E_{2},X)\right]\nonumber \\
 & = & Q^{2}e^{2}u_{\alpha}\frac{1}{8}(\frac{\mu^{2}}{\pi^{2}}+T^{2})\int\frac{\mathrm{d}\Omega}{4\pi}\frac{1}{q_{0}+|q_{\perp}|\cos\theta}-Q^{2}e^{2}\frac{1}{8}(\frac{\mu^{2}}{\pi^{2}}+T^{2})\int\frac{\mathrm{d}\Omega}{4\pi}\frac{\hat{q}_{1,\perp,\alpha}}{q_{0}+|q_{\perp}|\cos\theta}\nonumber \\
 & = & \frac{1}{8}(\frac{\mu^{2}}{\pi^{2}}+T^{2})Q^{2}e^{2}\left[\frac{u_{\alpha}}{|q_{\perp}|}\mathcal{Q}_{0}\left(\frac{q_{0}}{|q_{\perp}|}\right)+\frac{q_{\perp,\alpha}}{|q_{\perp}|^{2}}\mathcal{Q}_{1}\left(\frac{q_{0}}{|q_{\perp}|}\right)\right], 
 \label{eq:Im_Sigma_++_QED}
\end{eqnarray}
where we have used $\hat{q}_{1,\perp,\alpha}=\frac{q_{1,\perp,\alpha}}{|q_{1,\perp}|}\simeq\frac{q_{1,\perp,\alpha}}{E_{1}}$
and dropped the divergent terms in vacuum; we have also used the simple
angular symmetry. Because we select the real part of the integral,
$q^{\mu}$ is always timelike, $|q_{\perp}|<|q_{0}|$. We have also
introduced the real part of the Legendre function,
\begin{eqnarray}
\mathcal{Q}_{0}(x) & = & \frac{1}{2}\ln\left|\frac{x+1}{x-1}\right|,\qquad \mathcal{Q}_{1}(x)=x\mathcal{Q}_{0}(x)-1.
\end{eqnarray}

It is straightforward to generalize Eq.~(\ref{eq:Im_Sigma_++_QED}) to the QCD background, by substituting  $Q^{2}e^{2}\to g^{2}C_{\mathrm{F}}$ with $C_{\mathrm{F}}$ being the quadratic Casimir operator of $\mathrm{SU}(N)$ group, 
we get
\begin{eqnarray}
\mathrm{Im}\Sigma_{\mathrm{V},\alpha}^{++}(q,X) & = & m_{f}^{2}\left[\frac{u_{\alpha}}{|q_{\perp}|}\mathcal{Q}_{0}\left(\frac{q_{0}}{|q_{\perp}|}\right)+\frac{q_{\perp,\alpha}}{|q_{\perp}|^{2}}\mathcal{Q}_{1}\left(\frac{q_{0}}{|q_{\perp}|}\right)\right].\label{eq:Im_SE_QCD}
\end{eqnarray}
where the thermal fermion mass with QCD interaction is denoted as 
\begin{eqnarray}
m_{f}^{2} & = & C_{\mathrm{F}}\frac{g^{2}}{8}\left(\frac{\mu^{2}}{\pi^{2}}+T^{2}\right).
\end{eqnarray}
The gradient of the self-energy reads 
\begin{eqnarray}
& & \partial_{\beta}\mathrm{Im}\Sigma_{\mathrm{V},\alpha}^{++}(q,X) \nonumber \\
& = & C_{\mathrm{F}}\frac{g^{2}}{4}\left(\frac{\mu}{\pi^{2}}\partial_{\beta}\mu+T\partial_{\beta}T\right)\left[\frac{u_{\alpha}}{|q_{\perp}|}\mathcal{Q}_{0}\left(\frac{q_{0}}{|q_{\perp}|}\right)+\frac{q_{\perp,\alpha}}{|q_{\perp}|^{2}}\mathcal{Q}_{1}\left(\frac{q_{0}}{|q_{\perp}|}\right)\right]\nonumber \\
 &  & +m_{f}^{2}\left\{\frac{\partial_{\beta}u_{\alpha}}{|q_{\perp}|}\left[\mathcal{Q}_{0}\left(\frac{q_{0}}{|q_{\perp}|}\right)-\frac{q_{0}}{|q_{\perp}|}\mathcal{Q}_{1}\left(\frac{q_{0}}{|q_{\perp}|}\right)\right]-u_{\alpha}q_{\perp}^{\gamma}(\partial_{\beta}u_{\gamma})\frac{2}{|q_{\perp}|^{2}}\mathcal{Q}_{1}\left(\frac{q_{0}}{|q_{\perp}|}\right)\right.\nonumber \\
 &  & \left.+q_{\perp,\alpha}q_{\perp}^{\gamma}(\partial_{\beta}u_{\gamma})\frac{1}{|q_{\perp}|^{3}}\left[\mathcal{Q}_{0}\left(\frac{q_{0}}{|q_{\perp}|}\right)-3\frac{q_{0}}{|q_{\perp}|}\mathcal{Q}_{1}\left(\frac{q_{0}}{|q_{\perp}|}\right)\right]\right\},
\end{eqnarray}
where we have used
\begin{eqnarray}
\partial_{\beta}q_{\perp,\alpha} & = & -q_{0}\partial_{\beta}u_{\alpha}-u_{\alpha}q_{\perp}^{\gamma}\partial_{\beta}u_{\gamma},\qquad\partial_{\beta}|q_{\perp}|=\frac{q_{0}}{|q_{\perp}|}q_{\perp}^{\gamma}\partial_{\beta}u_{\gamma},\nonumber \\
\partial_{\beta}q_{0} & = & q_{\perp}^{\gamma}\partial_{\beta}u_{\gamma},\qquad\partial_{\beta}\frac{q_{0}}{|q_{\perp}|}=-\frac{q^{2}}{|q_{\perp}|^{3}}q_{\perp}^{\gamma}\partial_{\beta}u_{\gamma}.
\end{eqnarray}
In the non-relativistic limit with $|q_{\perp}|/m\to0^{+}$, we find 
\begin{eqnarray}
\partial_{\beta}\mathrm{Im}\Sigma_{\mathrm{V},\alpha}^{++}(q,X) & \to & C_{\mathrm{F}}\frac{g^{2}}{4}\left(\frac{\mu}{\pi^{2}}\partial_{\beta}\mu+T\partial_{\beta}T\right)\left[\frac{u_{\alpha}}{q_{0}}+\frac{q_{\perp,\alpha}}{q_{0}^{2}}\right]\nonumber \\
 &  & +m_{f}^{2}\left[\frac{2}{3q_{0}}\partial_{\beta}u_{\alpha}-u_{\alpha}q_{\perp}^{\gamma}(\partial_{\beta}u_{\gamma})\frac{2}{3q_{0}^{2}}+q_{\perp,\alpha}q_{\perp}^{\gamma}(\partial_{\beta}u_{\gamma})\frac{-4}{15q_{0}^{3}}\right]\nonumber \\
 & \to & C_{\mathrm{F}}\frac{g^{2}}{4}\left(\frac{\mu}{\pi^{2}}\partial_{\beta}\mu+T\partial_{\beta}T\right)\frac{u_{\alpha}}{m}+\frac{2m_{f}^{2}}{3m}\partial_{\beta}u_{\alpha}.
\end{eqnarray}
In this limit, we still have the contribution
to the spin polarization vector from the fluid vorticity or the temperature and chemical-potential gradients.

\subsubsection{Modification to spin polarization vector} \label{subsubsection:probe-s}
 We now consider the spin polarization of $s$ quarks in a QCD background with the contributions from the self-energy in Eq.~(\ref{eq:Im_SE_QCD}).  Up to the one quark-loop level, the new source of spin polarization for general distribution of $s$ quarks becomes  
\begin{eqnarray}
\delta\mathcal{A}_{\mu,\mathrm{SE}}^{<}(q,X) & = & -2\pi\epsilon(q_{0})\delta^{\prime}(q^{2}-m^{2})f_{\mathrm{V}}^{<}\frac{\hbar}{2}\epsilon_{\mu\nu\alpha\beta}q^{\nu}(2\nabla^{[\alpha}\overline{\Sigma}_{\mathrm{V}}^{\beta]})\nonumber \\
 & = & 2\pi\epsilon(q_{0})\partial_{q}^{\nu}\delta(q^{2}-m^{2})f_{\mathrm{V}}^{<}\frac{\hbar}{2}\epsilon_{\mu\nu\beta\alpha}\partial_{X}^{\beta}\mathrm{Im}\Sigma_{\mathrm{V}}^{++,\alpha},
\end{eqnarray}
where in the second line we have neglected the external electromagnetic field. Note that the vector distribution of the $s$ quarks $f^<_{\mathrm{V}}$ can be off-equilibrium in general.
Then inserting the derivative of the self-energies, we get \footnote{Such contribution to the spin polarization may not induce entropy
production \citep{Bellac:2011kqa} as the self-energy related to the
collision term $\mathcal{C}[f]$ are set to zero.}
\begin{eqnarray}
& & \delta\mathcal{A}_{\mathrm{SE}}^{<,\mu}(q,X) 
\nonumber \\
& = & 2\pi\epsilon(q_{0})[\partial_{q,\nu}\delta(q^{2}-m^{2})]f_{\mathrm{V}}^{<}\frac{\hbar^{2}}{2}\epsilon^{\mu\nu\beta\alpha} \nonumber \\
& & \times \left\{ C_{\mathrm{F}}\frac{g^{2}}{4}(\frac{\mu}{\pi^{2}}\partial_{\beta}\mu+T\partial_{\beta}T)\left[\frac{u_{\alpha}}{|q_{\perp}|}\mathcal{Q}_{0}\left(\frac{q_{0}}{|q_{\perp}|}\right)+\frac{q_{\perp,\alpha}}{|q_{\perp}|^{2}}\mathcal{Q}_{1}\left(\frac{q_{0}}{|q_{\perp}|}\right)\right]\right.\nonumber \\
 &  & +m_{f}^{2}\Bigg[\frac{\partial_{\beta}u_{\alpha}}{|q_{\perp}|}\left(\mathcal{Q}_{0}\left(\frac{q_{0}}{|q_{\perp}|}\right)-\frac{q_{0}}{|q_{\perp}|}\mathcal{Q}_{1}\left(\frac{q_{0}}{|q_{\perp}|}\right)\right)-u_{\alpha}q_{\perp}^{\gamma}(\partial_{\beta}u_{\gamma})\frac{2}{|q_{\perp}|^{2}}\mathcal{Q}_{1}\left(\frac{q_{0}}{|q_{\perp}|}\right)\nonumber \\
 &  & \left.+q_{\perp,\alpha}q_{\perp}^{\gamma}(\partial_{\beta}u_{\gamma})\frac{1}{|q_{\perp}|^{3}}\left(\mathcal{Q}_{0}\left(\frac{q_{0}}{|q_{\perp}|}\right)-3\frac{q_{0}}{|q_{\perp}|}\mathcal{Q}_{1}\left(\frac{q_{0}}{|q_{\perp}|}\right)\right) \Bigg]\right\} .
\end{eqnarray}
Since we are working in the $s$ quarks' rest frame when evaluating the spin polarization vector, the magnetization current term in the axial Wigner function (\ref{eq:Axial WF formal solution}) is, therefore, vanishing. 
On the other hand, note that we also decompose the particles' momentum by the fluid velocity $u^\mu$
as shown in Eqs.~(\ref{eq:momentum_decomposition}).  
The modification to the spin polarization pseudovector can be obtained
with a general off-equilibrium distribution function $f_{\mathrm{V}}^{<}$
for $s$ quark with $q_{0}=u\cdot{q}>0$ only, 
\begin{eqnarray}
 &  & \delta\mathcal{J}_{5,\mathrm{SE}}^{\mu}(\mathbf{q},x) =\int\frac{\mathrm{d}q_{0}}{2\pi}\delta\mathcal{A}_{\mathrm{SE}}^{<,\mu}(q,x) \nonumber \\
 &=&\frac{\hbar}{2}\epsilon^{\mu\nu\beta\alpha}\int_{0}^{\infty}\mathrm{d}q_{0}\partial_{q,\nu}\delta(q^{2}-m^{2})f_{\mathrm{V}}^{<}\left(\partial_{X,\beta}\mathrm{Im}\Sigma_{\mathrm{V},\alpha}^{++}\right)\nonumber \\
 & = & \frac{\hbar}{2}\epsilon^{\mu\nu\beta\alpha}\left[\frac{1}{2E_{q}}\left(\partial_{q_{\perp},\nu}f_{\mathrm{V}}^{<}(E_{q},\mathbf{q})+\frac{q_{\perp,\nu}}{E_{q}^{2}}f_{\mathrm{V}}^{<}(E_{q},\mathbf{q})\right)\left(\partial_{X,\beta}\mathrm{Im}\Sigma_{\mathrm{V},\alpha}^{++}\right)_{q_{0}=E_{q}}\right]\nonumber \\
 &  & +\frac{\hbar}{2}\epsilon^{\mu\nu\beta\alpha}\frac{f_{\mathrm{V}}^{<}(E_{q},\mathbf{q})}{2E_{q}}\partial_{q_{\perp},\nu}\left(\partial_{X,\beta}\mathrm{Im}\Sigma_{\mathrm{V},\alpha}^{++}\right)_{q_{0}=E_{q}}
 \nonumber \\
 &  &
 -\frac{\hbar}{2}\epsilon^{\mu\nu\beta\alpha}\frac{\left(\partial_{q,\nu}f_{\mathrm{V}}^{<}(q_{0},\mathbf{q})\right)_{q_{0}=E_{q}}}{2E_{q}}\left(\partial_{X,\beta}\mathrm{Im}\Sigma_{\mathrm{V},\alpha}^{++}\right)_{q_{0}=E_{q}}\nonumber \\
 &  & -\frac{\hbar}{2}\epsilon^{\mu\nu\beta\alpha}\frac{f_{\mathrm{V}}^{<}(E_{q},\mathbf{q})}{2E_{q}}\left[\partial_{q,\nu}\left(\partial_{X,\beta}\mathrm{Im}\Sigma_{\mathrm{V},\alpha}^{++}\right)\right]_{q_{0}=E_{q}},
\end{eqnarray}
where we have dropped the boundary terms when performing the integral and we decompose the momentum derivatives as 
\begin{eqnarray}
\partial_{q,\nu} & = & u_{\nu}\partial_{q_{0}}+\partial_{q_{\perp},\nu},
\end{eqnarray}
as in the comoving frame $u_{\mu}\approx(1,\mathbf{0})$ while $\partial_{\nu}u_{\mu}\neq0$.
By using
\begin{eqnarray}
\partial_{q_{\perp},\nu}E_{q} & = & -\frac{q_{\perp,\nu}}{E_{q}},\quad\partial_{q_{\perp},\nu}|q_{\perp}|=-\frac{q_{\perp,\nu}}{|q_{\perp}|},\quad\partial_{q_{\perp},\nu}\frac{E_{q}}{|q_{\perp}|}=\frac{m^{2}q_{\perp,\nu}}{|q_{\perp}|^{3}E_{q}},
\end{eqnarray}
we obtain
\begin{eqnarray}
 &  & \delta\mathcal{J}_{5,\mathrm{SE}}^{\mu}(\mathbf{q},x)\nonumber \\
 & = & \frac{\hbar}{2}\frac{\epsilon^{\mu\nu\beta\alpha}u_{\alpha}q_{\perp,\nu}}{2E_{q}^{2}}C_{\mathrm{F}}\frac{g^{2}}{4E_{q}^{2}}(\frac{\mu}{\pi^{2}}\partial_{\beta}\mu+T\partial_{\beta}T)G_{\mathrm{T}}(E_{q},\mathbf{q})+\frac{\hbar}{2}m_{f}^{2}\frac{\epsilon^{\mu\nu\beta\alpha}q_{\perp,\nu}u_{\beta}Du_{\alpha}}{2E_{q}^{4}}G_{\mathrm{T}}(E_{q},\mathbf{q})\nonumber \\
 &  & +\frac{\hbar}{2}m_{f}^{2}\frac{\epsilon^{\mu\nu\beta\alpha}u_{\nu}q_{\perp,\alpha}q_{\perp}^{\gamma}\sigma_{\beta\gamma}}{2E_{q}^{5}}G_{\omega_{1}}(E_{q},\mathbf{q})-\frac{\hbar}{2}m_{f}^{2}\frac{(q_{\perp}\cdot\omega)}{E_{q}^{4}}\left(\frac{q_{\perp}^{\mu}}{2E_{q}}G_{\omega_{1}}(E_{q},\mathbf{q})+u^{\mu}G_{\mathrm{T}}(E_{q},\mathbf{q})\right)\nonumber \\
 &  & +\frac{\hbar}{2}m_{f}^{2}\frac{\omega^{\mu}}{E_{q}^{3}}\left(-\frac{|q_{\perp}|^{2}}{2E_{q}^{2}}G_{\omega_{1}}(E_{q},\mathbf{q})+G_{\omega_{2}}(E_{q},\mathbf{q})\right),\label{eq:Additional J5}
\end{eqnarray}
where we have used the tensor decomposition for the gradient of the fluid velocity
(\ref{eq:fluid derivative}) and introduced the following dimensionless
coefficients, 
\begin{eqnarray}
G_{\mathrm{T}}(E_{q},\mathbf{q}) & = & f_{\mathrm{V}}^{<}(E_{q},\mathbf{q})\left[\mathcal{Q}_{0}\left(\frac{E_{q}}{|q_{\perp}|}\right)+\frac{E_{q}}{|q_{\perp}|}\mathcal{Q}_{1}\left(\frac{E_{q}}{|q_{\perp}|}\right)\right]\frac{E_{q}}{|q_{\perp}|}\nonumber \\
 &  & +(\partial_{E_{q}}f_{\mathrm{V}}^{<}(E_{q},\mathbf{q}))\frac{E_{q}^{2}}{|q_{\perp}|}\left[\frac{E_{q}}{|q_{\perp}|}\mathcal{Q}_{1}\left(\frac{E_{q}}{|q_{\perp}|}\right)-\mathcal{Q}_{0}\left(\frac{E_{q}}{|q_{\perp}|}\right)\right],\\
G_{\omega_{1}}(E_{q},\mathbf{q}) & = & f_{\mathrm{V}}^{<}(E_{q},\mathbf{q})\frac{2E_{q}^{2}}{|q_{\perp}|^{2}}\left[\frac{3E_{q}^{2}}{|q_{\perp}|^{2}}\mathcal{Q}_{1}\left(\frac{E_{q}}{|q_{\perp}|}\right)-1\right]\nonumber \\
 &  & -(\partial_{E_{q}}f_{\mathrm{V}}^{<}(E_{q},\mathbf{q}))\frac{E_{q}^{3}}{|q_{\perp}|^{2}}\left[\frac{E_{q}}{|q_{\perp}|}\mathcal{Q}_{0}\left(\frac{E_{q}}{|q_{\perp}|}\right)-\frac{2m^{2}+E_{q}^{2}}{|q_{\perp}|^{2}}\mathcal{Q}_{1}\left(\frac{E_{q}}{|q_{\perp}|}\right)\right],\\
G_{\omega_{2}}(E_{q},\mathbf{q}) & = & f_{\mathrm{V}}^{<}(E_{q},\mathbf{q})\frac{2E_{q}^{2}}{|q_{\perp}|^{2}}\mathcal{Q}_{1}\left(\frac{E_{q}}{|q_{\perp}|}\right)
\nonumber \\
& &-(\partial_{E_{q}}f_{\mathrm{V}}^{<}(E_{q},\mathbf{q}))\frac{E_{q}^{2}}{|q_{\perp}|}\left[\mathcal{Q}_{0}\left(\frac{E_{q}}{|q_{\perp}|}\right)-\frac{E_{q}}{|q_{\perp}|}\mathcal{Q}_{1}\left(\frac{E_{q}}{|q_{\perp}|}\right)\right].
\end{eqnarray}

In the non-relativistic limit  with $E_{q}/|q_{\perp}|\to+\infty$,
we observe that
\begin{eqnarray}
G_{\mathrm{T}}(E_{q},\mathbf{q}) & \to & \frac{4}{3}f_{\mathrm{V}}^{<}(E_{q},\mathbf{q})-\frac{2}{3}E_{q}\partial_{E_{q}}f_{\mathrm{V}}^{<}(E_{q},\mathbf{q}),\\
G_{\omega_{1}}(E_{q},\mathbf{q}) & \to & \frac{6}{5}f_{\mathrm{V}}^{<}(E_{q},\mathbf{q})-\frac{2}{5}E_{q}\partial_{E_{q}}f_{\mathrm{V}}^{<}(E_{q},\mathbf{q}),\\
G_{\omega_{2}}(E_{q},\mathbf{q}) & \to & \frac{2}{3}f_{\mathrm{V}}^{<}(E_{q},\mathbf{q})-\frac{2}{3}E_{q}\partial_{E_{q}}f_{\mathrm{V}}^{<}(E_{q},\mathbf{q}),
\end{eqnarray}
 where $\mathcal{Q}_{1}(x)\to\frac{1}{3x^{2}}+\frac{1}{5x^{4}}$ when $x\rightarrow \infty$.
In such a case, $\delta\mathcal{J}_{5,\mathrm{SE}}^{\mu}$ reduces to
\begin{eqnarray}
\delta\mathcal{J}_{5,\mathrm{SE}}^{\mu}(\mathbf{q},x) & \to & \frac{\hbar}{3}m_{f}^{2}\frac{\omega^{\mu}}{E_{q}^{3}}\left(f_{\mathrm{V}}^{<}(E_{q},\mathbf{q})-E_{q}\partial_{E_{q}}f_{\mathrm{V}}^{<}(E_{q},\mathbf{q})\right),
\end{eqnarray}
whereas all other terms in Eq.~(\ref{eq:Additional J5}) are relatively suppressed.

While, in the ultra-relativistic
case with large $|q_{\perp}|$ such that $E_{q}\to |q_{\perp}|$,
we find
\begin{eqnarray}
G_{\mathrm{T}}(E_{q},\mathbf{q}),G_{\omega_{2}}(E_{q},\mathbf{q}) & \to & 2\ln\frac{2|q_{\perp}|}{m}f_{\mathrm{V}}^{<}(E_{q},\mathbf{q})-E_{q}\partial_{E_{q}}f_{\mathrm{V}}^{<}(E_{q},\mathbf{q}).\\
G_{\omega_{1}}(E_{q},\mathbf{q}) & \to & 6\ln\frac{2|q_{\perp}|}{m}f_{\mathrm{V}}^{<}(E_{q},\mathbf{q})-E_{q}\partial_{E_{q}}f_{\mathrm{V}}^{<}(E_{q},\mathbf{q}),
\end{eqnarray}
and thus all of the different sources in Eq.~(\ref{eq:Additional J5}) are comparable for the ultra-relativistic quarks.

\subsection{Application to spin polarization and alignment}
\label{subsec:application}

Now, we discuss the possible corrections from self energies to the spin polarization spectrum and spin alignment. 
First, let us consider the modified covariant Cooper-Frye formula. 
The spin polarization pseudovector density in momentum space $\mathcal{P}^{\mu}(t,\mathbf{q})$ \citep{Becattini:2013fla,Fang:2016vpj, Hidaka:2017auj, Fu:2021pok,Fu:2022myl, Becattini:2021iol, Yi:2021ryh, Yi:2021unq, Wu:2022mkr, Yi:2023tgg} for quarks is then given by 
\begin{eqnarray}
\delta\mathcal{P}^{\mu}(t,\mathbf{q}) & = & \hbar\frac{\int_{\Sigma}q\cdot\mathrm{d}\sigma\int\frac{\mathrm{d}q_{0}}{2\pi}\delta\mathcal{A}^{<,\mu}(q,X)}{2\int_{\Sigma}q\cdot\mathrm{d}\sigma\int\frac{\mathrm{d}q_{0}}{2\pi}\mathcal{F}^{<}(q,X)}\nonumber \\
 & = & \delta\mathcal{P}_{\mathrm{therm}}^{\mu}(t,\mathbf{q})+\delta\mathcal{P}_{\mathrm{shear}}^{\mu}(t,\mathbf{q})+\delta\mathcal{P}_{\mathrm{chem}}^{\mu}(t,\mathbf{q})+\delta\mathcal{P}_{\mathrm{acc}}^{\mu}(t,\mathbf{q})+\delta\mathcal{P}_{\mathrm{vor}}^{\mu}(t,\mathbf{q}),\label{eq:Modified_C-F_formula}
\end{eqnarray}
where $\int_{\Sigma}d\sigma_{\mu}$ represents the integration over a freeze-out hypersurface. Here we have kept only the linear terms of self-energy.
The different sources of spin polarization, namely, the thermal
vorticity $\delta\mathcal{P}_{\mathrm{therm}}^{\mu}$, the shear tensor
$\delta\mathcal{P}_{\mathrm{shear}}^{\mu}$, the gradient of chemical
potential $\mu$ over temperature $T$, $\delta\mathcal{P}_{\mathrm{chem}}^{\mu}$, the fluid acceleration
$\delta\mathcal{P}_{\mathrm{acc}}^{\mu}$ and kinetic vorticity $\delta\mathcal{P}_{\mathrm{vor}}^{\mu}$
are defined as follows\modSF{\footnote{Apart from contributions from the thermal vorticity, we also obtain a coupling-dependent axial current term only related to the kinetic vorticity even in global equilibrium, such term may be associated with the interaction correction to the CVE \citep{Golkar:2012kb, Hou:2012xg} with finite mass since they are of same order in coupling constant.}},
\begin{eqnarray}\label{eq:delP_therm}
\delta\mathcal{P}_{\mathrm{therm}}^{\mu}(t,\mathbf{q}) & = & -\frac{\hbar^{2}}{2mN}\int_{\Sigma}q\cdot\mathrm{d}\sigma G_{\mathrm{T}}(E_{q},\mathbf{q})\frac{m_{f}^{2}T}{E_{q}^{3}}\epsilon^{\mu\nu\alpha\beta}q_{\nu}\partial_{\alpha}\left( \frac{u_{\beta}}{T}\right),\\\label{eq:delP_shear}
\delta\mathcal{P}_{\mathrm{shear}}^{\mu}(t,\mathbf{q}) & = & -\frac{\hbar^{2}}{4mN}\int_{\Sigma}q\cdot\mathrm{d}\sigma G_{\omega_{1}}(E_{q},\mathbf{q})\frac{m_{f}^{2}}{E_{q}^{3}}\frac{\epsilon^{\mu\nu\rho\sigma}q_{\rho}u_{\sigma}}{E_{q}}q^{\gamma}\sigma_{\nu\gamma},\\
\delta\mathcal{P}_{\mathrm{chem}}^{\mu}(t,\mathbf{q}) & = & -\frac{\hbar^{2}}{4mN}\int_{\Sigma}q\cdot\mathrm{d}\sigma G_{\mathrm{T}}(E_{q},\mathbf{q})\frac{C_{\mathrm{F}}g^{2}\mu T}{4\pi^{2}E_{q}^{2}}\frac{\epsilon^{\mu\nu\rho\sigma}q_{\rho}u_{\sigma}}{E_{q}}\nabla_{\nu}\left( \frac{\mu}{T}\right),\\\label{eq:delP_acc}
\delta\mathcal{P}_{\mathrm{acc}}^{\mu}(t,\mathbf{q}) & = & \frac{\hbar^{2}}{4mN}\int_{\Sigma}q\cdot\mathrm{d}\sigma G_{\mathrm{T}}(E_{q},\mathbf{q})\frac{3m_{f}^{2}}{E_{q}^{3}}\epsilon^{\mu\nu\rho\sigma}q_{\rho}u_{\sigma}Du_{\nu},
\\\label{eq:delP_vor}
\delta\mathcal{P}_{\mathrm{vor}}^{\mu}(t,\mathbf{q}) & = & \frac{\hbar^{2}}{4mN}\int_{\Sigma}q\cdot\mathrm{d}\sigma\frac{m_{f}^{2}}{E_{q}^{2}}\left[\omega^{\mu}\left(4G_{\mathrm{T}}(E_{q},\mathbf{q})-\frac{|q_{\perp}|^{2}}{E_{q}^{2}}G_{\omega_{1}}(E_{q},\mathbf{q})+2G_{\omega_{2}}(E_{q},\mathbf{q})\right)\right.\nonumber \\
 &  & \qquad\left.-\frac{(\omega\cdot q)}{E_{q}}\left(6u^{\mu}G_{\mathrm{T}}(E_{q},\mathbf{q})+\frac{q_{\perp}^{\mu}}{E_{q}}G_{\omega_{1}}(E_{q},\mathbf{q})\right)\right],
\end{eqnarray}
where $N\equiv\int_{\Sigma}q\cdot\mathrm{d}\sigma f_{\mathrm{V}}^{<}(E_{q},X)$. 

We would like to comment on the above corrections from self-energy of quarks. 
First, unlike the case in the global \citep{Becattini:2013fla,Fang:2016vpj} and local equilibrium \citep{Hidaka:2017auj, Fu:2021pok,Fu:2022myl, Becattini:2021iol, Yi:2021ryh, Yi:2021unq, Wu:2022mkr, Yi:2023tgg}, there exists an additional term proportional to the kinetic vorticity in Eq.~(\ref{eq:delP_vor}), which cannot be absorbed into Eq.~(\ref{eq:delP_therm}) led by thermal vorticity.
Second, we emphasize that Eqs.~(\ref{eq:delP_therm})-(\ref{eq:delP_vor}) are corrections for a generic $s$-quark probe, where $f_{\mathrm{V}}^{<}(E_{q},\mathbf{q})$ can be an out-of-equilibrium distribution function in general, even though they are also induced by the first-order gradients as the local-equilibrium contributions.

 Furthermore, except for Eq.~(\ref{eq:delP_shear}), we observe that Eqs.~(\ref{eq:delP_therm})-(\ref{eq:delP_acc}) all have the opposite signs compared with their counterparts in local equilibrium (also see the results in Refs.~\citep{Yi:2021ryh,Yi:2021unq,Wu:2022mkr}). Therefore, 
it is of importance to estimate the relative magnitude of these
corrections in Eq.~(\ref{eq:Modified_C-F_formula}) to the local equilibrium
spin polarization vector.  We choose the distribution function of $s$ quarks to be in equilibrium, with $f_{\mathrm{V}}^{<}=f_{\mathrm{F-D}}^{<}\equiv[e^{ (E_{q}-\mu)/T}+1]^{-1}$,
and work in the "$s$ equilibrium scenario" \citep{Fu:2021pok, Yi:2021ryh}, in which we assume that the spin polarization of the constituent s quark is smoothly transferred to the spin polarization of the $\Lambda$ and $\overline{\Lambda}$ hyperons. We consider the
ratio of $\delta\mathcal{J}_{i}^{5,\mu}$ to its local equilibrium
counterpart $\mathcal{J}_{i,\mathrm{leq}}^{5,\mu}$ under  different
momenta. 
From the local equilibrium form of Wigner function
\citep{Yi:2021ryh, Yi:2021unq, Wu:2022mkr, Yi:2023tgg}, the thermal vorticity and shear tensor induced
axial vector currents reads
\begin{eqnarray}
\mathcal{J}_{\mathrm{therm,leq}}^{5,\mu}(\mathbf{q},x) & = & \frac{\hbar}{8E_{q}}f_{\mathrm{V,leq}}^{<}f_{\mathrm{V,leq}}^{>}\epsilon^{\mu\nu\alpha\beta}q_{\nu}\partial_{\alpha}\left( \frac{u_{\beta}}{T}\right), \nonumber \\
\mathcal{J}_{\mathrm{shear,leq}}^{5,\mu}(\mathbf{q},x) & = & -\hbar\frac{1}{4E_{q}^{2}T}f_{\mathrm{V,leq}}^{<}f_{\mathrm{V,leq}}^{>}\epsilon^{\mu\nu\rho\sigma}q_{\rho}u_{\sigma}q^{\alpha}\sigma_{\nu\alpha},
\end{eqnarray}
where the 'leq' stands for the distribution function in the local equilibrium. Thus, we have 
\begin{eqnarray}
\left|\frac{\delta\mathcal{J}_{\mathrm{therm}}^{5,\mu}(\mathbf{q},x)}{\mathcal{J}_{\mathrm{therm,leq}}^{5,\mu}(\mathbf{q},x)} \right|& = & \frac{4m_{f}^{2}T}{E_{q}^{3}}\frac{G_{\mathrm{T}}(E_{q},\mathbf{q})}{f_{\mathrm{V,leq}}^{<}f_{\mathrm{V,leq}}^{>}},\\
\left| \frac{\delta\mathcal{J}_{\mathrm{shear}}^{5,\mu}(\mathbf{q},x)}{\mathcal{J}_{\mathrm{shear,leq}}^{5,\mu}(\mathbf{q},x)} \right| & = & \frac{m_{f}^{2}T}{E_{q}^{3}}\frac{G_{\omega_{1}}(E_{q},\mathbf{q})}{f_{\mathrm{V,leq}}^{<}f_{\mathrm{V,leq}}^{>}},\\
\left|\frac{\delta\mathcal{J}_{\mathrm{vor}}^{5,\mu}(\mathbf{q},x)}{\mathcal{J}_{\mathrm{therm,leq}}^{5,\mu}(\mathbf{q},x)} \right|& \simeq & \frac{2Tm_{f}^{2}}{E_{q}^{2}|q_{\perp}|}\frac{1}{f_{\mathrm{V,leq}}^{<}f_{\mathrm{V,leq}}^{>}}\Bigg[\left(4-\frac{6|q_{\perp}|}{E_{q}}\right)G_{\mathrm{T}}(E_{q},\mathbf{q})\nonumber \\
 &  & \qquad-\frac{2|q_{\perp}|^{2}}{E_{q}^{2}}G_{\omega_{1}}(E_{q},\mathbf{q})+2G_{\omega_{2}}(E_{q},\mathbf{q})\Bigg].
\end{eqnarray}
In the estimation of vorticity induced axial current from self-energy
corrections, we have assumed the gradients of fluid velocity are of
the same order.
\begin{table}
\begin{centering}
\begin{tabular}{|c|c|c|c|}
\hline 
 &
$|q_{\perp}|=0.5$ GeV &
$|q_{\perp}|=1.0$ GeV &
$|q_{\perp}|=2.0$ GeV\tabularnewline
\hline 
\hline 
$|\delta\mathcal{J}_{\mathrm{therm}}^{5,\mu}/\mathcal{J}_{\mathrm{therm},\mathrm{leq}}^{5,\mu}|$ &
0.325 &
0.098 &
0.024\tabularnewline
\hline 
$|\delta\mathcal{J}_{\mathrm{shear}}^{5,\mu}/\mathcal{J}_{\mathrm{shear},\mathrm{leq}}^{5,\mu}|$ &
0.081 &
0.028 &
0.007\tabularnewline
\hline 
$|\delta\mathcal{J}_{\mathrm{vor}}^{5,\mu}/\mathcal{J}_{\mathrm{therm},\mathrm{leq}}^{5,\mu}|$ &
0.177 &
0.103 &
0.030\tabularnewline
\hline 
\end{tabular}
\par\end{centering}
\caption{The relative magnitude of axial vector currents from self-energy corrections
to the local equilibrium axial vector currents at different momentum,
$|q_{\perp}|=0.5,1.0,2.0$ GeV. We have chosen $g^{2}=\frac{4\pi}{3}$ and $C_{\mathrm{F}}=\frac{4}{3}$
here; we set the temperature $T=0.165$ GeV, chemical potential $\mu=0.01$ GeV and constituent $s$ quark mass $m=0.3$ GeV. \label{tab:The-relative-magnitude}}
\end{table}
We list the numerical results for $|\delta\mathcal{J}^{5,\mu}/\mathcal{J}_{\mathrm{leq}}^{5,\mu}|$ at $|q_{\perp}|=0.5,1.0,2.0$ GeV in Table.\ref{tab:The-relative-magnitude}. 
We find the self-energy corrections can have
significant influence on the spin polarization at $s$ equilibrium scenarios
when the momentum $q\leq 5$ GeV even under our rough estimation.

Second, let us briefly remark on the potential application to the spin alignment of vector mesons based on the Eq.~(\ref{eq:Modified_C-F_formula}). Building upon the coalescence models in both non-relativistic \citep{Liang:2004ph, Liang:2004xn, Xia:2020tyd} and relativistic cases \citep{Sheng:2019kmk,Sheng:2022ffb,Kumar:2023ghs,Sheng:2023urn}, the $00$-component of spin density matrix, $\rho_{00}$, for vector mesons encompasses contributions arising from the spin polarization of quarks. 
For example, following Ref.~\cite{Kumar:2023ghs}, we may evaluate $\rho_{00}$ in the vector-meson rest frame in the non-relativistic limit for the constituent quark and antiquark with self-energy corrections through
\begin{eqnarray}
\rho_{00}\approx \frac{1-\hat{\Pi}^{yy}({\bf q=0})}{3-\sum_{i=x,y,z}\hat{\Pi}^{ii}({\bf q=0})},
\end{eqnarray}
where
\begin{eqnarray}
\hat{\Pi}^{ii}({\bf q})=\frac{4\int_{\Sigma} d\sigma\cdot q\left[\mathcal{J}_{5 q,\mathrm{a}}^{i}(\frac{\mathbf{q}}{2},x)+\delta\mathcal{J}_{5 q,\mathrm{SE}}^{i}(\frac{\mathbf{q}}{2},x)\right]\left[\mathcal{J}_{5\bar{q},\mathrm{a}}^{i}(\frac{\mathbf{q}}{2},x)+\delta\mathcal{J}_{5\bar{q},\mathrm{SE}}^{i}(\frac{\mathbf{q}}{2},x)\right]}{\int_{\Sigma} d\sigma\cdot q f^{<}_{{\rm V} q}(E_{\bf q/2},\frac{\mathbf{q}}{2})f^{<}_{{\rm V}\bar{q}}(E_{\bf q/2},\frac{\mathbf{q}}{2})} 
\end{eqnarray}
with the subscripts $q$ and $\bar{q}$ corresponding to the quark and antiquark forming the vector meson and we have chosen the $y$ direction as a spin-quantization axis. Here $\mathcal{J}_{5 q,\mathrm{a}}^{i}$ denotes the remanent spin polarization density 
pseudovector from the effective spin vector $a^\mu(q,x)$ in Eq.~(\ref{eq:Axial WF formal solution}) by solving the AKE (\ref{eq:Rest_frame_AKE}) in the rest frame.
One of the most important sources leading to spin alignment of $\phi$ mesons comes from the spin polarization of $s$ quarks caused by the quantum fluctuations of strong $\phi$ fields. 
Considering the scenario that self-energy corrections impact the spin polarization of 
$s$ quarks, these corrections consequently contribute to the spin alignment of $\phi$ mesons, supplementing the contribution discussed in Sec.~\ref{subsec:mean_field_contribution}.
Referring to Table.~\ref{tab:The-relative-magnitude}, we posit that self-energy corrections will also play a role in shaping the momentum spectrum of spin alignment for vector mesons in low $q$ cases.
The further numerical simulations for quantitative studies of both spin polarization and alignment induced by the self-energy corrections will be present somewhere else.


\section{Conclusions and outlook} \label{sec:Conclusions-and-outlook}

In this work, we derived the QKT for massive fermions with self-energy
corrections in addition to collisions and background electromagnetic
fields following the $\hbar$ power counting in Ref.~\citep{Yang:2020hri}. 
The inclusion of one-point potential and real part of the retarded self-energy modify the on-mass-shell condition, the perturbative solutions of Wigner functions and corresponding kinetic equations. 
We find that the gradients of self-energy, which play a similar role as the background electromagnetic fields, further induce quantum corrections especially upon the axial Wigner function and AKE for spin transport.

At the end, we implement our formalism  to spin polarization phenomena in relativistic heavy ion collisions. 

Interestingly, within our formalism, one can add the background effective meson fields into the spin polarization pseudovector (\ref{eq:MF_Modif_spin_vector}) from quantum field theory self-consistently.
Moreover, based on the axial Wigner function (\ref{eq:Axial WF formal solution}), we 
derive the modified Cooper-Frye formula for the spin polarization of $s$ quarks induced by the QCD thermal background shown in Eq.~(\ref{eq:Modified_C-F_formula}). We observe that the self-energy corrections result in the contributions from temperature gradient, vorticity, shear tensor, and gradient of $\mu/T$. 

Despite the extensive studies on off-equilibrium effects
upon the spin polarization in heavy ion collisions from the collisional effect \citep{Wang:2020pej,Wang:2021qnt,Sheng:2021kfc,Weickgenannt:2022zxs,Weickgenannt:2022qvh,Wagner:2022amr,Fang:2022ttm,Wang:2022yli,Lin:2022tma},
the self-energy correction is in general found to be more dominant in the weakly coupled scenario due to its lower order in gradients
and couplings. 
Given that our order-of-magnitude estimation for the self-energy gradient corrections on spin polarization of $s$ quarks at low momenta in the thermal QCD background is shown to be comparable to the local-equilibrium contribution in the $s$ equilibrium scenarios,
it is inevitable to include such corrections for the study of local spin polarization with future hydrodynamic simulations to further analyze the quantitative contributions from different sources and to address the azimuthal-angle dependence of longitudinal and transverse spin polarization of $\Lambda$ hyperons,  as well as the spin alignment of $\phi$ mesons. 

On the other hand, since our formalism is applicable for either quarks or spin-1/2 baryons, it is also intriguing to further study how self-energy gradient corrections could affect the spin polarization of baryons, e.g.  $\Lambda$ hyperons, through the hadronic interaction with effective models. In principle, such corrections should be incorporated when considering the local spin polarization in the $\Lambda$ equilibrium scenario \citep{Fu:2021pok,Yi:2021ryh, Wu:2022mkr} or isothermal equilibrium scenario \citep{Becattini:2021iol}. In addition,
the neglected self-energies $\overline{\Sigma}_{\mathrm{A}}^{\mu},\overline{\Sigma}_{\mathrm{T}}^{\mu\nu}$, which come from a polarized medium, 
could also make sizable contributions to
the spin polarization. But particularly due to the lack of the knowledge for Wigner functions of polarized gluons with quantum corrections in local equilibrium, we leave such an issue for the future studies.

\begin{acknowledgments}
The authors would like to thank B. M\"uller for useful discussions. This work is supported in part by the National Key Research and Development
Program of China under Contract No. 2022YFA1605500,  by the Chinese Academy of Sciences (CAS) under Grants No. YSBR-088 and by National Nature Science Foundation of China (NSFC) under Grants No. 12075235 and No. 12135011. D.-L. Yang is supported by National Science and Technology Council (Taiwan) under Grant
No. MOST 110-2112-M-001-070-MY3. 
\end{acknowledgments}

\appendix

\section{Master equations of Wigner functions\label{sec:Master-equations-of}}

In this part, we present the details for the deriving the master equations of Wigner functions with self-energy corrections. 

\subsection{General equations\label{subsec:General-equations}}

Inserting Eq.~(\ref{eq:Clifford_decomposition_WF}, \ref{eq:Clifford_decomposition_SE}),
yields,
\begin{eqnarray}
\{(\overrightarrow{\slashed{\Pi}}-m),S^{<}\} & = & (2\Pi_{\mu}\mathcal{V}^{<,\mu}-2m\mathcal{F}^{<})-2mi\mathcal{P}^{<}\gamma^{5}+2(\Pi_{\alpha}\mathcal{F}^{<}-m\mathcal{V}_{\alpha}^{<})\gamma^{\alpha}\nonumber \\
 &  & -(\epsilon_{\mu\alpha\rho\sigma}\Pi^{\alpha}\mathcal{S}^{<,\rho\sigma}+2m\mathcal{A}_{\mu}^{<})\gamma^{5}\gamma^{\mu}-(\epsilon_{\mu\nu\alpha\beta}\Pi^{\alpha}\mathcal{A}^{<,\beta}+m\mathcal{S}_{\mu\nu}^{<})\gamma^{\mu\nu},
\end{eqnarray}
where we have used
\begin{eqnarray}
\gamma^{5}\gamma_{\rho}\gamma_{\gamma} & = & \frac{1}{2}\epsilon_{\mu\nu\rho\gamma}\gamma^{\mu\nu}+\gamma^{5}\eta_{\gamma\rho},\\
\gamma^{\mu}\gamma^{\nu}\gamma^{\rho} & = & -i\epsilon^{\mu\nu\rho\sigma}\gamma^{5}\gamma_{\sigma}+\eta^{\mu\nu}\gamma^{\rho}+\eta^{\rho\nu}\gamma^{\mu}-\gamma^{\nu}\eta^{\rho\mu};
\end{eqnarray}
and
\begin{eqnarray}
[(\overrightarrow{\slashed{\Pi}}-m),S^{<}] & = & -2\Pi^{\mu}\mathcal{A}_{\mu}^{<}\gamma^{5}+2i\Pi^{\nu}\mathcal{S}_{\nu\mu}^{<}\gamma^{\mu}-2i\Pi_{\alpha}\mathcal{P}^{<}\gamma^{5}\gamma^{\alpha}-2i\Pi_{\mu}\mathcal{V}_{\nu}^{<}\gamma^{\mu\nu},
\\
\frac{i}{2}[\gamma^{\mu},\nabla_{\mu}S^{<}] & = & -i\nabla^{\mu}\mathcal{A}_{\mu}^{<}\gamma^{5}-\nabla^{\nu}\mathcal{S}_{\nu\mu}^{<}\gamma^{\mu}+\nabla_{\mu}\mathcal{P}^{<}\gamma^{5}\gamma^{\mu}+\nabla_{\mu}\mathcal{V}_{\nu}^{<}\gamma^{\mu\nu},\\
\frac{i}{2}\{\gamma^{\mu},\nabla_{\mu}S^{<}\} & = & i\nabla^{\mu}\mathcal{V}_{\mu}^{<}+i\nabla_{\alpha}\mathcal{F}^{<}\gamma^{\alpha}-\frac{i}{2}\epsilon_{\mu\nu\rho\sigma}\nabla^{\nu}\mathcal{S}^{<,\rho\sigma}\gamma^{5}\gamma^{\mu}-\frac{i}{2}\epsilon_{\mu\nu\rho\sigma}\nabla^{\rho}\mathcal{A}^{<,\sigma}\gamma^{\mu\nu}.
\end{eqnarray}

Then for the Moyal products, we find,  
\begin{eqnarray}
\{A,B\}_{\star} & = & A_{\alpha}\Gamma^{\alpha}\star B_{\beta}\Gamma^{\beta}+B_{\beta}\Gamma^{\beta}\star A_{\alpha}\Gamma^{\alpha}=\frac{1}{2}\{A_{\alpha},B_{\beta}\}_{\star}\{\Gamma^{\alpha},\Gamma^{\beta}\}+\frac{1}{2}[A_{\alpha},B_{\beta}]_{\star}[\Gamma^{\alpha},\Gamma^{\beta}], \nonumber \\{}
[A,B]_{\star} & = & A_{\alpha}\Gamma^{\alpha}\star B_{\beta}\Gamma^{\beta}-B_{\beta}\Gamma^{\beta}\star A_{\alpha}\Gamma^{\alpha}=\frac{1}{2}\{A_{\alpha},B_{\beta}\}_{\star}[\Gamma^{\alpha},\Gamma^{\beta}]+\frac{1}{2}[A_{\alpha},B_{\beta}]_{\star}\{\Gamma^{\alpha},\Gamma^{\beta}\}, \nonumber \\
\end{eqnarray}
with $\Gamma=\{1,\gamma^{5},\gamma^{\mu},\gamma^{5}\gamma^{\mu},\gamma^{\mu\nu}\}$.
We then get,
\begin{eqnarray}
 &  & \{\Sigma,S\}_{\star}\nonumber \\
 & = & \left(\{\Sigma_{\mathrm{F}},\mathcal{F}\}_{\star}+\{\Sigma_{\mathrm{V}}^{\mu},\mathcal{V}_{\mu}\}_{\star}-\{\Sigma_{\mathrm{P}},\mathcal{P}\}_{\star}-\{\Sigma_{\mathrm{A}}^{\mu},\mathcal{A}_{\mu}\}_{\star}+\frac{1}{2}\{\Sigma_{\mathrm{T}}^{\mu\nu},\mathcal{S}_{\mu\nu}\}_{\star}\right)\nonumber \\
 &  & +\left(\{\Sigma_{\mathrm{F}},\mathcal{P}\}_{\star}+\{\Sigma_{\mathrm{P}},\mathcal{F}\}_{\star}{\color{blue}-}\frac{1}{4}\{\Sigma_{\mathrm{T},\alpha\beta},\mathcal{S}_{\mu\nu}\}_{\star}\epsilon^{\mu\nu\alpha\beta}\right)i\gamma^{5}\nonumber \\
 &  & +\left(\{\Sigma_{\mathrm{F}},\mathcal{V}_{\mu}\}_{\star}+\{\Sigma_{\mathrm{V},\mu},\mathcal{F}\}_{\star}-\frac{1}{2}\epsilon_{\mu\nu\alpha\beta}(\{\Sigma_{\mathrm{A}}^{\nu},\mathcal{S}^{\alpha\beta}\}_{\star}+\{\Sigma_{\mathrm{T}}^{\alpha\beta},\mathcal{A}^{\nu}\}_{\star})\right)\gamma^{\mu}\nonumber \\
 &  & +\left(\{\Sigma_{\mathrm{F}},\mathcal{A}_{\mu}\}_{\star}+\{\Sigma_{\mathrm{A},\mu},\mathcal{F}\}_{\star}-\frac{1}{2}\epsilon_{\mu\nu\alpha\beta}(\{\Sigma_{\mathrm{T}}^{\alpha\beta},\mathcal{V}^{\nu}\}_{\star}+\{\Sigma_{\mathrm{V}}^{\nu},\mathcal{S}^{\alpha\beta}\}_{\star})\right)\gamma^{5}\gamma^{\mu}\nonumber \\
 &  & +\Big[\{\Sigma_{\mathrm{F}},\mathcal{S}_{\mu\nu}\}_{\star}+\{\Sigma_{\mathrm{T},\mu\nu},\mathcal{F}\}_{\star}-\frac{1}{2}\epsilon_{\mu\nu\alpha\beta}(\{\Sigma_{\mathrm{P}},\mathcal{S}^{\alpha\beta}\}_{\star}+\{\Sigma_{\mathrm{T}}^{\alpha\beta},\mathcal{P}\}_{\star})\nonumber \\
 &  & \qquad+\epsilon_{\alpha\beta\mu\nu}(-\{\Sigma_{\mathrm{V}}^{\alpha},\mathcal{A}^{\beta}\}_{\star}+\{\Sigma_{\mathrm{A}}^{\alpha},\mathcal{V}^{\beta}\}_{\star})\Big]\frac{\gamma^{\mu\nu}}{2}\nonumber \\
 &  & +i\left([\Sigma_{\mathrm{V}}^{\mu},\mathcal{A}_{\mu}]_{\star}-[\Sigma_{\mathrm{A}}^{\mu},\mathcal{V}_{\mu}]_{\star}\right)i\gamma^{5}+\left([\Sigma_{\mathrm{P}},\mathcal{A}_{\mu}]_{\star}+[\Sigma_{\mathrm{V}}^{\nu},\mathcal{S}_{\nu\mu}]_{\star}-[\Sigma_{\mathrm{A},\mu},\mathcal{P}]_{\star}+[\Sigma_{\mathrm{T},\mu\nu},\mathcal{V}^{\nu}]_{\star}\right)i\gamma^{\mu}\nonumber \\
 &  & +\left([\Sigma_{\mathrm{P}},\mathcal{V}_{\mu}]_{\star}+[\Sigma_{\mathrm{A}}^{\nu},\mathcal{S}_{\nu\mu}]_{\star}-[\Sigma_{\mathrm{V},\mu},\mathcal{P}]_{\star}+[\Sigma_{\mathrm{T},\mu\nu},\mathcal{A}^{\nu}]_{\star}\right)i\gamma^{5}\gamma^{\mu}\nonumber \\
 &  & +2\left(-[\Sigma_{\mathrm{V},[\mu},\mathcal{V}_{\nu]}]_{\star}+[\Sigma_{\mathrm{A},[\mu},\mathcal{A}_{\nu]}]_{\star}\right)i\frac{\gamma^{\mu\nu}}{2},
\end{eqnarray}
and 
\begin{eqnarray}
 &  & [\Sigma,S]_{\star}\nonumber \\
 & = & i\left(\{\Sigma_{\mathrm{V}}^{\mu},\mathcal{A}_{\mu}\}_{\star}-\{\Sigma_{\mathrm{A}}^{\mu},\mathcal{V}_{\mu}\}_{\star}\right)i\gamma^{5}+\left(\{\Sigma_{\mathrm{P}},\mathcal{A}_{\mu}\}_{\star}+\{\Sigma_{\mathrm{V}}^{\nu},\mathcal{S}_{\nu\mu}\}_{\star}-\{\Sigma_{\mathrm{A},\mu},\mathcal{P}\}_{\star}+\{\Sigma_{\mathrm{T},\mu\nu},\mathcal{V}^{\nu}\}_{\star}\right)i\gamma^{\mu}\nonumber \\
 &  & +\left(\{\Sigma_{\mathrm{P}},\mathcal{V}_{\mu}\}_{\star}-\{\Sigma_{\mathrm{V},\mu},\mathcal{P}\}_{\star}+\{\Sigma_{\mathrm{A}}^{\nu},\mathcal{S}_{\nu\mu}\}_{\star}+\{\Sigma_{\mathrm{T},\mu\nu},\mathcal{A}^{\nu}\}_{\star}\right)i\gamma^{5}\gamma^{\mu}\nonumber \\
 &  & +2\left(\{\Sigma_{\mathrm{A},[\mu},\mathcal{A}_{\nu]}\}_{\star}-\{\Sigma_{\mathrm{V},[\mu},\mathcal{V}_{\nu]}\}_{\star}\right)i\frac{\gamma^{\mu\nu}}{2}\nonumber \\
 &  & +\left([\Sigma_{\mathrm{F}},\mathcal{F}]_{\star}+[\Sigma_{\mathrm{V}}^{\mu},\mathcal{V}_{\mu}]_{\star}-[\Sigma_{\mathrm{P}},\mathcal{P}]_{\star}-[\Sigma_{\mathrm{A}}^{\mu},\mathcal{A}_{\mu}]_{\star}+\frac{1}{2}[\Sigma_{\mathrm{T}}^{\mu\nu},\mathcal{S}_{\mu\nu}]_{\star}\right)\nonumber \\
 &  & +\left([\Sigma_{\mathrm{F}},\mathcal{P}]_{\star}+[\Sigma_{\mathrm{P}},\mathcal{F}]_{\star}{\color{blue}-}\frac{1}{4}[\Sigma_{\mathrm{T},\alpha\beta},\mathcal{S}_{\mu\nu}]_{\star}\epsilon^{\mu\nu\alpha\beta}\right)i\gamma^{5}\nonumber \\
 &  & +\left([\Sigma_{\mathrm{F}},\mathcal{V}_{\mu}]_{\star}+[\Sigma_{\mathrm{V},\mu},\mathcal{F}]_{\star}-\frac{1}{2}\epsilon_{\mu\nu\alpha\beta}([\Sigma_{\mathrm{T}}^{\alpha\beta},\mathcal{A}^{\nu}]_{\star}+[\Sigma_{\mathrm{A}}^{\nu},\mathcal{S}^{\alpha\beta}]_{\star})\right)\gamma^{\mu}\nonumber \\
 &  & +\left([\Sigma_{\mathrm{F}},\mathcal{A}_{\mu}]_{\star}+[\Sigma_{\mathrm{A},\mu},\mathcal{F}]_{\star}-\frac{1}{2}\epsilon_{\mu\nu\alpha\beta}([\Sigma_{\mathrm{V}}^{\nu},\mathcal{S}^{\alpha\beta}]_{\star}+[\Sigma_{\mathrm{T}}^{\alpha\beta},\mathcal{V}^{\nu}]_{\star})\right)\gamma^{5}\gamma^{\mu}\nonumber \\
 &  & +\Big[[\Sigma_{\mathrm{F}},\mathcal{S}_{\mu\nu}]_{\star}+[\Sigma_{\mathrm{T},\mu\nu},\mathcal{F}]_{\star}-\frac{1}{2}\epsilon_{\mu\nu\alpha\beta}([\Sigma_{\mathrm{P}},\mathcal{S}^{\alpha\beta}]_{\star}+[\Sigma_{\mathrm{T}}^{\alpha\beta},\mathcal{P}]_{\star})\nonumber \\
 &  & \qquad+\epsilon_{\mu\nu\alpha\beta}(-[\Sigma_{\mathrm{V}}^{\alpha},\mathcal{A}^{\beta}]_{\star}+[\Sigma_{\mathrm{A}}^{\alpha},\mathcal{V}^{\beta}]_{\star})\Big]\frac{\gamma^{\mu\nu}}{2}.
\end{eqnarray}

By matching the coefficients in front of the basis $\{1,i\gamma^{5},\gamma^{\mu},\gamma^{5}\gamma^{\mu},\frac{1}{2}\gamma^{\mu\nu}\}$,
we can derive the master equations. For the addition part of KB equations (\ref{eq:Addition KB equation}):
the scalar part,
\begin{eqnarray}
0 & = & (2\Pi_{\mu}\mathcal{V}^{<,\mu}-2m\mathcal{F}^{<}) \nonumber \\
 &  &+\left(\{\overline{\Sigma}_{\mathrm{F}},\mathcal{F}^{<}\}_{\star}+\{\overline{\Sigma}_{\mathrm{V}}^{\mu},\mathcal{V}_{\mu}^{<}\}_{\star}-\{\overline{\Sigma}_{\mathrm{P}},\mathcal{P}^{<}\}_{\star}-\{\overline{\Sigma}_{\mathrm{A}}^{\mu},\mathcal{A}_{\mu}^{<}\}_{\star}+\frac{1}{2}\{\overline{\Sigma}_{\mathrm{T}}^{\mu\nu},\mathcal{S}_{\mu\nu}^{<}\}_{\star}\right)\nonumber \\
 &  & +\frac{i\hbar}{2}\left([\widehat{\Sigma_{\mathrm{F}},\mathcal{F}}]_{\star}+[\widehat{\Sigma_{\mathrm{V}}^{\mu},\mathcal{V}_{\mu}}]_{\star}-[\widehat{\Sigma_{\mathrm{P}},\mathcal{P}}]_{\star}-[\widehat{\Sigma_{\mathrm{A}}^{\mu},\mathcal{A}_{\mu}}]_{\star}+\frac{1}{2}[\widehat{\Sigma_{\mathrm{T}}^{\mu\nu},\mathcal{S}_{\mu\nu}}]_{\star}\right),
\end{eqnarray}
the pseudoscalar part,
\begin{eqnarray}
0 & = & -2m\mathcal{P}^{<}-\hbar\nabla^{\mu}\mathcal{A}_{\mu}^{<}+\left(\{\overline{\Sigma}_{\mathrm{F}},\mathcal{P}^{<}\}_{\star}+\{\overline{\Sigma}_{\mathrm{P}},\mathcal{F}^{<}\}_{\star}{\color{blue}-}\frac{1}{4}\{\overline{\Sigma}_{\mathrm{T},\alpha\beta},\mathcal{S}_{\mu\nu}^{<}\}_{\star}\epsilon^{\mu\nu\alpha\beta}\right) \nonumber \\
& & +i\left([\overline{\Sigma}_{\mathrm{V}}^{\mu},\mathcal{A}_{\mu}^{<}]_{\star}-[\overline{\Sigma}_{\mathrm{A}}^{\mu},\mathcal{V}_{\mu}^{<}]_{\star}\right)-\frac{\hbar}{2}\left(\{\widehat{\Sigma_{\mathrm{V}}^{\mu},\mathcal{A}_{\mu}}\}_{\star}-\{\widehat{\Sigma_{\mathrm{A}}^{\mu},\mathcal{V}_{\mu}}\}_{\star}\right)\nonumber \\
 &  & +\frac{i\hbar}{2}\left([\widehat{\Sigma_{\mathrm{F}},\mathcal{P}}]_{\star}+[\widehat{\Sigma_{\mathrm{P}},\mathcal{F}}]_{\star}{\color{blue}-}\frac{1}{4}[\widehat{\Sigma_{\mathrm{T},\alpha\beta},\mathcal{S}_{\mu\nu}}]_{\star}\epsilon^{\mu\nu\alpha\beta}\right),
\end{eqnarray}
the vector part,
\begin{eqnarray}
0 & = & 2(\Pi_{\mu}\mathcal{F}^{<}-m\mathcal{V}_{\mu}^{<})-\hbar\nabla^{\nu}\mathcal{S}_{\nu\mu}^{<} \nonumber \\ 
& & +\left(\{\overline{\Sigma}_{\mathrm{F}},\mathcal{V}_{\mu}^{<}\}_{\star}+\{\overline{\Sigma}_{\mathrm{V},\mu},\mathcal{F}^{<}\}_{\star}-\frac{1}{2}\epsilon_{\mu\nu\alpha\beta}(\{\overline{\Sigma}_{\mathrm{A}}^{\nu},\mathcal{S}^{<,\alpha\beta}\}_{\star}+\{\overline{\Sigma}_{\mathrm{T}}^{\alpha\beta},\mathcal{A}^{<\nu}\}_{\star})\right)\nonumber \\
 &  & +i\left([\overline{\Sigma}_{\mathrm{P}},\mathcal{A}_{\mu}^{<}]_{\star}+[\overline{\Sigma}_{\mathrm{V}}^{\nu},\mathcal{S}_{\nu\mu}^{<}]_{\star}-[\overline{\Sigma}_{\mathrm{A},\mu},\mathcal{P}^{<}]_{\star}+[\overline{\Sigma}_{\mathrm{T},\mu\nu},\mathcal{V}^{<,\nu}]_{\star}\right)\nonumber \\
 &  & -\frac{\hbar}{2}\left(\{\widehat{\Sigma_{\mathrm{P}},\mathcal{A}_{\mu}}\}_{\star}+\{\widehat{\Sigma_{\mathrm{V}}^{\nu},\mathcal{S}_{\nu\mu}}\}_{\star}-\{\widehat{\Sigma_{\mathrm{A},\mu},\mathcal{P}}\}_{\star}+\{\widehat{\Sigma_{\mathrm{T},\mu\nu},\mathcal{V}^{\nu}}\}_{\star}\right)\nonumber \\
 &  & +\frac{i\hbar}{2}\left([\widehat{\Sigma_{\mathrm{F}},\mathcal{V}_{\mu}}]_{\star}+[\widehat{\Sigma_{\mathrm{V},\mu},\mathcal{F}}]_{\star}-\frac{1}{2}\epsilon_{\mu\nu\alpha\beta}([\widehat{\Sigma_{\mathrm{T}}^{\alpha\beta},\mathcal{A}^{\nu}}]_{\star}+[\widehat{\Sigma_{\mathrm{A}}^{\nu},\mathcal{S}^{\alpha\beta}}]_{\star})\right),
\end{eqnarray}
the axial vector part,
\begin{eqnarray}
0 & = & -(\epsilon_{\mu\alpha\rho\sigma}\Pi^{\alpha}\mathcal{S}^{<,\rho\sigma}+2m\mathcal{A}_{\mu}^{<})+\hbar\nabla_{\mu}\mathcal{P}^{<}\nonumber \\
 &  & +\left(\{\overline{\Sigma}_{\mathrm{F}},\mathcal{A}_{\mu}^{<}\}_{\star}+\{\overline{\Sigma}_{\mathrm{A},\mu},\mathcal{F}^{<}\}_{\star}-\frac{1}{2}\epsilon_{\mu\nu\alpha\beta}(\{\overline{\Sigma}_{\mathrm{T}}^{\alpha\beta},\mathcal{V}^{<,\nu}\}_{\star}+\{\overline{\Sigma}_{\mathrm{V}}^{\nu},\mathcal{S}^{<,\alpha\beta}\}_{\star})\right)\nonumber \\
 &  & +i\left([\overline{\Sigma}_{\mathrm{P}},\mathcal{V}_{\mu}^{<}]_{\star}+[\overline{\Sigma}_{\mathrm{A}}^{\nu},\mathcal{S}_{\nu\mu}^{<}]_{\star}-[\overline{\Sigma}_{\mathrm{V},\mu},\mathcal{P}^{<}]_{\star}+[\overline{\Sigma}_{\mathrm{T},\mu\nu},\mathcal{A}^{<,\nu}]_{\star}\right)\nonumber \\
 &  & -\frac{\hbar}{2}\left(\{\widehat{\Sigma_{\mathrm{P}},\mathcal{V}_{\mu}}\}_{\star}-\{\widehat{\Sigma_{\mathrm{V},\mu},\mathcal{P}}\}_{\star}+\{\widehat{\Sigma_{\mathrm{A}}^{\nu},\mathcal{S}_{\nu\mu}}\}_{\star}+\{\widehat{\Sigma_{\mathrm{T},\mu\nu},\mathcal{A}^{\nu}}\}_{\star}\right)\nonumber \\
 &  & +\frac{i\hbar}{2}\left([\widehat{\Sigma_{\mathrm{F}},\mathcal{A}_{\mu}}]_{\star}+[\widehat{\Sigma_{\mathrm{A},\mu},\mathcal{F}}]_{\star}-\frac{1}{2}\epsilon_{\mu\nu\alpha\beta}([\widehat{\Sigma_{\mathrm{V}}^{\nu},\mathcal{S}^{\alpha\beta}}]_{\star}+[\widehat{\Sigma_{\mathrm{T}}^{\alpha\beta},\mathcal{V}^{\nu}}]_{\star})\right),
\end{eqnarray}
the tensor part,
\begin{eqnarray}
0 & = & -2(\epsilon_{\mu\nu\alpha\beta}\Pi^{\alpha}\mathcal{A}^{<,\beta}+m\mathcal{S}_{\mu\nu}^{<})+\hbar2\nabla_{[\mu}\mathcal{V}_{\nu]}^{<}+\{\overline{\Sigma}_{\mathrm{F}},\mathcal{S}_{\mu\nu}^{<}\}_{\star}+\{\overline{\Sigma}_{\mathrm{T},\mu\nu},\mathcal{F}^{<}\}_{\star}\nonumber \\
 &  & -\frac{1}{2}\epsilon_{\mu\nu\alpha\beta}(\{\overline{\Sigma}_{\mathrm{P}},\mathcal{S}^{<,\alpha\beta}\}_{\star}+\{\overline{\Sigma}_{\mathrm{T}}^{\alpha\beta},\mathcal{P}^{<}\}_{\star})+\epsilon_{\alpha\beta\mu\nu}(-\{\overline{\Sigma}_{\mathrm{V}}^{\alpha},\mathcal{A}^{<,\beta}\}_{\star}+\{\overline{\Sigma}_{\mathrm{A}}^{\alpha},\mathcal{V}^{<,\beta}\}_{\star})\nonumber \\
 &  & +2i\left(-[\overline{\Sigma}_{\mathrm{V},[\mu},\mathcal{V}_{\nu]}^{<}]_{\star}+[\overline{\Sigma}_{\mathrm{A},[\mu},\mathcal{A}_{\nu]}^{<}]_{\star}\right)-\hbar\left(\{\widehat{\Sigma_{\mathrm{A},[\mu},\mathcal{A}_{\nu]}}\}_{\star}-\{\widehat{\Sigma_{\mathrm{V},[\mu},\mathcal{V}_{\nu]}}\}_{\star}\right)\nonumber \\
 &  & +\frac{i\hbar}{2}\Big[[\widehat{\Sigma_{\mathrm{F}},\mathcal{S}_{\mu\nu}}]_{\star}+[\widehat{\Sigma_{\mathrm{T},\mu\nu},\mathcal{F}}]_{\star}-\frac{1}{2}\epsilon_{\mu\nu\alpha\beta}([\widehat{\Sigma_{\mathrm{P}},\mathcal{S}^{\alpha\beta}}]_{\star}+[\widehat{\Sigma_{\mathrm{T}}^{\alpha\beta},\mathcal{P}}]_{\star})\\
 &  & \qquad+\epsilon_{\mu\nu\alpha\beta}(-[\widehat{\Sigma_{\mathrm{V}}^{\alpha},\mathcal{A}^{\beta}}]_{\star}+[\widehat{\Sigma_{\mathrm{A}}^{\alpha},\mathcal{V}^{\beta}}]_{\star})\Big],
\end{eqnarray}
where the notation $\widehat{XY}$ is defined in Eqs.~(\ref{eq:def_widehat}).

Similarly, we can get the difference of the KB equations (\ref{eq:Difference KB equation})
in such Clifford basis. The scalar part,
\begin{eqnarray}
0 & = & i\hbar\nabla^{\mu}\mathcal{V}_{\mu}^{<}+\left([\overline{\Sigma}_{\mathrm{F}},\mathcal{F}^{<}]_{\star}+[\overline{\Sigma}_{\mathrm{V}}^{\mu},\mathcal{V}_{\mu}^{<}]_{\star}-[\overline{\Sigma}_{\mathrm{P}},\mathcal{P}^{<}]_{\star}-[\overline{\Sigma}_{\mathrm{A}}^{\mu},\mathcal{A}_{\mu}^{<}]_{\star}+\frac{1}{2}[\overline{\Sigma}_{\mathrm{T}}^{\mu\nu},\mathcal{S}_{\mu\nu}^{<}]_{\star}\right)\nonumber \\
 &  & +\frac{i\hbar}{2}\left(\{\widehat{\Sigma_{\mathrm{F}},\mathcal{F}}\}_{\star}+\{\widehat{\Sigma_{\mathrm{V}}^{\mu},\mathcal{V}_{\mu}}\}_{\star}-\{\widehat{\Sigma_{\mathrm{P}},\mathcal{P}}\}_{\star}-\{\widehat{\Sigma_{\mathrm{A}}^{\mu},\mathcal{A}_{\mu}}\}_{\star}+\frac{1}{2}\{\widehat{\Sigma_{\mathrm{T}}^{\mu\nu},\mathcal{S}_{\mu\nu}}\}_{\star}\right),
\end{eqnarray}
the pseudoscalar part,
\begin{eqnarray}
0 & = & 2i\Pi^{\mu}\mathcal{A}_{\mu}^{<}+i\left(\{\overline{\Sigma}_{\mathrm{V}}^{\mu},\mathcal{A}_{\mu}^{<}\}_{\star}-\{\overline{\Sigma}_{\mathrm{A}}^{\mu},\mathcal{V}_{\mu}^{<}\}_{\star}\right)+\left([\overline{\Sigma}_{\mathrm{F}},\mathcal{P}^{<}]_{\star}+[\overline{\Sigma}_{\mathrm{P}},\mathcal{F}^{<}]_{\star}{\color{blue}-}\frac{1}{4}\epsilon^{\mu\nu\alpha\beta}[\overline{\Sigma}_{\mathrm{T},\alpha\beta},\mathcal{S}_{\mu\nu}^{<}]_{\star}\right)\nonumber \\
 &  & +\frac{i\hbar}{2}\left(\{\widehat{\Sigma_{\mathrm{F}},\mathcal{P}}\}_{\star}+\{\widehat{\Sigma_{\mathrm{P}},\mathcal{F}}\}_{\star}{\color{blue}-}\frac{1}{4}\{\widehat{\Sigma_{\mathrm{T},\alpha\beta},\mathcal{S}_{\mu\nu}}\}_{\star}\epsilon^{\mu\nu\alpha\beta}\right)-\frac{\hbar}{2}\left([\widehat{\Sigma_{\mathrm{V}}^{\mu},\mathcal{A}_{\mu}}]_{\star}-[\widehat{\Sigma_{\mathrm{A}}^{\mu},\mathcal{V}_{\mu}}]_{\star}\right),
\end{eqnarray}
the vector part,
\begin{eqnarray}
0 & = & 2i\Pi^{\nu}\mathcal{S}_{\nu\mu}^{<}+i\hbar\nabla_{\mu}\mathcal{F}^{<}+i\left(\{\overline{\Sigma}_{\mathrm{P}},\mathcal{A}_{\mu}^{<}\}_{\star}+\{\overline{\Sigma}_{\mathrm{V}}^{\nu},\mathcal{S}_{\nu\mu}^{<}\}_{\star}-\{\overline{\Sigma}_{\mathrm{A},\mu},\mathcal{P}^{<}\}_{\star}+\{\overline{\Sigma}_{\mathrm{T},\mu\nu},\mathcal{V}^{<,\nu}\}_{\star}\right)\nonumber \\
 &  & +\left([\overline{\Sigma}_{\mathrm{F}},\mathcal{V}_{\mu}^{<}]_{\star}+[\overline{\Sigma}_{\mathrm{V},\mu},\mathcal{F}^{<}]_{\star}-\frac{1}{2}\epsilon_{\mu\nu\alpha\beta}([\overline{\Sigma}_{\mathrm{T}}^{\alpha\beta},\mathcal{A}^{<,\nu}]_{\star}+[\overline{\Sigma}_{\mathrm{A}}^{\nu},\mathcal{S}^{<,\alpha\beta}]_{\star})\right)\nonumber \\
 &  & +\frac{i\hbar}{2}\left(\{\widehat{\Sigma_{\mathrm{F}},\mathcal{V}_{\mu}}\}_{\star}+\{\widehat{\Sigma_{\mathrm{V},\mu},\mathcal{F}}\}_{\star}-\frac{1}{2}\epsilon_{\mu\nu\alpha\beta}(\{\widehat{\Sigma_{\mathrm{A}}^{\nu},\mathcal{S}^{\alpha\beta}}\}_{\star}+\{\widehat{\Sigma_{\mathrm{T}}^{\alpha\beta},\mathcal{A}^{\nu}}\}_{\star})\right)\nonumber \\
 &  & -\frac{\hbar}{2}\left([\widehat{\Sigma_{\mathrm{P}},\mathcal{A}_{\mu}}]_{\star}+[\widehat{\Sigma_{\mathrm{V}}^{\nu},\mathcal{S}_{\nu\mu}}]_{\star}-[\widehat{\Sigma_{\mathrm{A},\mu},\mathcal{P}}]_{\star}+[\widehat{\Sigma_{\mathrm{T},\mu\nu},\mathcal{V}^{\nu}}]_{\star}\right),
\end{eqnarray}
the axial vector part,
\begin{eqnarray}
0 & = & -2i\Pi_{\mu}\mathcal{P}^{<}-\frac{i\hbar}{2}\epsilon_{\mu\nu\rho\sigma}\nabla^{\nu}\mathcal{S}^{<,\rho\sigma} \nonumber \\ 
& & +i\left(\{\overline{\Sigma}_{\mathrm{P}},\mathcal{V}_{\mu}^{<}\}_{\star}-\{\overline{\Sigma}_{\mathrm{V},\mu},\mathcal{P}^{<}\}_{\star}+\{\overline{\Sigma}_{\mathrm{A}}^{\nu},\mathcal{S}_{\nu\mu}^{<}\}_{\star}+\{\overline{\Sigma}_{\mathrm{T},\mu\nu},\mathcal{A}^{<,\nu}\}_{\star}\right)\nonumber \\
 &  & +\left([\overline{\Sigma}_{\mathrm{F}},\mathcal{A}_{\mu}^{<}]_{\star}+[\overline{\Sigma}_{\mathrm{A},\mu},\mathcal{F}^{<}]_{\star}-\frac{1}{2}\epsilon_{\mu\nu\alpha\beta}([\overline{\Sigma}_{\mathrm{V}}^{\nu},\mathcal{S}^{<,\alpha\beta}]_{\star}+[\overline{\Sigma}_{\mathrm{T}}^{\alpha\beta},\mathcal{V}^{<,\nu}]_{\star})\right)\nonumber \\
 &  & +\frac{i\hbar}{2}\left(\{\widehat{\Sigma_{\mathrm{F}},\mathcal{A}_{\mu}}\}_{\star}+\{\widehat{\Sigma_{\mathrm{A},\mu},\mathcal{F}}\}_{\star}-\frac{1}{2}\epsilon_{\mu\nu\alpha\beta}(\{\widehat{\Sigma_{\mathrm{T}}^{\alpha\beta},\mathcal{V}^{\nu}}\}_{\star}+\{\widehat{\Sigma_{\mathrm{V}}^{\nu},\mathcal{S}^{\alpha\beta}}\}_{\star})\right)\nonumber \\
 &  & -\frac{\hbar}{2}\left([\widehat{\Sigma_{\mathrm{P}},\mathcal{V}_{\mu}}]_{\star}+[\widehat{\Sigma_{\mathrm{A}}^{\nu},\mathcal{S}_{\nu\mu}}]_{\star}-[\widehat{\Sigma_{\mathrm{V},\mu},\mathcal{P}}]_{\star}+[\widehat{\Sigma_{\mathrm{T},\mu\nu},\mathcal{A}^{\nu}}]_{\star}\right),
\end{eqnarray}
the tensor part,
\begin{eqnarray}
0 & = & -4i\Pi_{[\mu}\mathcal{V}_{\nu]}^{<}-i\hbar\epsilon_{\mu\nu\rho\sigma}\nabla^{\rho}\mathcal{A}^{<,\sigma} \nonumber \\
& & +2i\left(\{\overline{\Sigma}_{\mathrm{A},[\mu},\mathcal{A}_{\nu]}^{<}\}_{\star}-\{\overline{\Sigma}_{\mathrm{V},[\mu},\mathcal{V}_{\nu]}^{<}\}_{\star}\right)+[\overline{\Sigma}_{\mathrm{F}},\mathcal{S}_{\mu\nu}^{<}]_{\star}+[\overline{\Sigma}_{\mathrm{T},\mu\nu},\mathcal{F}^{<}]_{\star}\nonumber \\
 &  & -\frac{1}{2}\epsilon_{\mu\nu\alpha\beta}([\overline{\Sigma}_{\mathrm{P}},\mathcal{S}^{<,\alpha\beta}]_{\star}+[\overline{\Sigma}_{\mathrm{T}}^{\alpha\beta},\mathcal{P}^{<}]_{\star})+\epsilon_{\mu\nu\alpha\beta}(-[\overline{\Sigma}_{\mathrm{V}}^{\alpha},\mathcal{A}^{<,\beta}]_{\star}+[\overline{\Sigma}_{\mathrm{A}}^{\alpha},\mathcal{V}^{<,\beta}]_{\star})\nonumber \\
 &  & +\frac{i\hbar}{2}\Big[\{\widehat{\Sigma_{\mathrm{F}},\mathcal{S}_{\mu\nu}}\}_{\star}+\{\widehat{\Sigma_{\mathrm{T},\mu\nu},\mathcal{F}}\}_{\star}-\frac{1}{2}\epsilon_{\mu\nu\alpha\beta}(\{\widehat{\Sigma_{\mathrm{P}},\mathcal{S}^{\alpha\beta}}\}_{\star}+\{\widehat{\Sigma_{\mathrm{T}}^{\alpha\beta},\mathcal{P}}\}_{\star})\nonumber \\
 &  & \qquad+\epsilon_{\alpha\beta\mu\nu}(-\{\widehat{\Sigma_{\mathrm{V}}^{\alpha},\mathcal{A}^{\beta}}\}_{\star}+\{\widehat{\Sigma_{\mathrm{A}}^{\alpha},\mathcal{V}^{\beta}}\}_{\star})\Big]-\hbar\left(-[\widehat{\Sigma_{\mathrm{V},[\mu},\mathcal{V}_{\nu]}}]_{\star}+[\widehat{\Sigma_{\mathrm{A},[\mu},\mathcal{A}_{\nu]}}]_{\star}\right).
\end{eqnarray}

\subsection{Keep up to $\mathcal{O}(\hbar^{2})$} \label{subsec:Keep-up-to}

By inserting the Moyal products in the $\widehat{AB}$ terms and keep
up to $\mathcal{O}(\hbar^{2})$, e.g. 
\begin{eqnarray}
[\widehat{\Sigma_{\mathrm{F}},\mathcal{F}}]_{\star} & = & i\hbar[\widehat{\Sigma_{\mathrm{F}}\mathcal{F}}]_{\mathrm{P.B.}}^{F}+\mathcal{O}(\hbar^{2})+i\mathcal{O}(\hbar^{3}),\\
\{\widehat{\Sigma_{\mathrm{P}},\mathcal{A}_{\mu}}\}_{\star} & = & 2\widehat{\Sigma_{\mathrm{P}}\mathcal{A}_{\mu}}+\mathcal{O}(\hbar^{2})+i\mathcal{O}(\hbar^{3})+\mathcal{O}(\hbar^{4}).
\end{eqnarray}
with $[\widehat{\Sigma_{\mathrm{F}},\mathcal{F}}]_{\mathrm{P.B.}}=[\Sigma_{\mathrm{F}}^{>}\mathcal{F}^{<}]_{\mathrm{P.B.}}-[\Sigma_{\mathrm{F}}^{<}\mathcal{F}^{>}]_{\mathrm{P.B.}}$ as a shorthand
notation.

Then we have
\begin{eqnarray}
\widetilde{m}\mathcal{F}^{<} & = & \widetilde{\Pi}_{\mu}\mathcal{V}^{<,\mu}-\overline{\Sigma}_{\mathrm{P}}\mathcal{P}^{<}-\overline{\Sigma}_{\mathrm{A}}^{\mu}\mathcal{A}_{\mu}^{<}+\frac{1}{2}\overline{\Sigma}_{\mathrm{T}}^{\mu\nu}\mathcal{S}_{\mu\nu}^{<}\nonumber \\
 &  & -\frac{\hbar^{2}}{4}[\widehat{\Sigma_{\mathrm{F}}\mathcal{F}}+\widehat{\Sigma_{\mathrm{V}}^{\mu}\mathcal{V}_{\mu}}-\widehat{\Sigma_{\mathrm{P}}\mathcal{P}}-\widehat{\Sigma_{\mathrm{A}}^{\mu}\mathcal{A}_{\mu}}+\frac{1}{2}\widehat{\Sigma_{\mathrm{T}}^{\mu\nu}\mathcal{S}_{\mu\nu}}]_{\mathrm{P.B.}}^{F}+\mathcal{O}(\hbar^{3}).\label{eq:Master eq 1-1}
\end{eqnarray}
Notice that in the leading orders of $\hbar$, the equation is real,
so the imaginary can be decoupled from our real part, and the $\mathcal{O}(\hbar^{2}\overline{\Sigma})$
is imaginary, so we can collect the higher real $\overline{\Sigma}$
terms as $\mathcal{O}(\hbar^{3})$ in our interested part. And we
should also match the imaginary and real parts in both sides just
as what we do for the chiral master equations. Similarly, we can derive the master equations of
other components,
\begin{eqnarray}
\widetilde{m}\mathcal{P}^{<} &   = & -\frac{\hbar}{2}(\widetilde{\mathcal{D}}^{\mu}\mathcal{A}_{\mu}^{<}-\widehat{\Sigma_{\mathrm{A}}^{\mu}\mathcal{V}_{\mu}})+\overline{\Sigma}_{\mathrm{P}}\mathcal{F}^{<}{\color{blue}-}\frac{1}{4}\epsilon^{\mu\nu\alpha\beta}\overline{\Sigma}_{\mathrm{T},\alpha\beta}\mathcal{S}_{\mu\nu}^{<}+\frac{\hbar}{2}[\overline{\Sigma}_{\mathrm{A}}^{\mu}\mathcal{V}_{\mu}^{<}]_{\mathrm{P.B.}}^{F}\nonumber \\
 &  & -\frac{\hbar^{2}}{4}\left([\widehat{\Sigma_{\mathrm{F}}\mathcal{P}}+\widehat{\Sigma_{\mathrm{P}}\mathcal{F}}{\color{blue}-}\frac{1}{4}\epsilon^{\mu\nu\alpha\beta}\widehat{\Sigma_{\mathrm{T},\alpha\beta}\mathcal{S}_{\mu\nu}}]_{\mathrm{P.B.}}^{F}\right)+\mathcal{O}(\hbar^{3}),\label{eq:Master eq 1-2}\\
 - 2\widetilde{\Pi}_{\mu}\mathcal{F}^{<} & = & -2\widetilde{m}\mathcal{V}_{\mu}^{<}-\hbar\widetilde{\mathcal{D}}^{\nu}S_{\nu\mu}^{<}-\hbar\left(\widehat{\Sigma_{\mathrm{P}}\mathcal{A}_{\mu}}-\widehat{\Sigma_{\mathrm{A},\mu}\mathcal{P}}+\widehat{\Sigma_{\mathrm{T},\mu\nu}\mathcal{V}^{\nu}}\right)\nonumber \\
 &  & -\epsilon_{\mu\nu\alpha\beta}(\overline{\Sigma}_{\mathrm{A}}^{\nu}\mathcal{S}^{<,\alpha\beta}+\overline{\Sigma}_{\mathrm{T}}^{\alpha\beta}\mathcal{A}^{<\nu})-\hbar[\overline{\Sigma}_{\mathrm{P}}\mathcal{A}_{\mu}^{<}-\overline{\Sigma}_{\mathrm{A},\mu}\mathcal{P}^{<}+\overline{\Sigma}_{\mathrm{T},\mu\nu}\mathcal{V}^{<,\nu}]_{\mathrm{P.B.}}^{F}\nonumber \\
 &  & - \frac{\hbar^{2}}{2}\left[\widehat{\Sigma_{\mathrm{F}}\mathcal{V}_{\mu}}+\widehat{\Sigma_{\mathrm{V},\mu}\mathcal{F}}-\frac{1}{2}\epsilon_{\mu\nu\alpha\beta}(\widehat{\Sigma_{\mathrm{T}}^{\alpha\beta}\mathcal{A}^{\nu}}+\widehat{\Sigma_{\mathrm{A}}^{\nu}\mathcal{S}^{\alpha\beta}})\right]_{\mathrm{P.B.}}^{F}+\mathcal{O}(\hbar^{3}),\label{eq:Master eq 1-3}\\
 - \hbar\widetilde{\mathcal{D}}_{\mu}\mathcal{P}^{<}& = & -\epsilon_{\mu\nu\alpha\beta}\widetilde{\Pi}^{\nu}\mathcal{S}^{<,\alpha\beta}-2\widetilde{m}\mathcal{A}_{\mu}^{<}-\hbar\left(\widehat{\Sigma_{\mathrm{P}}\mathcal{V}_{\mu}}+\widehat{\Sigma_{\mathrm{A}}^{\nu}\mathcal{S}_{\nu\mu}}+\widehat{\Sigma_{\mathrm{T},\mu\nu}\mathcal{A}^{\nu}}\right)\nonumber \\
 &  & +2\overline{\Sigma}_{\mathrm{A},\mu}\mathcal{F}^{<}-\epsilon_{\mu\nu\alpha\beta}\overline{\Sigma}_{\mathrm{T}}^{\alpha\beta}\mathcal{V}^{<,\nu}-\hbar\left[\overline{\Sigma}_{\mathrm{P}}\mathcal{V}_{\mu}^{<}+\overline{\Sigma}_{\mathrm{A}}^{\nu}\mathcal{S}_{\nu\mu}^{<}+\overline{\Sigma}_{\mathrm{T},\mu\nu}\mathcal{A}^{<,\nu}\right]_{\mathrm{P.B.}}^{F}\nonumber \\
 &  & - \frac{\hbar^{2}}{2}\left[\widehat{\Sigma_{\mathrm{F}}\mathcal{A}_{\mu}}+\widehat{\Sigma_{\mathrm{A},\mu}\mathcal{F}}-\frac{1}{2}\epsilon_{\mu\nu\alpha\beta}(\widehat{\Sigma_{\mathrm{V}}^{\nu}\mathcal{S}^{\alpha\beta}}+\widehat{\Sigma_{\mathrm{T}}^{\alpha\beta}\mathcal{V}^{\nu}})\right]_{\mathrm{P.B.}}^{F}+\mathcal{O}(\hbar^{3}),\label{eq:Master eq 1-4}\\
 - \widetilde{m}\mathcal{S}_{\mu\nu}^{<} & = & \epsilon_{\mu\nu\alpha\beta}\widetilde{\Pi}^{\alpha}\mathcal{A}^{<,\beta}-\hbar(\widetilde{\mathcal{D}}_{[\mu}\mathcal{V}_{\nu]}^{<}-\widehat{\Sigma_{\mathrm{A},[\mu},\mathcal{A}_{\nu]}})\nonumber \\
 &  & -\overline{\Sigma}_{\mathrm{T},\mu\nu}\mathcal{F}^{<}+\frac{1}{2}\epsilon_{\mu\nu\alpha\beta}(\overline{\Sigma}_{\mathrm{P}}\mathcal{S}^{<,\alpha\beta}+\overline{\Sigma}_{\mathrm{T}}^{\alpha\beta}\mathcal{P}^{<})-\epsilon_{\mu\nu\alpha\beta}\overline{\Sigma}_{\mathrm{A}}^{\alpha}\mathcal{V}^{<,\beta}+\hbar\left[\overline{\Sigma}_{\mathrm{A},[\mu}\mathcal{A}_{\nu]}^{<}\right]_{\mathrm{P.B.}}^{F}\nonumber \\
 &  & + \frac{\hbar^{2}}{4}\left[\widehat{\Sigma_{\mathrm{F}}\mathcal{S}_{\mu\nu}}+\widehat{\Sigma_{\mathrm{T},\mu\nu}\mathcal{F}}-\frac{1}{2}\epsilon_{\mu\nu\alpha\beta}(\widehat{\Sigma_{\mathrm{P}}\mathcal{S}^{\alpha\beta}}+\widehat{\Sigma_{\mathrm{T}}^{\alpha\beta}\mathcal{P}})+\epsilon_{\mu\nu\alpha\beta}(-\widehat{\Sigma_{\mathrm{V}}^{\alpha}\mathcal{A}^{\beta}}+\widehat{\Sigma_{\mathrm{A}}^{\alpha}\mathcal{V}^{\beta}})\right]_{\mathrm{P.B.}}^{F} \nonumber \\
 & & +\mathcal{O}(\hbar^{3}),\label{eq:Master eq 1-5}
\end{eqnarray}
and
\begin{eqnarray}
 \widetilde{\mathcal{D}}^{\mu}\mathcal{V}_{\mu}^{<} & = & -\left(\widehat{\Sigma_{\mathrm{F}}\mathcal{F}}-\widehat{\Sigma_{\mathrm{P}}\mathcal{P}}-\widehat{\Sigma_{\mathrm{A}}^{\mu}\mathcal{A}_{\mu}}+\frac{1}{2}\widehat{\Sigma_{\mathrm{T}}^{\mu\nu}\mathcal{S}_{\mu\nu}}\right)\nonumber \\
 &  & -\left[\overline{\Sigma}_{\mathrm{F}}\mathcal{F}^{<}-\overline{\Sigma}_{\mathrm{P}}\mathcal{P}^{<}-\overline{\Sigma}_{\mathrm{A}}^{\mu}\mathcal{A}_{\mu}^{<}+\frac{1}{2}\overline{\Sigma}_{\mathrm{T}}^{\mu\nu}\mathcal{S}_{\mu\nu}^{<}\right]_{\mathrm{P.B.}}^{F}+\mathcal{O}(\hbar^{2}),\label{eq:Master eq 2-1}\\
 2\widetilde{\Pi}^{\mu}\mathcal{A}_{\mu}^{<} & = & -\hbar\left(\widehat{\Sigma_{\mathrm{F}}\mathcal{P}}+\widehat{\Sigma_{\mathrm{P}}\mathcal{F}}{\color{blue}-}\frac{1}{4}\widehat{\Sigma_{\mathrm{T},\alpha\beta}\mathcal{S}_{\mu\nu}}\epsilon^{\mu\nu\alpha\beta}\right)+2\overline{\Sigma}_{\mathrm{A}}^{\mu}\mathcal{V}_{\mu}^{<}\nonumber \\
 &  & -\hbar\left[\overline{\Sigma}_{\mathrm{F}}\mathcal{P}^{<}+\overline{\Sigma}_{\mathrm{P}}\mathcal{F}^{<}{\color{blue}-}\frac{1}{4}\epsilon^{\mu\nu\alpha\beta}\overline{\Sigma}_{\mathrm{T},\alpha\beta}\mathcal{S}_{\mu\nu}^{<}\right]_{\mathrm{P.B.}}^{F}+\frac{\hbar^{2}}{2}\left[\widehat{\Sigma_{\mathrm{V}}^{\mu}\mathcal{A}_{\mu}}-\widehat{\Sigma_{\mathrm{A}}^{\mu}\mathcal{V}_{\mu}}\right]_{\mathrm{P.B.}}^{F} \nonumber \\ 
 & & +\mathcal{O}(\hbar^{3}),\label{eq:Master eq 2-2}\\
 - 2\widetilde{\Pi}^{\nu}\mathcal{S}_{\nu\mu}^{<} & = & \hbar\widetilde{\mathcal{D}}_{\mu}\mathcal{F}^{<}+\hbar\left(\widehat{\Sigma_{\mathrm{F}}\mathcal{V}_{\mu}}-\frac{1}{2}\epsilon_{\mu\nu\alpha\beta}(\widehat{\Sigma_{\mathrm{A}}^{\nu}\mathcal{S}^{\alpha\beta}}+\widehat{\Sigma_{\mathrm{T}}^{\alpha\beta}\mathcal{A}^{\nu}})\right)\nonumber \\
 &  & +2\left(\overline{\Sigma}_{\mathrm{P}}\mathcal{A}_{\mu}^{<}-\overline{\Sigma}_{\mathrm{A},\mu}\mathcal{P}^{<}+\overline{\Sigma}_{\mathrm{T},\mu\nu}\mathcal{V}^{<,\nu}\right)
 \nonumber \\ 
 & & +\hbar\left[\overline{\Sigma}_{\mathrm{F}}\mathcal{V}_{\mu}^{<}-\frac{1}{2}\epsilon_{\mu\nu\alpha\beta}(\overline{\Sigma}_{\mathrm{T}}^{\alpha\beta}\mathcal{A}^{<,\nu}+\overline{\Sigma}_{\mathrm{A}}^{\nu}\mathcal{S}^{<,\alpha\beta})\right]_{\mathrm{P.B.}}^{F}\nonumber \\
 &  & - \frac{\hbar^{2}}{2}\left[\widehat{\Sigma_{\mathrm{P}}\mathcal{A}_{\mu}}+\widehat{\Sigma_{\mathrm{V}}^{\nu}\mathcal{S}_{\nu\mu}}-\widehat{\Sigma_{\mathrm{A},\mu}\mathcal{P}}+\widehat{\Sigma_{\mathrm{T},\mu\nu}\mathcal{V}^{\nu}}\right]_{\mathrm{P.B.}}^{F}+\mathcal{O}(\hbar^{3}),\label{eq:Master eq 2-3}\\
 - 2\widetilde{\Pi}_{\mu}\mathcal{P}^{<} & = & +\frac{\hbar}{2}\epsilon_{\mu\nu\rho\sigma}(\widetilde{\mathcal{D}}^{\nu}\mathcal{S}^{<,\rho\sigma}+\widehat{\Sigma_{\mathrm{T}}^{\rho\sigma}\mathcal{V}^{\nu}}) -\hbar\left(\widehat{\Sigma_{\mathrm{F}}\mathcal{A}_{\mu}}+\widehat{\Sigma_{\mathrm{A},\mu}\mathcal{F}}\right)
\nonumber \\  & & -2\left(\overline{\Sigma}_{\mathrm{P}}\mathcal{V}_{\mu}^{<}+\overline{\Sigma}_{\mathrm{A}}^{\nu}\mathcal{S}_{\nu\mu}^{<}+\overline{\Sigma}_{\mathrm{T},\mu\nu}\mathcal{A}^{<,\nu}\right)\nonumber \\
 &  & -\hbar\left[\overline{\Sigma}_{\mathrm{F}}\mathcal{A}_{\mu}^{<}+\overline{\Sigma}_{\mathrm{A},\mu}\mathcal{F}^{<}-\frac{1}{2}\epsilon_{\mu\nu\alpha\beta}\overline{\Sigma}_{\mathrm{T}}^{\alpha\beta}\mathcal{V}^{<,\nu}\right]_{\mathrm{P.B.}}^{F}\nonumber \\
 &  &  + \frac{\hbar^{2}}{2}\left[\widehat{\Sigma_{\mathrm{P}}\mathcal{V}_{\mu}}+\widehat{\Sigma_{\mathrm{A}}^{\nu}\mathcal{S}_{\nu\mu}}-\widehat{\Sigma_{\mathrm{V},\mu}\mathcal{P}}+\widehat{\Sigma_{\mathrm{T},\mu\nu}\mathcal{A}^{\nu}}\right]_{\mathrm{P.B.}}^{F}+\mathcal{O}(\hbar^{3}),\label{eq:Master eq 2-4}\\
 - 2\widetilde{\Pi}_{[\mu}\mathcal{V}_{\nu]}^{<} & = & +\frac{\hbar}{2}\epsilon_{\mu\nu\rho\sigma}\left(\widetilde{\mathcal{D}}^{\rho}\mathcal{A}^{<,\sigma}-\widehat{\Sigma_{\mathrm{A}}^{\rho}\mathcal{V}^{\sigma}}\right) -2\overline{\Sigma}_{\mathrm{A},[\mu}\mathcal{A}_{\nu]}^{<} \nonumber \\
 & &
 -\frac{\hbar}{2}\left(\widehat{\Sigma_{\mathrm{F}}\mathcal{S}_{\mu\nu}}+\widehat{\Sigma_{\mathrm{T},\mu\nu}\mathcal{F}}-\frac{1}{2}\epsilon_{\mu\nu\alpha\beta}(\widehat{\Sigma_{\mathrm{P}}\mathcal{S}^{\alpha\beta}}+\widehat{\Sigma_{\mathrm{T}}^{\alpha\beta}\mathcal{P}})\right)\nonumber \\
 &  & -\frac{\hbar}{2}\left[\overline{\Sigma}_{\mathrm{F}}\mathcal{S}_{\mu\nu}^{<}+\overline{\Sigma}_{\mathrm{T},\mu\nu}\mathcal{F}^{<}-\frac{1}{2}\epsilon_{\mu\nu\alpha\beta}(\overline{\Sigma}_{\mathrm{P}}\mathcal{S}^{<,\alpha\beta}+\overline{\Sigma}_{\mathrm{T}}^{\alpha\beta}\mathcal{P}^{<})+\epsilon_{\mu\nu\alpha\beta}\overline{\Sigma}_{\mathrm{A}}^{\alpha}\mathcal{V}^{<,\beta}\right]_{\mathrm{P.B.}}^{F}\nonumber \\
 &  & +\frac{\hbar^{2}}{2}\left[-\widehat{\Sigma_{\mathrm{V},[\mu},\mathcal{V}_{\nu]}}+\widehat{\Sigma_{\mathrm{A},[\mu},\mathcal{A}_{\nu]}}\right]_{\mathrm{P.B.}}^{F}+\mathcal{O}(\hbar^{3}).\label{eq:Master eq 2-5}
\end{eqnarray}
These are the master equations which contains the kinetic equations
of the Wigner functions and their constraints.

\subsection{Linear order of self-energy $\Sigma$} \label{subsec:Linear-order-of}

Following Ref.~\citep{Yang:2020hri}, we can eliminate the $\mathcal{F},\mathcal{P},\mathcal{S}_{\mu\nu}$
using $\mathcal{V}^{\mu},\mathcal{A}^{\mu}$. But as we can see, $\widetilde{m}\mathcal{F},\widetilde{m}\mathcal{P},\widetilde{m}\mathcal{S}_{\mu\nu}$
are equivalent to the terms like $\widehat{\Sigma_{\mathrm{F}}\mathcal{F}},\widehat{\Sigma_{\mathrm{F}}\mathcal{P}},\widehat{\Sigma_{\mathrm{F}}\mathcal{S}_{\mu\nu}}$,
so we cannot express them as an explicit function of $\mathcal{V}^{\mu},\mathcal{A}^{\mu}$.
But now we assume that the interaction is weak enough, all the non-linear
self-energy couplings are dropped, from Eqs.~(\ref{eq:Master eq 1-1}, \ref{eq:Master eq 1-2}, \ref{eq:Master eq 1-5}),we
can get, 
\begin{eqnarray}
\mathcal{F}^{<} & = & \frac{\widetilde{\Pi}_{\mu}}{\widetilde{m}}\mathcal{V}^{<,\mu}-\frac{\hbar^{2}}{4\widetilde{m}}[\widehat{\Sigma_{\mathrm{F}}\frac{\widetilde{q}_{\mu}}{\widetilde{m}}\mathcal{V}^{\mu}}+\widehat{\Sigma_{\mathrm{V}}^{\mu}\mathcal{V}_{\mu}}-\widehat{\Sigma_{\mathrm{A}}^{\mu}\mathcal{A}_{\mu}}-\frac{1}{2}\widehat{\Sigma_{\mathrm{T}}^{\mu\nu}(\frac{1}{\widetilde{m}}\epsilon_{\mu\nu\alpha\beta}\widetilde{\Pi}^{\alpha}\mathcal{A}^{\beta})}]_{\mathrm{P.B.}}^{F}\nonumber \\
 &  & -\frac{1}{\widetilde{m}}\left[-\overline{\Sigma}_{\mathrm{P}}\frac{\hbar}{2\widetilde{m}}\widetilde{\mathcal{D}}^{\mu}\mathcal{A}_{\mu}^{<}+\overline{\Sigma}_{\mathrm{A}}^{\mu}\mathcal{A}_{\mu}^{<}-\frac{1}{2}\overline{\Sigma}_{\mathrm{T}}^{\mu\nu}\left(-\frac{1}{\widetilde{m}}\epsilon_{\mu\nu\alpha\beta}\widetilde{\Pi}^{\alpha}\mathcal{A}^{<,\beta}+\frac{\hbar}{\widetilde{m}}\widetilde{\mathcal{D}}_{[\mu}\mathcal{V}_{\nu]}^{<}\right)\right] \nonumber \\ 
 & & +\mathcal{O}(\hbar^{3})\\
\mathcal{P}^{<} & = & -\frac{\hbar}{2\widetilde{m}}(\widetilde{\mathcal{D}}^{\mu}\mathcal{A}_{\mu}^{<}-\widehat{\Sigma_{\mathrm{A}}^{\mu}\mathcal{V}_{\mu}})-\frac{\hbar^{2}}{4\widetilde{m}}[\widehat{\Sigma_{\mathrm{P}}\frac{\widetilde{q}_{\mu}\mathcal{V}^{\mu}}{\widetilde{m}}}-\widehat{\Sigma_{\mathrm{T},\alpha\beta}(\frac{1}{\widetilde{m}}\widetilde{q}^{[\alpha}\mathcal{A}^{<,\beta]})}]_{\mathrm{P.B.}}^{F}+\frac{\hbar}{2\widetilde{m}}[\overline{\Sigma}_{\mathrm{A}}^{\mu}\mathcal{V}_{\mu}^{<}]_{\mathrm{P.B.}}^{F}\nonumber \\
 &  & +\frac{1}{\widetilde{m}}\left[\overline{\Sigma}_{\mathrm{P}}\frac{\widetilde{\Pi}_{\mu}}{\widetilde{m}}\mathcal{V}^{<,\mu}{\color{blue}-}\frac{1}{4}\epsilon^{\mu\nu\alpha\beta}\overline{\Sigma}_{\mathrm{T},\alpha\beta}\left(-\frac{1}{\widetilde{m}}\epsilon_{\mu\nu\rho\sigma}\widetilde{\Pi}^{\rho}\mathcal{A}^{<,\sigma}+\frac{\hbar}{\widetilde{m}}\widetilde{\mathcal{D}}_{[\mu}\mathcal{V}_{\nu]}^{<}\right)\right] \nonumber \\ 
 & &
 +\mathcal{O}(\hbar^{3}),\\
\mathcal{S}_{\mu\nu}^{<} & = & -\frac{1}{\widetilde{m}}\epsilon_{\mu\nu\alpha\beta}\widetilde{\Pi}^{\alpha}\mathcal{A}^{<,\beta}+\frac{\hbar}{\widetilde{m}}\left(\widetilde{\mathcal{D}}_{[\mu}\mathcal{V}_{\nu]}^{<}-\widehat{\Sigma_{\mathrm{A},[\mu},\mathcal{A}_{\nu]}}\right)\nonumber \\
 &  & -\frac{1}{\widetilde{m}}\Bigg\{-\overline{\Sigma}_{\mathrm{T},\mu\nu}\frac{\widetilde{\Pi}_{\alpha}}{\widetilde{m}}\mathcal{V}^{<,\alpha}+\frac{1}{2}\epsilon_{\mu\nu\alpha\beta}(\overline{\Sigma}_{\mathrm{P}}\left(-\frac{1}{\widetilde{m}}\epsilon^{\alpha\beta\lambda\gamma}\widetilde{\Pi}_{\lambda}\mathcal{A}_{\gamma}^{<}+\frac{\hbar}{\widetilde{m}}\widetilde{\mathcal{D}}^{[\alpha}\mathcal{V}^{<,\beta]}\right)\nonumber \\
 &  & \qquad-\frac{\hbar}{2\widetilde{m}}\overline{\Sigma}_{\mathrm{T}}^{\alpha\beta}\widetilde{\mathcal{D}}^{\mu}\mathcal{A}_{\mu}^{<})-\epsilon_{\mu\nu\alpha\beta}\overline{\Sigma}_{\mathrm{A}}^{\alpha}\mathcal{V}^{<,\beta}\Bigg\}-\frac{\hbar}{\widetilde{m}}\left[\overline{\Sigma}_{\mathrm{A},[\mu}\mathcal{A}_{\nu]}^{<}\right]_{\mathrm{P.B.}}^{F} \nonumber \\
 &  & -\frac{\hbar^{2}}{4\widetilde{m}}\left[-\widehat{\Sigma_{\mathrm{F}}(\frac{1}{\widetilde{m}}\epsilon_{\mu\nu\alpha\beta}\widetilde{q}^{\alpha}\mathcal{A}^{\beta})}+\widehat{\Sigma_{\mathrm{T},\mu\nu}(\frac{\widetilde{q}_{\lambda}\mathcal{V}^{\lambda}}{\widetilde{m}})}-2\widehat{\Sigma_{\mathrm{P}}(\frac{1}{\widetilde{m}}\widetilde{q}_{[\mu}\mathcal{A}_{\nu]})} \right. \nonumber \\ 
 & & \qquad \left.+\epsilon_{\mu\nu\alpha\beta}(-\widehat{\Sigma_{\mathrm{V}}^{\alpha}\mathcal{A}^{\beta}}+\widehat{\Sigma_{\mathrm{A}}^{\alpha}\mathcal{V}^{\beta}})\right]_{\mathrm{P.B.}}^{F} +\mathcal{O}(\hbar^{3}),
\end{eqnarray}
where we have defined 
\begin{eqnarray}
\widetilde{q}_{\mu} & = & q_{\mu}+\overline{\Sigma}_{\mathrm{V},\mu}.
\end{eqnarray}

\subsection{Redundancy of Eq.~(\ref{eq:Redundant master eq})}\label{app:redundancy}
We can check Eq.~(\ref{eq:Redundant master eq}) is redundant: we start
from Eq.~(\ref{eq:Redundant master eq}), and firstly calculate,
\begin{eqnarray}
\widetilde{\nabla}_{\mu}\widetilde{q}_{\nu} & \equiv & \nabla_{\mu}\widetilde{q}_{\nu}-(\nabla_{\alpha}\overline{\Sigma}_{\mathrm{V},\mu})\partial_{q}^{\alpha}\widetilde{q}_{\nu}+(\partial_{q,\alpha}\overline{\Sigma}_{\mathrm{V},\mu})\partial_{X}^{\alpha}\widetilde{q}_{\nu}\nonumber \\
 & \simeq & QeF_{\mu\nu}+2\nabla_{[\mu}\overline{\Sigma}_{\mathrm{V},\nu]}+\mathcal{O}(\overline{\Sigma}^{2}),
\end{eqnarray}
which is antisymmetric w.r.t. the $\mu,\nu$ indices, and
\begin{eqnarray}
\widetilde{\nabla}_{\mu}\widetilde{m} & = & -\widetilde{\nabla}_{\mu}\overline{\Sigma}_{\mathrm{F}},
\end{eqnarray}
where we have kept up to the linear terms of the self-energies $\overline{\Sigma}$.
We then find, 
\begin{eqnarray}
\widetilde{\mathcal{D}}_{\mu}(\frac{1}{\widetilde{m}}\widetilde{q}_{\nu}\mathcal{V}^{<,\nu}) & = & \widetilde{\nabla}_{\mu}(\frac{\widetilde{q}_{\nu}}{\widetilde{m}})\mathcal{V}^{<,\nu}+\frac{\widetilde{q}_{\nu}}{\widetilde{m}}\widetilde{\mathcal{D}}_{\mu}\mathcal{V}^{<,\nu}.
\end{eqnarray}
Now Eq.~(\ref{eq:Redundant master eq}) can be also expressed as, 
\begin{eqnarray*}
0 & = & \widetilde{q}\cdot\widetilde{\mathcal{D}}\mathcal{V}_{\mu}^{<}+(\widetilde{\nabla}_{\mu}\widetilde{q}_{\nu})\mathcal{V}^{<,\nu}-\frac{1}{\widetilde{m}}(\widetilde{\nabla}_{\mu}\widetilde{m})\widetilde{q}_{\nu}\mathcal{V}^{<,\nu}+\widetilde{m}\widehat{\Sigma_{\mathrm{F}}\mathcal{V}_{\mu}} \nonumber \\ & &
+\widetilde{m}\left[-(\nabla_{\alpha}\overline{\Sigma}_{\mathrm{F}})(\partial_{q}^{\alpha}\mathcal{V}_{\mu}^{<})+(\partial_{q,\alpha}\overline{\Sigma}_{\mathrm{F}})(\partial_{X}^{\alpha}\mathcal{V}_{\mu}^{<})\right].
\end{eqnarray*}
On the other hand, acting the derivative $\widetilde{\mathcal{D}}_{\mu}$
on Eq.~(\ref{eq:Vector constrain 1}) and using the SKE
(\ref{eq:Vector kinetic eq}), we get, 
\begin{eqnarray*}
0 & = & \widetilde{\mathcal{D}}^{\mu}(\widetilde{q}_{\mu}\mathcal{V}_{\nu}^{<}-\widetilde{q}_{\nu}\mathcal{V}_{\mu}^{<})\\
 & = & \widetilde{q}\cdot\widetilde{\mathcal{D}}\mathcal{V}_{\nu}^{<}+(\widetilde{\nabla}_{\nu}\widetilde{q}_{\mu})\mathcal{V}^{<,\mu}+\widetilde{m}\widehat{\Sigma_{\mathrm{F}}\mathcal{V}_{\nu}}-(\nabla_{\alpha}\overline{\Sigma}_{\mathrm{F}})\partial_{q}^{\alpha}(\widetilde{m}\mathcal{V}_{\nu}^{<})+(\partial_{q,\alpha}\overline{\Sigma}_{\mathrm{F}})\partial_{X}^{\alpha}(\widetilde{m}\mathcal{V}_{\nu}^{<})\\
 &  & +\frac{1}{\widetilde{m}}\widetilde{q}_{\mu}\mathcal{V}^{<,\mu}(\nabla_{\alpha}\overline{\Sigma}_{\mathrm{F}})\partial_{q}^{\alpha}\widetilde{q}_{\nu}-\frac{1}{\widetilde{m}}\widetilde{q}_{\mu}\mathcal{V}^{<,\mu}(\partial_{q,\alpha}\overline{\Sigma}_{\mathrm{F}})\partial_{X}^{\alpha}\widetilde{q}_{\nu}+\mathcal{O}(\hbar^{2})\\
 & = & \widetilde{q}\cdot\widetilde{\mathcal{D}}\mathcal{V}_{\nu}^{<}+(\widetilde{\nabla}_{\nu}\widetilde{q}_{\mu})\mathcal{V}^{<,\mu}-\frac{\widetilde{\nabla}_{\nu}\widetilde{m}}{\widetilde{m}}\widetilde{q}_{\mu}\mathcal{V}^{<,\mu}+\widetilde{m}\widehat{\Sigma_{\mathrm{F}}\mathcal{V}_{\nu}} \nonumber \\ & &
 +\widetilde{m}\left[-(\nabla_{\alpha}\overline{\Sigma}_{\mathrm{F}})\partial_{q}^{\alpha}\mathcal{V}_{\nu}^{<}+(\partial_{q,\alpha}\overline{\Sigma}_{\mathrm{F}})\partial_{X}^{\alpha}\mathcal{V}_{\nu}^{<}\right]+\mathcal{O}(\hbar^{2}),
\end{eqnarray*}
and therefore Eq.~(\ref{eq:Redundant master eq}) is indeed redundant.

\section{The mean field contributions in Abelian gauge theory}\label{sec:The- mean-field- contributions-in }

Let us return to the l.h.s. of the Dyson-Schwinger equation (\ref{eq:GF_D-S eq 1}).
Considering the QED interaction,
\begin{eqnarray}
i\hbar\int_{\mathrm{C}}\mathrm{d}^{4}z\Sigma(x,z)G(z,y) & = & Qe\langle\widetilde{T}\overline{\psi}(y)\gamma^{\mu}a_{\mu}(x)\psi(x)\rangle,
\end{eqnarray}
and conducting the mean field approximation (or the pairing approximation) such that
all the connected $n>3$ point functions are set to zero, we can
approximate the self-energy as
\begin{eqnarray}
Qe\langle\widetilde{T}\overline{\psi}(y)\gamma^{\mu}a_{\mu}(x)\psi(x)\rangle & = & Qe\gamma_{\alpha\beta}^{\mu}\langle\widetilde{T}\overline{\psi}_{\alpha}(y)\psi_{\beta}(x)\rangle\langle a_{\mu}(x)\rangle,
\end{eqnarray}
which yields
\begin{eqnarray}
i\hbar\int_{\mathrm{C}}\mathrm{d}^{4}z\Sigma(x,z)G(z,y) & = & Qe\langle a_{\mu}(x)\rangle\gamma^{\mu}G(x,y)=\int_{\mathrm{C}}\mathrm{d}^{4}z\delta_{\mathrm{C}}^{(4)}(x-z)\Sigma_{\delta}(x)G(z,y),
\end{eqnarray}
and thus
\begin{eqnarray}
\Sigma_{\delta}(x) & = & Qe\gamma^{\mu}\langle a_{\mu}(x)\rangle.
\end{eqnarray}
It is obviously gauge invariant as expected. First, in the background field gauge \citep{Peskin:1995ev}, the quantum part of the gauge fields transform trivially under local gauge transformation which means our expressions of tadpole self-energy is also gauge invariant. Second, one can find $\Sigma_\delta$ is gauge invariant from the EoMs of gauge invariant Green's function in Eqs.~(\ref{eq:EoM_Gauge_inv_GF_1},\ref{eq:EoM_Gauge_inv_GF_2}). But since we have
separated the classical part and quantum part of the gauge fields
in our previous discussions, we have $\langle a_{\mu}(x)\rangle=0$.
Notice that one could also define a gauge-dependent Wigner
functions and derive the corresponding EoMs, and the background electromagnetic
field in the solutions can be generated dynamically in such approach.

\section{Feynman vector Wigner function\label{sec:Feynman-vector-Wigner}}

In this section, we derive the Feynman vector Wigner function up to
leading order in $\hbar$ and coupling. 

We have,
\begin{eqnarray}
& & \mathcal{V}^{++,\beta}(q,X) \nonumber \\
& = & \int\mathrm{d}^{4}Ye^{i\frac{q\cdot Y}{\hbar}}\left(\theta(Y_{0})\mathcal{V}^{>,\beta}(x,y)-\theta(-Y_{0})\mathcal{V}^{<,\beta}(x,y)\right)\nonumber \\
 & = & i\hbar \int\mathrm{d}^{4}Ye^{i\frac{q\cdot Y}{\hbar}}\int_{-\infty}^{+\infty}\frac{\mathrm{d}k_{0}}{2\pi\hbar}\frac{\mathrm{d}^{4}q^{\prime}}{(2\pi\hbar)^{4}}e^{-i\frac{q^{\prime}\cdot Y}{\hbar}}e^{-i\frac{k_{0}Y_{0}}{\hbar}}\left(\frac{1}{k_{0}+i\eta}\mathcal{V}^{>,\beta}(q^{\prime},X)+\frac{1}{k_{0}-i\eta}\mathcal{V}^{<,\beta}(q^{\prime},X)\right)\nonumber \\
 & = & i\hbar\int\frac{\mathrm{d}q_{0}^{\prime}}{2\pi\hbar}\left(\frac{1}{q_{0}-q_{0}^{\prime}+i\eta}\mathcal{V}^{>,\beta}(q_{0}^{\prime},\mathbf{q};X)+\frac{1}{q_{0}-q_{0}^{\prime}-i\eta}\mathcal{V}^{<,\beta}(q_{0}^{\prime},\mathbf{q};X)\right)\nonumber \\
 & = & i\hbar\int\frac{\mathrm{d}q_{0}^{\prime}}{2\pi\hbar}2\pi\epsilon(q_{0}^{\prime})\delta(\widetilde{q}^{\prime,2}-\widetilde{m}^{2})\widetilde{q}^{\prime,\beta}\left(\frac{1}{q_{0}-q_{0}^{\prime}+i\eta}f_{\mathrm{V}}^{>}(q_{0}^{\prime},\mathbf{q};X)+\frac{1}{q_{0}-q_{0}^{\prime}-i\eta}f_{\mathrm{V}}^{<}(q_{0}^{\prime},\mathbf{q};X)\right), \nonumber \\
\end{eqnarray}
where we have used, 
\begin{eqnarray}
\theta(\pm t) & = & \pm i\int_{-\infty}^{+\infty}\frac{\mathrm{d}k_{0}^{\prime}}{2\pi}\frac{e^{-ik_{0}^{\prime}t}}{k_{0}^{\prime}\pm i\eta},\quad t>0, 
\end{eqnarray}
and in the last step we have substituted, 
\begin{eqnarray}
\mathcal{V}^{<,\beta}(q_{0},\mathbf{q};X) & = & 2\pi\epsilon(q_{0})\delta(\widetilde{q}^{2}-\widetilde{m}^{2})\widetilde{q}^{\mu}f_{\mathrm{V}}^{<}(q,X),
\end{eqnarray}
for conventional reason. Then inserting, 
\begin{eqnarray}
\frac{1}{x+i\epsilon} & = & \mathrm{P.V.}\frac{1}{x}-i\pi\delta(x),
\end{eqnarray}
we can get,
\begin{eqnarray}
\mathcal{V}^{++,\beta}(q,X) & = & i\int\mathrm{d}q_{0}^{\prime}\epsilon(q_{0}^{\prime})\frac{\delta(\widetilde{q}^{\prime,2}-\widetilde{m}^{2})\widetilde{q}^{\prime\beta}}{q_{0}-q_{0}^{\prime}} \nonumber \\
& &+\pi\epsilon(q_{0})\delta(\widetilde{q}^{2}-\widetilde{m}^{2})\widetilde{q}^{\beta}\left(f_{\mathrm{V}}^{>}(q,X)-f_{\mathrm{V}}^{<}(q,X)\right).
\end{eqnarray}

At the leading order in coupling, we can simply reduce $\widetilde{q}\to q$,
and it can greatly simplify our result ,
\begin{eqnarray}
\mathcal{V}^{++,\beta}(q,X) & = & \frac{iq^{\beta}}{q^{2}-m^{2}}+\pi\epsilon(q_{0})\delta(q^{2}-m^{2})q^{\beta}\left(f_{\mathrm{V}}^{>}(q,X)-f_{\mathrm{V}}^{<}(q,X)\right)\nonumber \\
 & = & \frac{iq^{\beta}}{q^{2}-m^{2}+i\eta}-2\pi\delta(q^{2}-m^{2})q^{\beta}\widetilde{n}_{\mathrm{V}}^{<}(q,X),\label{eq:Vector WF++}
\end{eqnarray}
with $E_{\mathbf{q}}=\mathbf{q}^{2}+m^{2}$ and here we have used,
\begin{eqnarray}
\epsilon(q_{0})f_{\mathrm{V}}^{<}(q,X) & = & \widetilde{n}_{\mathrm{V}}^{<}(q,X)-\theta(-q_{0}), \nonumber \\
\widetilde{n}_{\mathrm{V}}^{<}(q,X) & = & \theta(q_{0})n_{\mathrm{V},+}^{<}(q,X)+\theta(-q_{0})n_{\mathrm{V},-}^{<}(q,X).
\end{eqnarray}
In the local equilibrium case, we have, 
\begin{eqnarray}
f_{\mathrm{V}}^{<}(q,X) & = & \frac{1}{e^{(q_{0}-\mu)/T}+1},
\end{eqnarray}
so that, 
\begin{eqnarray}
f_{\mathrm{V}}^{<}(q,X) & = & \theta(q_{0})n_{\mathrm{V},+}^{<}(q,X)+\theta(-q_{0})n_{\mathrm{V},-}^{>}(q,X),
\end{eqnarray}
with 
\begin{eqnarray}
n_{\mathrm{V},+}^{<}(q,X)=\frac{1}{e^{(|q_{0}|-\mu)/T}+1}, \qquad n_{\mathrm{V},-}^{<}(q,X)=\frac{1}{e^{(|q_{0}|+\mu)/T}+1}.
\end{eqnarray}
Here we find that, in Eq.~(\ref{eq:Vector WF++}), the first term of Eq.~(\ref{eq:Vector WF++})
is related to the $T=0$ part while the second term is the finite
temperature part which is of our interest. And Eq.~(\ref{eq:Vector WF++})
also coincides with the real-time thermal equilibrium Green's function
as shown in Ref.~\citep{Bellac:2011kqa},
\begin{eqnarray}
S_{11}^{\mathrm{F}}(p) & = & (\gamma^{\mu}p_{\mu}+m)\Bigg[\frac{i}{p^{2}-m^{2}+i\eta}\nonumber \\
 &  & \qquad-2\pi\delta(p^{2}-m^{2})\left(\widetilde{f}_{\mathrm{F-D}}^{<}(E_{\mathbf{p}}-\mu)\theta(p_{0})+\widetilde{f}_{\mathrm{F-D}}^{<}(E_{\mathbf{p}}+\mu)\theta(-p_{0})\right)\Bigg].
\end{eqnarray}
It inspires us to directly write down the photonic Wigner function
in the Feynman gauge up to leading order in $\hbar$ and coupling
constant from textbook as, 
\begin{eqnarray}
D_{\mathrm{F},\mu\nu}^{11}(q,X) & = & \frac{-i\eta_{\mu\nu}}{q^{2}+i\eta}-2\pi\eta_{\mu\nu}n_{\mathrm{V}}(q,X)\delta(q^{2}),
\end{eqnarray}
where $n(q,X)$ is the photonic distribution function and in local
equilibrium it reads, $n_{\mathrm{V}}(q,X)=(e^{|q_{0}|/T}-1)^{-1}$.


\bibliographystyle{h-physrev}
\bibliography{qkt-ref}

\begin{thebibliography}{100}

\bibitem{Rischke:2003mt}
D.~H. Rischke,
\newblock Prog. Part. Nucl. Phys. {\bf 52}, 197 (2004), nucl-th/0305030.

\bibitem{Gyulassy:2004vg}
M.~Gyulassy,
\newblock {The QGP discovered at RHIC},
\newblock in {\em {NATO Advanced Study Institute: Structure and Dynamics of Elementary Matter}}, pp. 159--182, 2004, nucl-th/0403032.

\bibitem{Shuryak:2004cy}
E.~V. Shuryak,
\newblock Nucl. Phys. A {\bf 750}, 64 (2005), hep-ph/0405066.

\bibitem{Kharzeev:2015znc}
D.~E. Kharzeev, J.~Liao, S.~A. Voloshin, and G.~Wang,
\newblock Prog. Part. Nucl. Phys. {\bf 88}, 1 (2016), 1511.04050.

\bibitem{Vilenkin:1980fu}
A.~Vilenkin,
\newblock Phys. Rev. {\bf D22}, 3080 (1980).

\bibitem{Nielsen:1983rb}
H.~B. Nielsen and M.~Ninomiya,
\newblock Phys. Lett. B {\bf 130}, 389 (1983).

\bibitem{Kharzeev:2004ey}
D.~Kharzeev,
\newblock Phys. Lett. {\bf B633}, 260 (2006), hep-ph/0406125.

\bibitem{Kharzeev:2007jp}
D.~E. Kharzeev, L.~D. McLerran, and H.~J. Warringa,
\newblock Nucl. Phys. {\bf A803}, 227 (2008), 0711.0950.

\bibitem{Fukushima:2008xe}
K.~Fukushima, D.~E. Kharzeev, and H.~J. Warringa,
\newblock Phys. Rev. {\bf D78}, 074033 (2008), 0808.3382.

\bibitem{Huang:2015oca}
X.-G. Huang,
\newblock Rept. Prog. Phys. {\bf 79}, 076302 (2016), 1509.04073.

\bibitem{Kharzeev:2020jxw}
D.~E. Kharzeev and J.~Liao,
\newblock Nature Rev. Phys. {\bf 3}, 55 (2021), 2102.06623.

\bibitem{Gao:2020vbh}
J.-H. Gao, G.-L. Ma, S.~Pu, and Q.~Wang,
\newblock Nucl. Sci. Tech. {\bf 31}, 90 (2020), 2005.10432.

\bibitem{Hidaka:2022dmn}
Y.~Hidaka, S.~Pu, Q.~Wang, and D.-L. Yang,
\newblock Prog. Part. Nucl. Phys. {\bf 127}, 103989 (2022), 2201.07644.

\bibitem{Deng:2012pc}
W.-T. Deng and X.-G. Huang,
\newblock Phys. Rev. C {\bf 85}, 044907 (2012), 1201.5108.

\bibitem{Tuchin:2013ie}
K.~Tuchin,
\newblock Adv. High Energy Phys. {\bf 2013}, 490495 (2013), 1301.0099.

\bibitem{Roy:2015kma}
V.~Roy, S.~Pu, L.~Rezzolla, and D.~Rischke,
\newblock Phys. Lett. {\bf B750}, 45 (2015), 1506.06620.

\bibitem{Pu:2016ayh}
S.~Pu, V.~Roy, L.~Rezzolla, and D.~H. Rischke,
\newblock Phys. Rev. {\bf D93}, 074022 (2016), 1602.04953.

\bibitem{Pu:2016bxy}
S.~Pu and D.-L. Yang,
\newblock Phys. Rev. {\bf D93}, 054042 (2016), 1602.04954.

\bibitem{Siddique:2019gqh}
I.~Siddique, R.-j. Wang, S.~Pu, and Q.~Wang,
\newblock Phys. Rev. {\bf D99}, 114029 (2019), 1904.01807.

\bibitem{Peng:2022cya}
H.-H. Peng, S.~Wu, R.-j. Wang, D.~She, and S.~Pu,
\newblock Phys. Rev. D {\bf 107}, 096010 (2023), 2211.11286.

\bibitem{Manton:1983nd}
N.~S. Manton,
\newblock Phys. Rev. D {\bf 28}, 2019 (1983).

\bibitem{Klinkhamer:1984di}
F.~R. Klinkhamer and N.~S. Manton,
\newblock Phys. Rev. D {\bf 30}, 2212 (1984).

\bibitem{McLerran:1990de}
L.~D. McLerran, E.~Mottola, and M.~E. Shaposhnikov,
\newblock Phys. Rev. D {\bf 43}, 2027 (1991).

\bibitem{Arnold:1996dy}
P.~B. Arnold, D.~Son, and L.~G. Yaffe,
\newblock Phys. Rev. D {\bf 55}, 6264 (1997), hep-ph/9609481.

\bibitem{Liang:2004ph}
Z.-T. Liang and X.-N. Wang,
\newblock Phys. Rev. Lett. {\bf 94}, 102301 (2005), nucl-th/0410079,
\newblock [Erratum: Phys.Rev.Lett. 96, 039901 (2006)].

\bibitem{Liang:2004xn}
Z.-T. Liang and X.-N. Wang,
\newblock Phys. Lett. B {\bf 629}, 20 (2005), nucl-th/0411101.

\bibitem{Gao:2007bc}
J.-H. Gao {\em et~al.},
\newblock Phys. Rev. C {\bf 77}, 044902 (2008), 0710.2943.

\bibitem{Becattini:2007sr}
F.~Becattini, F.~Piccinini, and J.~Rizzo,
\newblock Phys. Rev. C {\bf 77}, 024906 (2008), 0711.1253.

\bibitem{Vilenkin:1979ui}
A.~Vilenkin,
\newblock Phys. Rev. {\bf D20}, 1807 (1979).

\bibitem{Banerjee:2008th}
N.~Banerjee {\em et~al.},
\newblock JHEP {\bf 01}, 094 (2011), 0809.2596.

\bibitem{Erdmenger:2008rm}
J.~Erdmenger, M.~Haack, M.~Kaminski, and A.~Yarom,
\newblock JHEP {\bf 01}, 055 (2009), 0809.2488.

\bibitem{Torabian:2009qk}
M.~Torabian and H.-U. Yee,
\newblock JHEP {\bf 08}, 020 (2009), 0903.4894.

\bibitem{Son:2009tf}
D.~T. Son and P.~Surowka,
\newblock Phys. Rev. Lett. {\bf 103}, 191601 (2009), 0906.5044.

\bibitem{Pu:2010as}
S.~Pu, J.-h. Gao, and Q.~Wang,
\newblock Phys. Rev. D {\bf 83}, 094017 (2011), 1008.2418.

\bibitem{Gao:2012ix}
J.-H. Gao, Z.-T. Liang, S.~Pu, Q.~Wang, and X.-N. Wang,
\newblock Phys. Rev. Lett. {\bf 109}, 232301 (2012), 1203.0725.

\bibitem{Son:2012zy}
D.~T. Son and N.~Yamamoto,
\newblock Phys. Rev. D {\bf 87}, 085016 (2013), 1210.8158.

\bibitem{Son:2012wh}
D.~T. Son and N.~Yamamoto,
\newblock Phys. Rev. Lett. {\bf 109}, 181602 (2012), 1203.2697.

\bibitem{Lin:2019ytz}
S.~Lin and A.~Shukla,
\newblock JHEP {\bf 06}, 060 (2019), 1901.01528.

\bibitem{Stephanov:2012ki}
M.~A. Stephanov and Y.~Yin,
\newblock Phys. Rev. Lett. {\bf 109}, 162001 (2012), 1207.0747.

\bibitem{Chen:2013iga}
J.-W. Chen, J.-y. Pang, S.~Pu, and Q.~Wang,
\newblock Phys. Rev. D {\bf 89}, 094003 (2014), 1312.2032.

\bibitem{Chen:2014cla}
J.-Y. Chen, D.~T. Son, M.~A. Stephanov, H.-U. Yee, and Y.~Yin,
\newblock Phys. Rev. Lett. {\bf 113}, 182302 (2014), 1404.5963.

\bibitem{Chen:2015gta}
J.-Y. Chen, D.~T. Son, and M.~A. Stephanov,
\newblock Phys. Rev. Lett. {\bf 115}, 021601 (2015), 1502.06966.

\bibitem{Chen:2012ca}
J.-W. Chen, S.~Pu, Q.~Wang, and X.-N. Wang,
\newblock Phys. Rev. Lett. {\bf 110}, 262301 (2013), 1210.8312.

\bibitem{Hidaka:2016yjf}
Y.~Hidaka, S.~Pu, and D.-L. Yang,
\newblock Phys. Rev. D {\bf 95}, 091901 (2017), 1612.04630.

\bibitem{Hidaka:2017auj}
Y.~Hidaka, S.~Pu, and D.-L. Yang,
\newblock Phys. Rev. D {\bf 97}, 016004 (2018), 1710.00278.

\bibitem{Huang:2018wdl}
A.~Huang, S.~Shi, Y.~Jiang, J.~Liao, and P.~Zhuang,
\newblock Phys. Rev. D {\bf 98}, 036010 (2018), 1801.03640.

\bibitem{Mueller:2017arw}
N.~Mueller and R.~Venugopalan,
\newblock Phys. Rev. D {\bf 96}, 016023 (2017), 1702.01233.

\bibitem{Mueller:2017lzw}
N.~Mueller and R.~Venugopalan,
\newblock Phys. Rev. D {\bf 97}, 051901 (2018), 1701.03331.

\bibitem{Copinger:2020nyx}
P.~Copinger and S.~Pu,
\newblock Int. J. Mod. Phys. A {\bf 35}, 2030015 (2020), 2008.03635.

\bibitem{Manuel:2013zaa}
C.~Manuel and J.~M. Torres-Rincon,
\newblock Phys. Rev. D {\bf 89}, 096002 (2014), 1312.1158.

\bibitem{Manuel:2014dza}
C.~Manuel and J.~M. Torres-Rincon,
\newblock Phys. Rev. D {\bf 90}, 076007 (2014), 1404.6409.

\bibitem{Carignano:2018gqt}
S.~Carignano, C.~Manuel, and J.~M. Torres-Rincon,
\newblock Phys. Rev. D {\bf 98}, 076005 (2018), 1806.01684.

\bibitem{Carignano:2019zsh}
S.~Carignano, C.~Manuel, and J.~M. Torres-Rincon,
\newblock Phys. Rev. D {\bf 102}, 016003 (2020), 1908.00561.

\bibitem{Lin:2018aon}
S.~Lin and L.~Yang,
\newblock Phys. Rev. D {\bf 98}, 114022 (2018), 1810.02979.

\bibitem{Pu:2014fva}
S.~Pu, S.-Y. Wu, and D.-L. Yang,
\newblock Phys. Rev. {\bf D91}, 025011 (2015), 1407.3168.

\bibitem{Pu:2014cwa}
S.~Pu, S.-Y. Wu, and D.-L. Yang,
\newblock Phys. Rev. {\bf D89}, 085024 (2014), 1401.6972.

\bibitem{Chen:2016xtg}
J.-W. Chen, T.~Ishii, S.~Pu, and N.~Yamamoto,
\newblock Phys. Rev. D {\bf 93}, 125023 (2016), 1603.03620.

\bibitem{Gorbar:2017toh}
E.~V. Gorbar, D.~O. Rybalka, and I.~A. Shovkovy,
\newblock Phys. Rev. D {\bf 95}, 096010 (2017), 1702.07791.

\bibitem{Dayi:2017xrr}
O.~F. Dayi and E.~Kilin\c{c}arslan,
\newblock Phys. Rev. D {\bf 96}, 043514 (2017), 1705.01267.

\bibitem{Ebihara:2017suq}
S.~Ebihara, K.~Fukushima, and S.~Pu,
\newblock Phys. Rev. D {\bf 96}, 016016 (2017), 1705.08611.

\bibitem{Huang:2017tsq}
A.~Huang, Y.~Jiang, S.~Shi, J.~Liao, and P.~Zhuang,
\newblock Phys. Lett. B {\bf 777}, 177 (2018), 1703.08856.

\bibitem{Hidaka:2018ekt}
Y.~Hidaka and D.-L. Yang,
\newblock Phys. Rev. D {\bf 98}, 016012 (2018), 1801.08253.

\bibitem{Yang:2018lew}
D.-L. Yang,
\newblock Phys. Rev. D {\bf 98}, 076019 (2018), 1807.02395.

\bibitem{Liu:2020ymh}
Y.-C. Liu and X.-G. Huang,
\newblock Nucl. Sci. Tech. {\bf 31}, 56 (2020), 2003.12482.

\bibitem{Sun:2016nig}
Y.~Sun, C.~M. Ko, and F.~Li,
\newblock Phys. Rev. {\bf C94}, 045204 (2016), 1606.05627.

\bibitem{Sun:2016mvh}
Y.~Sun and C.~M. Ko,
\newblock Phys. Rev. {\bf C95}, 034909 (2017), 1612.02408.

\bibitem{Sun:2017xhx}
Y.~Sun and C.~M. Ko,
\newblock Phys. Rev. {\bf C96}, 024906 (2017), 1706.09467.

\bibitem{Sun:2018bjl}
Y.~Sun and C.~M. Ko,
\newblock Phys. Rev. {\bf C99}, 011903 (2019), 1810.10359.

\bibitem{Liu:2019krs}
S.~Y.~F. Liu, Y.~Sun, and C.~M. Ko,
\newblock Phys. Rev. Lett. {\bf 125}, 062301 (2020), 1910.06774.

\bibitem{Zhou:2018rkh}
W.-H. Zhou and J.~Xu,
\newblock Phys. Rev. {\bf C98}, 044904 (2018), 1810.01030.

\bibitem{Zhou:2019jag}
W.-H. Zhou and J.~Xu,
\newblock Phys. Lett. {\bf B798}, 134932 (2019), 1904.01834.

\bibitem{Liu:2018xip}
Y.-C. Liu, L.-L. Gao, K.~Mameda, and X.-G. Huang,
\newblock Phys. Rev. D {\bf 99}, 085014 (2019), 1812.10127.

\bibitem{Hayata:2020sqz}
T.~Hayata, Y.~Hidaka, and K.~Mameda,
\newblock JHEP {\bf 05}, 023 (2021), 2012.12494.

\bibitem{Yamamoto:2020zrs}
N.~Yamamoto and D.-L. Yang,
\newblock Astrophys. J. {\bf 895}, 56 (2020), 2002.11348.

\bibitem{Yang:2020mtz}
S.-Z. Yang, J.-H. Gao, Z.-T. Liang, and Q.~Wang,
\newblock Phys. Rev. D {\bf 102}, 116024 (2020), 2003.04517.

\bibitem{Mameda:2023ueq}
K.~Mameda,
\newblock Phys. Rev. D {\bf 108}, 016001 (2023), 2305.02134.

\bibitem{Kamada:2022nyt}
K.~Kamada, N.~Yamamoto, and D.-L. Yang,
\newblock Prog. Part. Nucl. Phys. {\bf 129}, 104016 (2023), 2207.09184.

\bibitem{Yamamoto:2015gzz}
N.~Yamamoto,
\newblock Phys. Rev. D {\bf 93}, 065017 (2016), 1511.00933.

\bibitem{Yamamoto:2021hjs}
N.~Yamamoto and D.-L. Yang,
\newblock Phys. Rev. D {\bf 104}, 123019 (2021), 2103.13159.

\bibitem{Yamamoto:2022yva}
N.~Yamamoto and D.-L. Yang,
\newblock Phys. Rev. Lett. {\bf 131}, 012701 (2023), 2211.14465.

\bibitem{Yamamoto:2023okm}
N.~Yamamoto and D.-L. Yang,
\newblock (2023), 2308.08257.

\bibitem{Armitage:2017cjs}
N.~P. Armitage, E.~J. Mele, and A.~Vishwanath,
\newblock Rev. Mod. Phys. {\bf 90}, 015001 (2018), 1705.01111.

\bibitem{Gorbar:2021ebc}
E.~V. Gorbar, V.~A. Miransky, I.~A. Shovkovy, and P.~O. Sukhachov,
\newblock {\em {Electronic Properties of Dirac and Weyl Semimetals}} (World Scientific, Singapore, 2021).

\bibitem{Gao:2019znl}
J.-H. Gao and Z.-T. Liang,
\newblock Phys. Rev. D {\bf 100}, 056021 (2019), 1902.06510.

\bibitem{Weickgenannt:2019dks}
N.~Weickgenannt, X.-L. Sheng, E.~Speranza, Q.~Wang, and D.~H. Rischke,
\newblock Phys. Rev. D {\bf 100}, 056018 (2019), 1902.06513.

\bibitem{Weickgenannt:2020aaf}
N.~Weickgenannt, E.~Speranza, X.-l. Sheng, Q.~Wang, and D.~H. Rischke,
\newblock Phys. Rev. Lett. {\bf 127}, 052301 (2021), 2005.01506.

\bibitem{Hattori:2019ahi}
K.~Hattori, Y.~Hidaka, and D.-L. Yang,
\newblock Phys. Rev. D {\bf 100}, 096011 (2019), 1903.01653.

\bibitem{Yang:2020hri}
D.-L. Yang, K.~Hattori, and Y.~Hidaka,
\newblock JHEP {\bf 07}, 070 (2020), 2002.02612.

\bibitem{Liu:2020flb}
Y.-C. Liu, K.~Mameda, and X.-G. Huang,
\newblock Chin. Phys. C {\bf 44}, 094101 (2020), 2002.03753,
\newblock [Erratum: Chin.Phys.C 45, 089001 (2021)].

\bibitem{Weickgenannt:2021cuo}
N.~Weickgenannt, E.~Speranza, X.-l. Sheng, Q.~Wang, and D.~H. Rischke,
\newblock Phys. Rev. D {\bf 104}, 016022 (2021), 2103.04896.

\bibitem{Sheng:2021kfc}
X.-L. Sheng, N.~Weickgenannt, E.~Speranza, D.~H. Rischke, and Q.~Wang,
\newblock Phys. Rev. D {\bf 104}, 016029 (2021), 2103.10636.

\bibitem{Wang:2019moi}
Z.~Wang, X.~Guo, S.~Shi, and P.~Zhuang,
\newblock Phys. Rev. D {\bf 100}, 014015 (2019), 1903.03461.

\bibitem{Manuel:2021oah}
C.~Manuel and J.~M. Torres-Rincon,
\newblock Phys. Rev. D {\bf 103}, 096022 (2021), 2101.05832.

\bibitem{Lin:2021mvw}
S.~Lin,
\newblock Phys. Rev. D {\bf 105}, 076017 (2022), 2109.00184.

\bibitem{Yang:2021fea}
D.-L. Yang,
\newblock JHEP {\bf 06}, 140 (2022), 2112.14392.

\bibitem{Ma:2022ins}
S.-X. Ma and J.-H. Gao,
\newblock (2022), 2209.10737.

\bibitem{Huang:2020kik}
X.-G. Huang, P.~Mitkin, A.~V. Sadofyev, and E.~Speranza,
\newblock JHEP {\bf 10}, 117 (2020), 2006.03591.

\bibitem{Hattori:2020gqh}
K.~Hattori, Y.~Hidaka, N.~Yamamoto, and D.-L. Yang,
\newblock JHEP {\bf 02}, 001 (2021), 2010.13368.

\bibitem{Mameda:2022ojk}
K.~Mameda, N.~Yamamoto, and D.-L. Yang,
\newblock Phys. Rev. D {\bf 105}, 096019 (2022), 2203.08449.

\bibitem{Mrowczynski:1992hq}
S.~Mrowczynski and U.~W. Heinz,
\newblock Annals Phys. {\bf 229}, 1 (1994).

\bibitem{Blaizot:1999xk}
J.-P. Blaizot and E.~Iancu,
\newblock Nucl. Phys. B {\bf 557}, 183 (1999), hep-ph/9903389.

\bibitem{Blaizot:2001nr}
J.-P. Blaizot and E.~Iancu,
\newblock Phys. Rept. {\bf 359}, 355 (2002), hep-ph/0101103.

\bibitem{Becattini:2020ngo}
F.~Becattini and M.~A. Lisa,
\newblock Ann. Rev. Nucl. Part. Sci. {\bf 70}, 395 (2020), 2003.03640.

\bibitem{Becattini:2022zvf}
F.~Becattini,
\newblock Rept. Prog. Phys. {\bf 85}, 122301 (2022), 2204.01144.

\bibitem{STAR:2017ckg}
STAR, L.~Adamczyk {\em et~al.},
\newblock Nature {\bf 548}, 62 (2017), 1701.06657.

\bibitem{STAR:2019erd}
STAR, J.~Adam {\em et~al.},
\newblock Phys. Rev. Lett. {\bf 123}, 132301 (2019), 1905.11917.

\bibitem{ALICE:2019aid}
ALICE, S.~Acharya {\em et~al.},
\newblock Phys. Rev. Lett. {\bf 125}, 012301 (2020), 1910.14408.

\bibitem{STAR:2020xbm}
STAR, J.~Adam {\em et~al.},
\newblock Phys. Rev. Lett. {\bf 126}, 162301 (2021), 2012.13601.

\bibitem{STAR:2022fan}
STAR, M.~S. Abdallah {\em et~al.},
\newblock Nature {\bf 614}, 244 (2023), 2204.02302.

\bibitem{ALICE:2023jad}
ALICE, S.~Acharya {\em et~al.},
\newblock Phys. Rev. Lett. {\bf 131}, 042303 (2023).

\bibitem{Becattini:2007nd}
F.~Becattini and F.~Piccinini,
\newblock Annals Phys. {\bf 323}, 2452 (2008), 0710.5694.

\bibitem{Becattini:2013fla}
F.~Becattini, V.~Chandra, L.~Del~Zanna, and E.~Grossi,
\newblock Annals Phys. {\bf 338}, 32 (2013), 1303.3431.

\bibitem{Becattini:2020sww}
F.~Becattini,
\newblock Lect. Notes Phys. {\bf 987}, 15 (2021), 2004.04050.

\bibitem{Fang:2016vpj}
R.-h. Fang, L.-g. Pang, Q.~Wang, and X.-n. Wang,
\newblock Phys. Rev. C {\bf 94}, 024904 (2016), 1604.04036.

\bibitem{Yi:2021ryh}
C.~Yi, S.~Pu, and D.-L. Yang,
\newblock Phys. Rev. C {\bf 104}, 064901 (2021), 2106.00238.

\bibitem{Liu:2021nyg}
Y.-C. Liu and X.-G. Huang,
\newblock Sci. China Phys. Mech. Astron. {\bf 65}, 272011 (2022), 2109.15301.

\bibitem{Karpenko:2016jyx}
I.~Karpenko and F.~Becattini,
\newblock Eur. Phys. J. C {\bf 77}, 213 (2017), 1610.04717.

\bibitem{Becattini:2017gcx}
F.~Becattini and I.~Karpenko,
\newblock Phys. Rev. Lett. {\bf 120}, 012302 (2018), 1707.07984.

\bibitem{Xie:2017upb}
Y.~Xie, D.~Wang, and L.~P. Csernai,
\newblock Phys. Rev. C {\bf 95}, 031901 (2017), 1703.03770.

\bibitem{Pang:2016igs}
L.-G. Pang, H.~Petersen, Q.~Wang, and X.-N. Wang,
\newblock Phys. Rev. Lett. {\bf 117}, 192301 (2016), 1605.04024.

\bibitem{Li:2017slc}
H.~Li, L.-G. Pang, Q.~Wang, and X.-L. Xia,
\newblock Phys. Rev. C {\bf 96}, 054908 (2017), 1704.01507.

\bibitem{Wei:2018zfb}
D.-X. Wei, W.-T. Deng, and X.-G. Huang,
\newblock Phys. Rev. C {\bf 99}, 014905 (2019), 1810.00151.

\bibitem{Ryu:2021lnx}
S.~Ryu, V.~Jupic, and C.~Shen,
\newblock Phys. Rev. C {\bf 104}, 054908 (2021), 2106.08125.

\bibitem{Shi:2017wpk}
S.~Shi, K.~Li, and J.~Liao,
\newblock Phys. Lett. B {\bf 788}, 409 (2019), 1712.00878.

\bibitem{Fu:2020oxj}
B.~Fu, K.~Xu, X.-G. Huang, and H.~Song,
\newblock Phys. Rev. C {\bf 103}, 024903 (2021), 2011.03740.

\bibitem{Wu:2022mkr}
X.-Y. Wu, C.~Yi, G.-Y. Qin, and S.~Pu,
\newblock Phys. Rev. C {\bf 105}, 064909 (2022), 2204.02218.

\bibitem{Alzhrani:2022dpi}
S.~Alzhrani, S.~Ryu, and C.~Shen,
\newblock Phys. Rev. C {\bf 106}, 014905 (2022), 2203.15718.

\bibitem{Xu:2022hql}
K.~Xu, F.~Lin, A.~Huang, and M.~Huang,
\newblock Phys. Rev. D {\bf 106}, L071502 (2022), 2205.02420.

\bibitem{Xia:2018tes}
X.-L. Xia, H.~Li, Z.-B. Tang, and Q.~Wang,
\newblock Phys. Rev. C {\bf 98}, 024905 (2018), 1803.00867.

\bibitem{Liu:2020dxg}
S.~Y.~F. Liu and Y.~Yin,
\newblock Phys. Rev. D {\bf 104}, 054043 (2021), 2006.12421.

\bibitem{Liu:2021uhn}
S.~Y.~F. Liu and Y.~Yin,
\newblock JHEP {\bf 07}, 188 (2021), 2103.09200.

\bibitem{Becattini:2021suc}
F.~Becattini, M.~Buzzegoli, and A.~Palermo,
\newblock Phys. Lett. B {\bf 820}, 136519 (2021), 2103.10917.

\bibitem{Becattini:2021iol}
F.~Becattini, M.~Buzzegoli, G.~Inghirami, I.~Karpenko, and A.~Palermo,
\newblock Phys. Rev. Lett. {\bf 127}, 272302 (2021), 2103.14621.

\bibitem{Fu:2021pok}
B.~Fu, S.~Y.~F. Liu, L.~Pang, H.~Song, and Y.~Yin,
\newblock Phys. Rev. Lett. {\bf 127}, 142301 (2021), 2103.10403.

\bibitem{Florkowski:2021xvy}
W.~Florkowski, A.~Kumar, A.~Mazeliauskas, and R.~Ryblewski,
\newblock Phys. Rev. C {\bf 105}, 064901 (2022), 2112.02799.

\bibitem{Sun:2021nsg}
Y.~Sun, Z.~Zhang, C.~M. Ko, and W.~Zhao,
\newblock Phys. Rev. C {\bf 105}, 034911 (2022), 2112.14410.

\bibitem{Sheng:2019kmk}
X.-L. Sheng, L.~Oliva, and Q.~Wang,
\newblock Phys. Rev. D {\bf 101}, 096005 (2020), 1910.13684.

\bibitem{Sheng:2020ghv}
X.-L. Sheng, Q.~Wang, and X.-N. Wang,
\newblock Phys. Rev. D {\bf 102}, 056013 (2020), 2007.05106.

\bibitem{Sheng:2022ffb}
X.-L. Sheng, L.~Oliva, Z.-T. Liang, Q.~Wang, and X.-N. Wang,
\newblock (2022), 2206.05868.

\bibitem{Sheng:2022wsy}
X.-L. Sheng, L.~Oliva, Z.-T. Liang, Q.~Wang, and X.-N. Wang,
\newblock Phys. Rev. Lett. {\bf 131}, 042304 (2023), 2205.15689.

\bibitem{Sheng:2023urn}
X.-L. Sheng, S.~Pu, and Q.~Wang,
\newblock Phys. Rev. C {\bf 108}, 054902 (2023), 2308.14038.

\bibitem{Muller:2021hpe}
B.~M\"uller and D.-L. Yang,
\newblock Phys. Rev. D {\bf 105}, L011901 (2022), 2110.15630,
\newblock [Erratum: Phys.Rev.D 106, 039904 (2022)].

\bibitem{Kumar:2022ylt}
A.~Kumar, B.~M\"uller, and D.-L. Yang,
\newblock Phys. Rev. D {\bf 107}, 076025 (2023), 2212.13354.

\bibitem{Kumar:2023ghs}
A.~Kumar, B.~M\"uller, and D.-L. Yang,
\newblock Phys. Rev. D {\bf 108}, 016020 (2023), 2304.04181.

\bibitem{Sheng:2022ssp}
X.-L. Sheng, S.-Y. Yang, Y.-L. Zou, and D.~Hou,
\newblock (2022), 2209.01872.

\bibitem{Wagner:2022gza}
D.~Wagner, N.~Weickgenannt, and E.~Speranza,
\newblock Phys. Rev. Res. {\bf 5}, 013187 (2023), 2207.01111.

\bibitem{Fu:2023qht}
B.~Fu, F.~Gao, Y.~Liu, and H.~Song,
\newblock (2023), 2308.07936.

\bibitem{Li:2022vmb}
F.~Li and S.~Y.~F. Liu,
\newblock (2022), 2206.11890.

\bibitem{Wang:2020pej}
Z.~Wang, X.~Guo, and P.~Zhuang,
\newblock Eur. Phys. J. C {\bf 81}, 799 (2021), 2009.10930.

\bibitem{Wang:2021qnt}
Z.~Wang and P.~Zhuang,
\newblock (2021), 2105.00915.

\bibitem{Weickgenannt:2022zxs}
N.~Weickgenannt, D.~Wagner, E.~Speranza, and D.~H. Rischke,
\newblock Phys. Rev. D {\bf 106}, 096014 (2022), 2203.04766.

\bibitem{Weickgenannt:2022qvh}
N.~Weickgenannt, D.~Wagner, E.~Speranza, and D.~H. Rischke,
\newblock Phys. Rev. D {\bf 106}, L091901 (2022), 2208.01955.

\bibitem{Wagner:2022amr}
D.~Wagner, N.~Weickgenannt, and D.~H. Rischke,
\newblock Phys. Rev. D {\bf 106}, 116021 (2022), 2210.06187.

\bibitem{Fang:2022ttm}
S.~Fang, S.~Pu, and D.-L. Yang,
\newblock Phys. Rev. D {\bf 106}, 016002 (2022), 2204.11519.

\bibitem{Wang:2022yli}
Z.~Wang,
\newblock Phys. Rev. D {\bf 106}, 076011 (2022), 2205.09334.

\bibitem{Lin:2022tma}
S.~Lin and Z.~Wang,
\newblock JHEP {\bf 12}, 030 (2022), 2206.12573.

\bibitem{Peskin:1995ev}
M.~E. Peskin and D.~V. Schroeder,
\newblock {\em {An Introduction to quantum field theory}} (Addison-Wesley, Reading, USA, 1995).

\bibitem{Elze:1986qd}
H.~T. Elze, M.~Gyulassy, and D.~Vasak,
\newblock Nucl. Phys. {\bf B276}, 706 (1986).

\bibitem{Vasak:1987um}
D.~Vasak, M.~Gyulassy, and H.~T. Elze,
\newblock Annals Phys. {\bf 173}, 462 (1987).

\bibitem{Itzykson:1980rh}
C.~Itzykson and J.~B. Zuber,
\newblock {\em {Quantum Field Theory}}International Series In Pure and Applied Physics (McGraw-Hill, New York, 1980).

\bibitem{BLT_spin}
V.~Bargmann, L.~Michel, and V.~L. Telegdi,
\newblock Phys. Rev. Lett. {\bf 2}, 435 (1959).

\bibitem{Serot:1992ti}
B.~D. Serot,
\newblock Rept. Prog. Phys. {\bf 55}, 1855 (1992).

\bibitem{Serot:1997xg}
B.~D. Serot and J.~D. Walecka,
\newblock Int. J. Mod. Phys. E {\bf 6}, 515 (1997), nucl-th/9701058.

\bibitem{Muller:2018ibh}
B.~M\"uller and A.~Sch\"afer,
\newblock Phys. Rev. D {\bf 98}, 071902 (2018), 1806.10907.

\bibitem{Guo:2019joy}
Y.~Guo, S.~Shi, S.~Feng, and J.~Liao,
\newblock Phys. Lett. B {\bf 798}, 134929 (2019), 1905.12613.

\bibitem{Buzzegoli:2022qrr}
M.~Buzzegoli,
\newblock Nucl. Phys. A {\bf 1036}, 122674 (2023), 2211.04549.

\bibitem{Bellac:2011kqa}
M.~L. Bellac,
\newblock {\em {Thermal Field Theory}}Cambridge Monographs on Mathematical Physics (Cambridge University Press, 2011).

\bibitem{Fu:2022myl}
B.~Fu, L.~Pang, H.~Song, and Y.~Yin,
\newblock (2022), 2201.12970.

\bibitem{Yi:2021unq}
C.~Yi, S.~Pu, J.-H. Gao, and D.-L. Yang,
\newblock (2021), 2112.15531.

\bibitem{Yi:2023tgg}
C.~Yi {\em et~al.},
\newblock (2023), 2304.08777.

\bibitem{Golkar:2012kb}
S.~Golkar and D.~T. Son,
\newblock JHEP {\bf 02}, 169 (2015), 1207.5806.

\bibitem{Hou:2012xg}
D.-F. Hou, H.~Liu, and H.-c. Ren,
\newblock Phys. Rev. D {\bf 86}, 121703 (2012), 1210.0969.

\bibitem{Xia:2020tyd}
X.-L. Xia, H.~Li, X.-G. Huang, and H.~Zhong~Huang,
\newblock Phys. Lett. B {\bf 817}, 136325 (2021), 2010.01474.

\end{thebibliography}

\end{document}